\documentclass[a4paper,fleqn,usenatbib]{mnras}

\usepackage[T1]{fontenc}
\usepackage{ae,aecompl}
\usepackage{epsfig}
\usepackage{epstopdf}

\usepackage{graphicx}	% Including figure files
\usepackage{amsmath}	% Advanced maths commands
\usepackage{amssymb}	% Extra maths symbols
\usepackage{float}
\usepackage{caption}
\usepackage{subcaption}
\usepackage{color}

\usepackage{epstopdf}
\usepackage{float}
\restylefloat{table}
\usepackage{amsmath}
\usepackage{mathtools}
\newcommand{\be}{\begin{equation}}  
\newcommand{\ee}{\end{equation}}
\newcommand{\ba}{\begin{eqnarray*}}\newcommand{\ea}{\end{eqnarray*}}
\newcommand{\bm}[1]{\mbox{\boldmath{$#1$}}}   %this is bold italic for MNRAS

\usepackage{siunitx}    % for numbers with units
\DeclareSIUnit{\degree}{deg} % we want degrees written with 'deg' and
\DeclareSIUnit{\arcmin}{arcmin} % we want degrees written  'arcmin' and
\DeclareSIUnit{\arcsec}{arcsec} % we want arcsec written  'arcsec' and
\definecolor{purple}{RGB}{76, 0,153}
\newcommand{\am}[1]{\textcolor{black}{ #1}}
\newcommand{\ch}[1]{\textcolor{black}{ #1}}

\title[KiDS-$i$-800]{KiDS-i-800: Comparing weak gravitational lensing measurements from same-sky surveys}

\author[A. Amon et al.]
{A. Amon$^1$\thanks{Email: aamon@roe.ac.uk}, 
C. Heymans$^1$, 
D. Klaes$^2$,
T. Erben$^2$,
C. Blake$^3$,
H. Hildebrandt$^2$,
\newauthor
H. Hoekstra$^4$,
K. Kuijken$^4$,
L. Miller$^{5}$,
C.B. Morrison$^{6}$,
A. Choi$^{7}$,
J.T.A. de Jong$^{4,8}$,
\newauthor
K. Glazebrook$^{3}$,
N. Irisarri$^4$,
B. Joachimi$^9$,
S. Joudaki$^{5}$,
A. Kannawadi$^4$,
\newauthor
C. Lidman$^{10}$,
N. Napolitano$^{11}$,
D. Parkinson$^{12}$,
P. Schneider$^2$,
E. van Uitert$^9$,
\newauthor
M. Viola$^4$,
and
C. Wolf$^{13}$
\\ 
$^1$Institute for Astronomy, University of Edinburgh, Royal Observatory, Blackford Hill, Edinburgh EH9 3HJ, UK\\
$^2$Argelander-Institut f\"ur Astronomie, Auf dem H\"ugel 71, 53121 Bonn, Germany\\
$^3$Centre for Astrophysics \& Supercomputing, Swinburne University of Technology, PO Box 218, Hawthorn, VIC 3122, Australia\\
$^4$Leiden Observatory, Leiden University, Niels Bohrweg 2, 2333 CA Leiden, the Netherlands\\
$^5$Department of Physics, University of Oxford, Denys Wilkinson Building, Keble Road, Oxford OX1 3RH, UK\\
$^{6}$Department of Astronomy, University of Washington, Box 351580, Seattle, WA 98195, USA\\
$^{7}$Center for Cosmology and AstroParticle Physics, The Ohio State University, 191 West Woodruff Avenue, Columbus, OH 43210, USA\\
$^{8}$Kapteyn Astronomical Institute, University of Groningen, 9700AD Groningen, the Netherlands\\
$^9$Department of Physics and Astronomy, University College London, Gower Street, London WC1E 6BT, UK\\
$^{10}$Australian Astronomical Observatory, PO Box 915, North Ryde, NSW 1670, Australia\\
$^{11}$INAF -- Osservatorio Astronomico di Capodimonte, Via Moiariello 16, 80131 Napoli, Italy\\
$^{12}$School of Mathematics and Physics, University of Queensland, Brisbane, QLD 4072, Australia\\
$^{13}$Research School of Astronomy and Astrophysics, Australian National University, Canberra, ACT 2611, Australia\\
\vspace{-0.5cm}  %necessary so a LaTeX crash doesn't occur...
}

\date{Accepted XXX. Received YYY; in original form ZZZ}

\pubyear{2017}

\begin{document}
\label{firstpage}
\pagerange{\pageref{firstpage}--\pageref{lastpage}}
\maketitle

% Abstract of the paper
\begin{abstract}
We present a weak gravitational lensing analysis of 815 square degree of $i$-band imaging from the Kilo-Degree Survey (KiDS-$i$-800).   
In contrast to the deep $r$-band observations, which take priority during excellent seeing conditions and form the primary KiDS dataset (KiDS-$r$-450), the complementary yet shallower KiDS-$i$-800 spans a wide range of observing conditions. The overlapping KiDS-$i$-800 and KiDS-$r$-450 imaging therefore provides a unique opportunity to assess the robustness of weak lensing measurements. In our analysis we introduce two new `null' tests.  The `nulled' two-point shear correlation function uses a matched catalogue to show that the calibrated KiDS-$i$-800 and KiDS-$r$-450 shear measurements agree at the level of $1 \pm 4$\%.  We use five galaxy lens samples to determine a `nulled' galaxy-galaxy lensing signal from the full KiDS-$i$-800 and KiDS-$r$-450 surveys and find that the measurements agree to $7 \pm 5$\% when the KiDS-$i$-800 source redshift distribution is calibrated using either spectroscopic redshifts, or the 30-band photometric redshifts from the COSMOS survey.
\end{abstract}

\begin{keywords}
gravitational lensing: weak -- surveys, cosmology: observations -- galaxies: photometry  

\vspace{0.4cm}
\end{keywords}

%%%%%%%%%%%%%%%%%%%%%%%%%%%%%%%%%%%%%%%%%%%%%%%%%%

%%%%%%%%%%%%%%%%% BODY OF PAPER %%%%%%%%%%%%%%%%%%

\section{Introduction}
Weak gravitational lensing provides a powerful way to measure the total matter distribution. Light rays from background `source' galaxies are deflected by massive foreground structures and the statistical measurement of these distortions allows for the detection of the gravitational potential of the foreground `lenses'. This gives information about cosmic geometry and the growth of large-scale structures in the Universe, without any prior assumptions about dark matter or galaxy bias \citep{HoekstraJain2008, Kilbinger2015}.   

As the lensing distortion of a single galaxy is typically much smaller than the intrinsic ellipticity, measurements require wide-area, deep, high-quality optical images. Some large optical surveys that have been exploited for weak lensing studies in the last decade are the Sloan Digital Sky Survey \citep[SDSS;][]{Mandelbaum/etal:2005}, the Canada-France-Hawaii Telescope Legacy Survey \citep[CFHTLenS;][]{Heymans2012}, the Deep Lens Survey \citep[DLS;][]{Wittman2002} and the Red Sequence Cluster Survey \citep[RCS and RCSLenS;][]{vanuitert/etal:2011, Hildebrandt/etal:2016}, as well as the on-going Dark Energy Survey \citep[DES;][]{jarvis/etal:2016}, the Hyper Supreme-Cam Survey \citep[HSC;][]{HSC2017} and the Kilo-Degree Survey \citep[KiDS;][]{Kuijken/etal:2015}. The non-trivial nature of weak lensing measurements, owing to their susceptibility to various systematics, stimulates a need for consistency checks between the lensing signals derived from unique datasets.

\begin{figure*}
	\includegraphics[width=\textwidth]{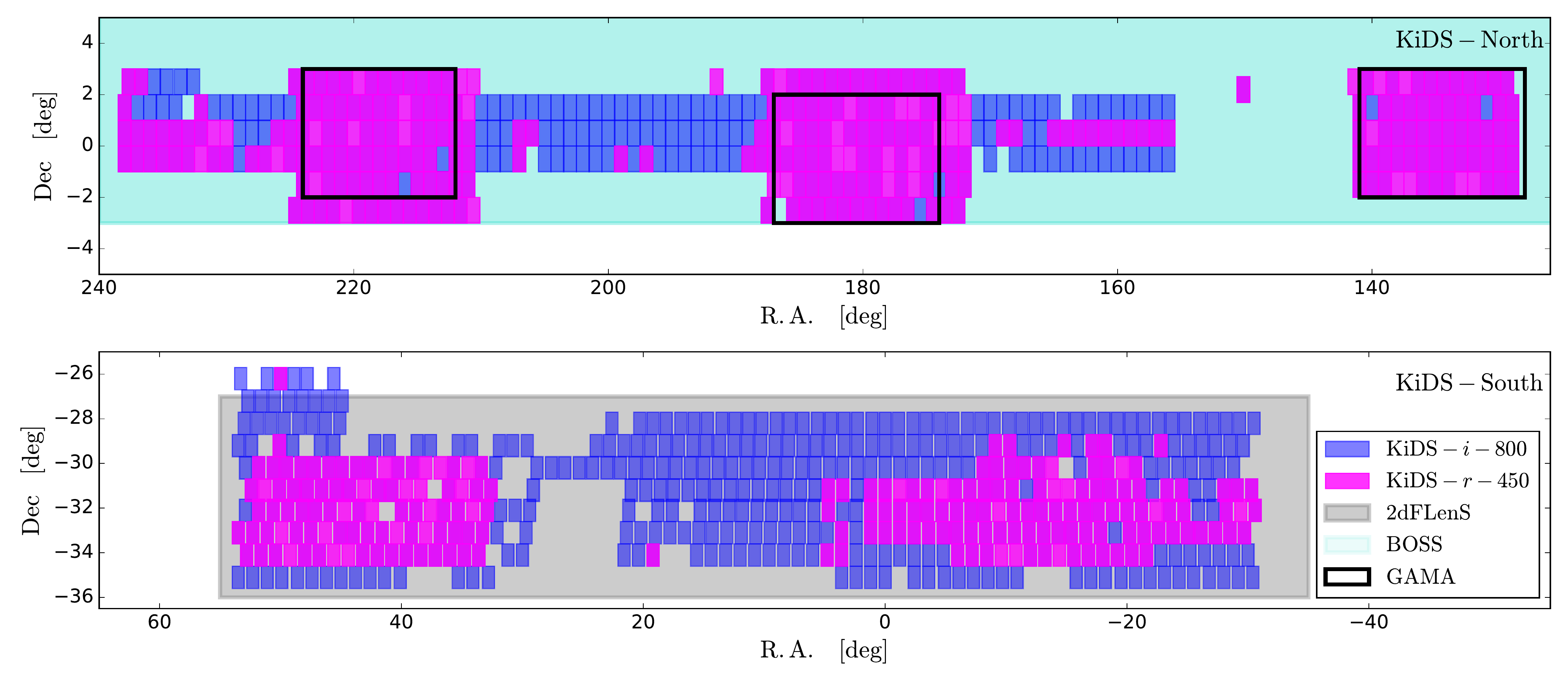}
    \caption{\label{fig:map}KiDS-$i$-800 survey footprint. Each purple box corresponds to a single KiDS $i$-band pointing of \SI{1}{\square\degree} and for comparison, each overplotted pink box corresponds to a KiDS-$r$-450 pointing. The cyan region indicates the BOSS spectroscopic coverage and the grey region in the South indicates the 2dFLenS spectroscopic coverage. The black outlined rectangles are the GAMA spectroscopic fields that overlap with the KiDS North field.}
\end{figure*}

This paper presents the first lensing results using \SI{815}{\square\degree} of KiDS $i$-band imaging (hereafter referred to as KiDS-$i$-800), along with the first large-scale lensing analysis of two overlapping imaging surveys, where we make a detailed comparison to lensing measurements from \SI{450}{\square\degree} of $r$-band imaging (hereafter referred to as KiDS-$r$-450). KiDS is a multi-band, large-scale, imaging survey that seeks to unveil the properties of the evolving dark universe by tracing the density of clustered matter using weak lensing tomography.   Its observations are taken in four broad-band filters (\textit{ugri}) using the OmegaCAM at the VLT Survey Telescope (VST) at the European Southern Observatory's Paranal Observatory \citep{deJong/etal:2013, Kuijken/etal:2015}. Details of the KiDS-$r$-450 data reduction and subsequent cosmic shear analysis are presented in \cite{Hildebrandt/etal:2017}.

The KiDS observing strategy is fashioned to provide optimal imaging for shape measurements in the $r$-band where the data are homogeneous in terms of limiting depth and low atmospheric seeing.  In contrast, the $i$-band imaging encompasses a wide range of depth owing to its varied seeing conditions and sky brightness. 
Though these $i$-band images are highly variable in quality, \am{if the redshift distribution can be sufficiently calibrated, the cosmological range in scale probed by the data available could be useful for cross-correlation studies} such as galaxy-shear cross correlation, or galaxy-galaxy lensing \citep{HoekstraYeeGladder2004,Mandelbaum/etal:2005} and galaxy-CMB lensing \citep[for an application of this technique see][]{Hand/etal:2013}. In addition, galaxy-galaxy lensing can be combined with galaxy clustering to shed light on the growth of structure \citep{Leauthaud2017, Kwan2017}, as well as with redshift-space distortions to test gravity \citep{Blake2016eg, Alam2017}. 

Furthermore, the areal overlap between these two shape catalogues allows for a unique consistency test of our shear and redshift estimates across different observing conditions and depths. \am{The galaxy-galaxy lensing measurement of the excess surface mass density is invariant to the redshift of the source samples.  As this measurement is also essentially insensitive to the assumed cosmology, this allows for a powerful systematic test \citep{Mandelbaum/etal:2005, Heymans2012}.  The excess surface mass density statistic is, however, sensitive to both errors in the shear calibration and redshift distributions and therefore cannot distinguish between these two sources of systematic error, providing only a joint calibration of the two effects.   As such we employ a complementary `nulled' two-point shear correlation test using a matched galaxy sample to independently identify calibration errors in the shear measurement.}

The paper is organised as follows. Section \ref{sec:data} presents the survey outline, details the shape measurement pipeline and reviews the $i$-band data quality. An outline of the various methods for estimating the redshift distribution is given in Section \ref{sec:zdata}. Section \ref{sec:ir} compares the KiDS-$i$-800 dataset to the KiDS-$r$-450 dataset in terms of the nulled two-point shear correlation function and the the nulled galaxy-galaxy lensing signal of the datasets. That is, we explore the difference in shear only for galaxies measured in both bands, as well as the shape and photometry of all galaxies in each band. Finally, we summarise the outcomes of this study and the outlook in Section \ref{sec:conc}. In the Appendices we detail the differences in the data reduction process between KiDS-$r$-450 and KiDS-$i$-800 (Appendix~\ref{app:QC}), the selection criteria we apply for galaxy-galaxy lensing (Appendix~\ref{app:cuts}), a comparison of our star selection with the {\it Gaia} survey (Appendix~\ref{app:Gaia}), the corrections applied to the galaxy-galaxy lensing signal (Appendix~\ref{app:gglcorr}) and the computation of the analytical covariance for the nulled two-point shear correlation function (Appendix~\ref{app:covderiv}).

\section{Shear data} \label{sec:data} Both the OmegaCAM and the VST are \am{specifically} designed to be
optimally suited for uniform and high-quality images over the
one-square degree field of view. For a particular field in any of the
$(u)gri$ filters, observations comprise (four) five dithered exposures
in immediate succession.

The KiDS deep $r$-band images are observed in dark time with a total
exposure time of 1800 seconds during the best-seeing conditions
with FWHM$<$0.9 arcsec and a median FWHM of 0.66 arcsec 
\citep[for the public data release, see][]{deJong2017}. The $r$-band
observations thus provide the primary images for weak lensing analyses
\citep{Kuijken/etal:2015, Hildebrandt/etal:2017}. The $u$-band and
$g$-band also use dark time with weaker seeing constraints. In
contrast the $i$-band data is observed in bright time, with a shorter
total exposure time of 1200 seconds, over a range of seeing
conditions satisfying FWHM$<$ 1.2 arcsec, in this case with a
median FWHM of 0.79 arcsec. The data collection rate for this
variable seeing bright time data therefore surpasses that of the $ugr$
data. At present, the full 1500 square degree KiDS footprint is
essentially complete in $i$-band, in contrast to the completed $ugr$
imaging which, as of January 2018, spans seventy percent of the final survey
area. This enhanced areal $i$-band coverage, in comparison to the
multi-band imaging, thus motivated our investigation into its use for
weak lensing analyses.

The KiDS-$i$-800 dataset consists of all fields observed in the
\textit{i}-band filter before December 14th, 2014. These fields were
analysed and subjected to a series of strict quality-control tests
during the data reduction, as presented in Appendix~\ref{app:QC}. This
selection resulted in a dataset of 815 fields, hence the name
`KiDS-$i$-800'. Out of these 815 fields, 381 have also undergone a
weak lensing analysis in the $r$-band as part of the KiDS-$r$-450 data
release.

Figure~\ref{fig:map} shows the KiDS-$i$-800 coverage and the
overlapping spectroscopic area with the Baryon Oscillation
Spectroscopic Survey \citep[][BOSS]{Dawson/etal:2013} and the Galaxy
and Mass Assembly survey \citep[][GAMA]{Driver/etal:2011} in the
North. In the South, the 2-degree Field Lensing Survey
\citep[][2dFLenS]{Blake/etal:2016} is specifically designed as the
spectroscopic follow-up of KiDS. The complete spectroscopic overlap
between these datasets renders KiDS-$i$-800 an optimal survey for
cross-correlation studies, such as galaxy-galaxy lensing.

\subsection{Data reduction and Object Detection}

The \textsc{theli} pipeline \citep{Erben2005, Schirmer2013}, developed
from CARS \citep{Erben/etal:2009} and CFHTLenS \citep{Erben/etal:2013}
and fully described in \cite{Kuijken/etal:2015}, was used for a
lensing-quality reduction of the KiDS-$i$-800 dataset. The basis of
our \textsc{theli} processing starts with the removal of the
instrumental signatures of OmegaCAM data provided by the ESO archive.
\am{Next, photometric zero-points, atmospheric extinction coefficients and colour terms are estimated per complete processing run and where necessary, we correct the OmegaCAM data for any evidence of electronic cross-talk between detectors on the images and fringing.  Finally the sky is subtracted from each CCD in every exposure. Individual sky background models are created by \textsc{SExtractor}, adopting a filtering scale (BACKSIZE) of 512 pixels.}
All images from each KiDS pointing are astrometrically calibrated
against the SDSS Data Release 12 \citep{Alam/etal:2015} where available
and the 2MASS catalogue \citep{Skrutskie/etal:2006} otherwise. These calibrated
images are co-added with a weighted mean algorithm.
\textsc{SExtractor} \citep{Bertin/etal:1996} is run on the co-added
images to generate the source catalogue for the lensing measurements.
Masks that cover image defects, reflections and ghosts, are also
created \citep[see Section 3.4 of][for more
details]{Kuijken/etal:2015}. An account of the differences between the
data reduction for KiDS-$i$-800 and KiDS-$r$-450 is given in
Appendix~\ref{app:QC}. After masking and accounting for overlap
between the tiles, the KiDS \textit{i}-800 dataset spans an effective
area of 733 square degree.

\subsection{Modelling the Point Spread Function}
\label{sec:psf}

\begin{figure*}
	\includegraphics[width=\textwidth]{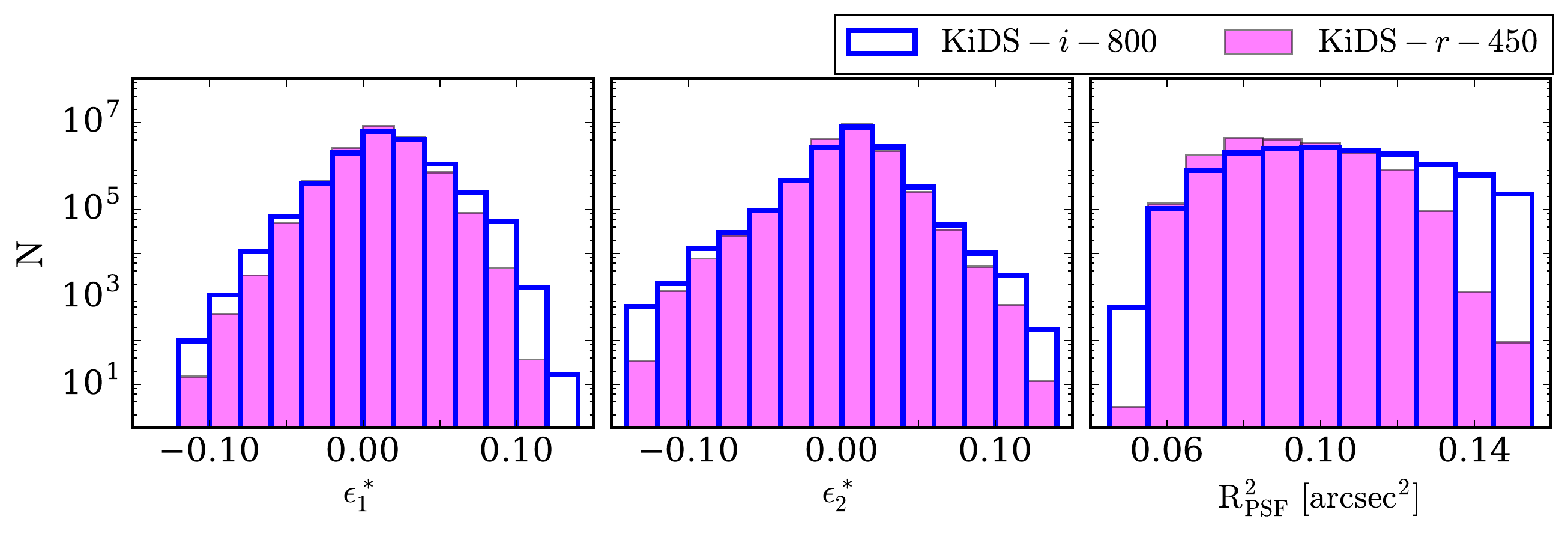}
    \caption{\label{fig:psf}Comparison of the properties of the PSF model reconstructed at the position of each resolved galaxy in KiDS-$i$-800 (blue) and KiDS-$r$-450 (pink): The left-hand and middle panels show the distribution of each component of PSF ellipticity.  The width of the KiDS-$i$-800 PSF ellipticity distribution is comparable to that of KiDS-$r$-450. The right-hand panel shows the distribution of the local PSF size illustrating the wider range of seeing conditions with KiDS-$i$-800 observations. Note that all panels have a log scaling to highlight the differences in the distribution of the KiDS $i$ and $r$-band data in the extremes.}
\end{figure*}

Galaxy images are smeared as photons travel through the Earth's
atmosphere and further distorted due to telescope optics and detector
imperfections. This gives rise to a spatially and temporally variable
point spread function (PSF) that can be characterised and corrected
for using star catalogues.

With high-resolution KiDS $r$-band imaging, star-galaxy separation can
be reliably determined by inspecting the size and `peakiness' of each
object in each exposure. A star catalogue is then assembled by
selecting the objects that group together in a distinct stellar peak
and appear in three or more of the five exposures \citep[see Section
3.2 of][for details]{Kuijken/etal:2015}. For the variable seeing
$i$-band imaging, however, we found this method to be unreliable, as
in very poor seeing the stellar peak is no longer as distinct from the
galaxy sample.

For KiDS-$i$-800 we first select stellar candidates automatically in the size-magnitude plane \citep[see Section 4 of][for details]{Erben/etal:2013}.   We estimate the complex ellipticity of each stellar candidate, from each exposure, in terms of its weighted second order quadrupole moments $Q_{ij}$,
\be
Q_{ij} =\frac{\int \mathrm{d}^2{\bf x} \,W({\bf |x|}) \, I({\bf x}) \, x_i \, x_j}{\int \mathrm{d}^2{\bf x} \, W({\bf |x|}) \, I({\bf x})} \, ,
\ee
where $I({\bf x})$ is the surface brightness of the object at position ${\bf x}$, measured from the \textsc{SExtractor} position and $W({\bf |x|})$ is a Gaussian weighting function with dispersion of three pixels, \citep[following][]{Kuijken/etal:2015}, which we employ to suppress noise at large scales. The complex stellar ellipticity is then calculated from,
\be
\epsilon^{\mathrm{*}}=\epsilon_1^{\mathrm{*}} + \mathrm{i} \epsilon_2^{\mathrm{*}} = \frac{Q_{11}-Q_{22}+2\mathrm{i} Q_{12}}{Q_{11}+Q_{22}+2 \sqrt{Q_{11}Q_{22}-Q_{12}^2}} \, .
	\label{eq:psf}
\ee
In the case of a perfect ellipse, the unweighted complex ellipticity $\epsilon$ (where $W({\bf |x|}) =1$ for all ${\bf |x|}$), is related to the axial ratio $q$ and orientation of the ellipse $\phi$ as,
\be
\epsilon = \epsilon_1 + \mathrm{i} \epsilon_2 = \left( \frac{1-q}{1+q} \right) \mathrm{e}^{2\mathrm{i}\phi} \, .
\label{eqn:ellipticity}
\ee
Using a second-order polynomial model, the spatially varying stellar ellipticity, or PSF, is modelled across each exposure.  Outliers are rejected from the candidate sample if their measured ellipticities differ by more than $3 \sigma$ from the local PSF model, where $\sigma^2$ is the variance of the PSF model ellipticity across the field of view.  A final $i$-band star catalogue is then assembled from the cleaned stellar candidate lists by again requiring that the stellar object has been selected in three or more exposures.

In Appendix~\ref{app:Gaia} we investigate the robustness of our two different star-galaxy selection methods in both the $i$ and $r$-bands by comparing our star catalogues to the stellar catalogues published by the {\it Gaia} mission in their first data release \citep{Gaia/etal:2016}.  We find that, considering objects brighter than $i<20$, our $i$-band stellar selection rejects 14 percent of unsaturated {\it Gaia} sources compared to our $r$-band stellar selection which rejects 10 percent.

In principle, our star selection could yield an unrepresentative sample of stars, leading to an error in the PSF model. In order to inspect the quality of the PSF modelling for the exposures of each field, we therefore compute the residual PSF ellipticty, $\delta\epsilon^*=\epsilon^*(\rm{model})-\epsilon^*(\rm{data})$. For an accurate PSF model, this should be dominated by photon noise and therefore be uncorrelated between neighbouring stars. An investigation into the two-point $i$-band PSF residual ellipticity correlation function, $\langle \delta\epsilon^*\overline{\delta\epsilon}^*\rangle$, where the bar denotes the complex conjugate, revealed that this statistic was consistent with zero between the angular scales of 0.8 arcmin to 60 arcmin.  From this we can conclude that the PSF model accurately predicts the amplitude and angular dependence of the two-point PSF ellipticity correlation function.  The same conclusion was drawn in the assessment of the $r$-band imaging in \citet{Kuijken/etal:2015}.

Figure~\ref{fig:psf} compares the PSF model properties of the KiDS-$i$-800 and KiDS-$r$-450 data.  The left and middle panels show the number of resolved galaxies, in each dataset, as a function of the model PSF ellipticity $\epsilon^*$ at the location of the galaxy.  We find that the spread of PSF ellipticities in the $i$-band is comparable to that of KiDS-$r$-450, with slightly more instances of higher-ellipticity PSFs in the tails of the distribution.

The right panel of Figure~\ref{fig:psf} shows the distribution of the local PSF size at the positions of resolved galaxies, where the PSF size is determined in terms of the quadrupole moments, $Q_{ij}$, as
\be
{R_{\rm{PSF}}^2}=\sqrt{Q_{11}Q_{22}-Q_{12}^2 } \, .
\label{eq:psfsize}
\ee
This panel illustrates the wider range of seeing conditions within the $i$-band dataset, in comparison to the more homogeneous KiDS-$r$-450 data. Note that we examined how the ellipticity of the $i$-band PSF varied with worsening seeing conditions but found that these two quantities were largely uncorrelated.

\subsection{Galaxy shape measurement and selection}\label{sec:shape}
\label{sec:selection}

Galaxy shapes were measured using \emph{lens}fit, a likelihood based model-fitting method that fits PSF-convolved bulge-plus-disk galaxy models to each exposure simultaneously in order to estimate the shear \citep{miller/etal:2013}.   In this analysis, we adopt the latest `self-calibrating' version of \emph{lens}fit \citep{fenech-conti/etal:2016}. As any single point measurement of galaxy ellipticity is biased by pixel noise in the image, this upgraded version is designed to mitigate these effects based on the actual measurements and an extensive suite of image simulations .  In addition, weights are recalibrated in order to correct for biases that arise due to the relative orientation of the PSF and the galaxy, as highlighted by \citet{miller/etal:2013}, and a revised de-blending algorithm is adopted in order to reject fewer galaxies that are too close to their nearest neighbour. We refer the reader to Section 2.5 of \citet{Hildebrandt/etal:2017} for a comprehensive list of the advances on the version of the algorithm used in previous analyses, such as \citet{Kuijken/etal:2015}. This version of \emph{lens}fit leaves a percent-level residual multiplicative noise bias, which we parametrise using image simulations. It was demonstrated in \cite{fenech-conti/etal:2016} that model bias contributes at the per mille level for a KiDS-like survey when tested with simulations of COSMOS galaxies \citep{Voigt2010}.

We account for the intrinsic differences between the $i$ and $r$-band galaxy populations by adopting different priors on galaxy size for the $i$ and $r$-band \emph{lens}fit analyses \citep{Kuijken/etal:2015}.    We do, however, assume the distribution of galaxy ellipticities and the bulge-to-disk ratio are the same for both bands.  \citet{Hildebrandt/etal:2016} found that using an $i$-band size prior to analyse $r$-band data using \emph{lens}fit resulted in an average change in the observed galaxy ellipticity of less than 1 percent.  This demonstrates that we do not require high levels of accuracy in the determination of the galaxy size prior in each band.

Using an extensive suite of $r$-band image simulations, \citet{fenech-conti/etal:2016} show that \emph{lens}fit provides shear estimates that are accurate at the percent level.  We use these results to calibrate a possible residual multiplicative shear measurement bias, $m$, in the $i$-band observations.  We note two important caveats, however, that the \citet{fenech-conti/etal:2016} image simulations did not explore: the extreme PSF sizes found in KiDS-$i$-800 and an $i$-band galaxy population.  As the calibration corrections are determined as a function of galaxy resolution, that is, the ratio of the galaxy size and the PSF size, and because the $r$-band galaxy population is similar to the $i$-band population, we expect the conclusions from \citet{fenech-conti/etal:2016} to apply to $i$-band observations.  We note that any high- accuracy science, for example cosmic shear, using KiDS-$i$-800 would, however, require independent verification of the $i$-band calibration corrections adopted in this analysis.

The \citet{fenech-conti/etal:2016} image simulation analysis was limited to galaxies fainter than $r>20$.  Providing a calibration correction below this magnitude would require an extension to the image simulation pipeline, as these bright galaxies typically extend beyond the standard simulated postage stamp size.  By comparing galaxies in $r-i$ colour space we determined an equivalent $i$-band limit to be $i>19.4$, limiting our $i$-band analysis to galaxies fainter than this threshold.

Each \emph{lens}fit ellipticity measurement is accompanied by an inverse variance weight that is set to zero when the object is unresolved or point-like, for example.  Requiring that shapes have a non-zero \textit{lens}fit weight therefore effectively removes stars and faint unresolved galaxies. The 0.01$\%$ of objects that were deemed by their `fitclass' value to be poorly fit by a bulge-plus-disk galaxy model were also removed, effectively removing any image defects that entered the object detection catalogue \citep[see Section D1 of][for details]{Hildebrandt/etal:2017}. We note that without multi-colour information we were unable to detect and remove faint satellite or asteroid trails in the $i$-band, or identify any moving sources from the individual exposures, which were shown in \citet{Hildebrandt/etal:2017} to be a significant contaminating source for some fields of the $r$-band data analysis. While this would be important for the case of cosmic shear, these artefacts have a negligible effect for cross-correlation studies.

We investigated how the average ellipticity of the galaxy sample varied when applying progressively more conservative cuts on our de-blending parameter, the contamination radius. This is a measure of the distance to neighbouring galaxies and therefore the contaminating light in the image of the main galaxy.  We found that the average ellipticity of the full sample converged when galaxies with a contamination radius greater than 4.25 pixels were selected.  \citet{Hildebrandt/etal:2017} also concluded that a de-blending selection criterion of 4.25 pixels was optimal for the $r$-band imaging.

\subsection{Calibrating KiDS galaxy shapes} \label{sec:sys}
Observed galaxy images are convolved with the PSF and pixellated.  They are also inherently noisy and in order to deal with the residual noise bias, shear measurements typically require calibration corrections with a suite of image simulations.  Corrections to the observed shear estimator, $\epsilon^{\mathrm{obs}}$ can be modelled in terms of a multiplicative shear term $m$, a multiplicative PSF model term $\alpha\epsilon^* = \alpha_1 \epsilon_1^* + \mathrm{i} \alpha_2 \epsilon_2^*$, a PSF modelling error term $ \beta \delta\epsilon^{\mathrm{*}}$, and an additive term, $c = c_1 + \mathrm{i} c_2$, that is uncorrelated with the PSF, such that
\be
\epsilon^{\mathrm{obs}} = \left( \frac{\epsilon^{\mathrm{int}} + \gamma}{1+\bar{\gamma}\epsilon^{\mathrm{int}} }\right)(1+m)+ \epsilon^{\mathrm{n}} + \alpha \epsilon^{\mathrm{*}}+ \beta\, \delta\epsilon^{\mathrm{*}}+c \, .
	\label{eq:eobs}
\ee
Here all quantities are complex (see equation~\ref{eqn:ellipticity}), with the exception of the multiplicative calibration scalars $m$ and $\beta$.  The first bracketed term transforms the galaxy's intrinsic ellipticity $\epsilon^{\mathrm{int}}$ by $\gamma$, the reduced lensing-induced shear that we wish to detect \citep{seitz/schneider:1997}.  In this analysis we take the weak lensing approximation that the reduced shear and the shear are equal and use the notation $\bar{\gamma}$, to indicate a complex conjugate.  $\epsilon^{\mathrm{n}}$ is the random noise on the measured galaxy ellipticity which will increase as the signal-to-noise of the galaxy decreases \citep{Viola2014}, and $\epsilon^{\mathrm{*}}$ is the ellipticity of the true PSF.  For a perfect shape measurement method, $m, c$ and $\alpha\epsilon^{\mathrm{*}}$ would all be zero and for a perfect PSF model $\beta\, \delta\epsilon^{\mathrm{*}}$ would also be zero \citep{Hoekstra2004, Heymans2006}.

In this analysis we use the PSF model as a proxy for the true PSF, in which case the $\beta$ becomes subsumed into \am{$\alpha$, the PSF contamination}.  This is appropriate given that the measured PSF ellipticity residual correlation function $\langle \delta\epsilon^*\overline{\delta\epsilon}^*\rangle$, was found to be consistent with zero (see Section~\ref{sec:psf}). The additive calibration correction $c$ and PSF term $\alpha$ can then be estimated empirically by fitting the model in equation~\ref{eq:eobs} directly to the data assuming that the data volume is sufficiently large such that the average $\langle \gamma+ \epsilon^{\mathrm{int}} \rangle = 0$.  For KiDS-$i$-800 we find that $c_1 = -0.0011 \pm 0.0001$, $c_2 = 0.0018 \pm 0.0001$, $\alpha_1 = 0.067 \pm 0.006$ and $\alpha_2 = 0.074 \pm 0.006$.  As with a similar analysis for KiDS-$r$-450, we find measurements of $\alpha$ to be uncorrelated with $c$.

\begin{figure}
	\includegraphics[width=\columnwidth]{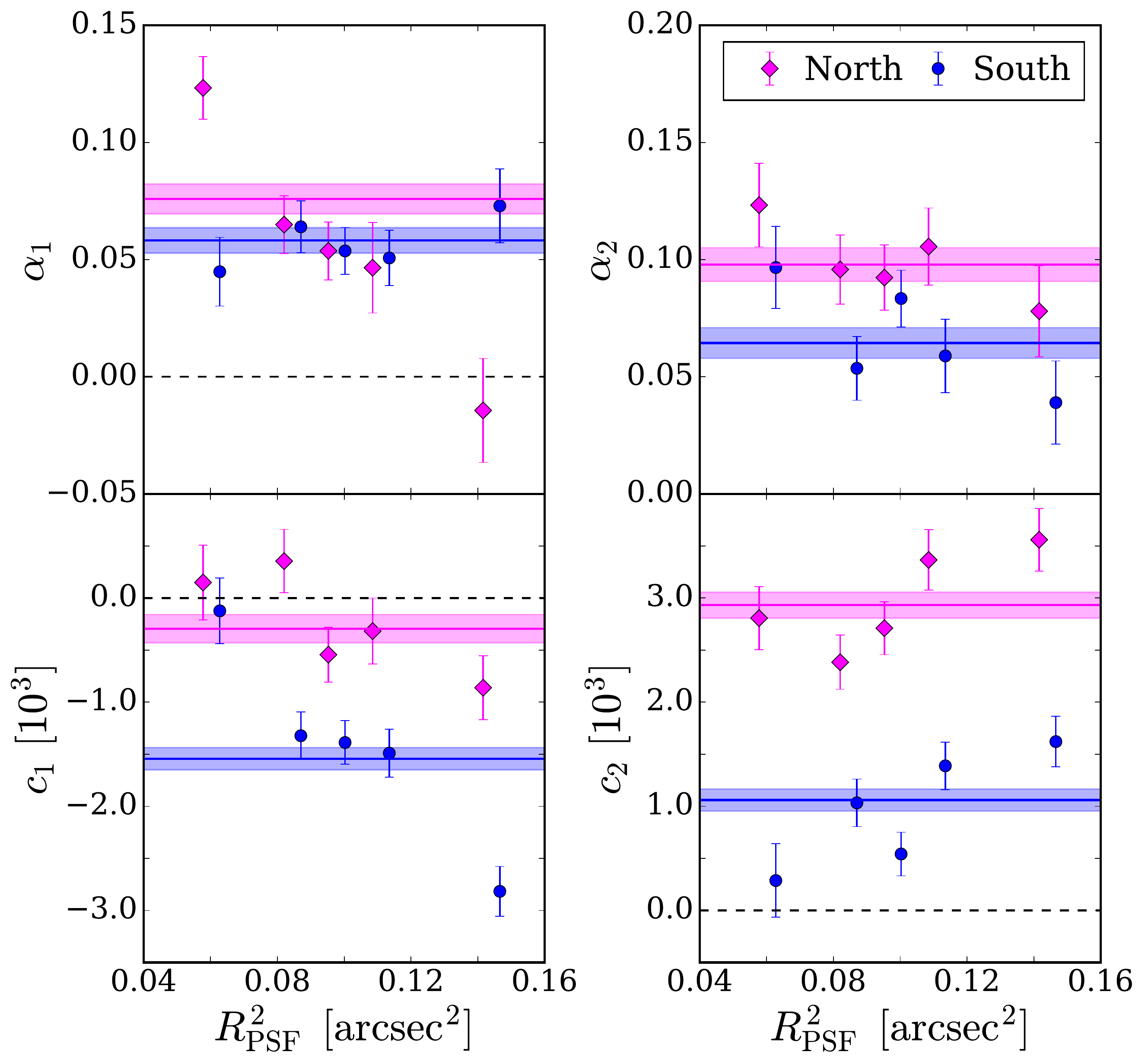}
    \caption{\label{fig:calpha}The variation of the additive bias term, $c$ (lower panel) and the multiplicative PSF model term, $\alpha$ (upper panel) with the size of the PSF. The analysis of the Northern fields are shown in pink and the Southern fields in blue. The solid line represents the mean of the data points and the coloured bands indicate a 1$\sigma$ deviation.}
\end{figure}

In Figure~\ref{fig:calpha} we show the measured additive calibration correction $c$ and PSF term $\alpha$ for the Northern and Southern KiDS-$i$-800 patches as a function of the observed PSF size, $R_{\rm{PSF}}^2$ (equation~\ref{eq:psfsize}).  
\am{We find that the $i$-band PSF contamination is significant (even when the $i$-band data are restricted to the same seeing range as the $r$-band), in comparison to the case of KiDS-r-450, where the PSF contamination in each tomographic bin and survey patch ranged between $ -0.03 < \alpha < 0.02$ with an error $\sim0.01$  (for further details see Section D4 of Hildebrandt et al. 2017).}
As the PSF ellipticity distributions between the two bands are comparable (see Figure~\ref{fig:psf}), the fact that we find different levels of PSF contamination between the $i$ and $r$-band images could lead to a better understanding of how differences in the data reduction and analysis lead to a PSF error. The primary difference between the KiDS-$i$-800 and KiDS-$r$-450 data reduction in the Southern field is the method used to determine the astrometric solution.  In KiDS-$i$-800, this was determined for each pointing individually, whereas an improved full global solution was derived for the $r$-band. In the Northern patch, however, astrometry for both KiDS-$i$-800 and KiDS-$r$-450 was tied to SDSS \citep{Alam/etal:2015}.  With similar levels of PSF contamination in the Northern and Southern KiDS-$i$-800 patches as demonstrated in Figure~\ref{fig:calpha}, we can conclude that astrometry is likely not to be at the root of this issue.  The method to determine a stellar catalogue also differed (see Section~\ref{sec:psf}).  Our comparison to stellar catalogues from {\it Gaia} in Appendix~\ref{app:Gaia} suggests that a selection bias could have been introduced during star selection.  With PSF residuals shown to be consistent with zero in Section~\ref{sec:psf}, however, we can also conclude that PSF modelling is likely not to be at the root of this issue. 
\am{The third main difference between the data sets is a non-negligible level of residual fringing in the KiDS-i-800 images (see the discussion in Appendix~\ref{app:QC}).  Residual fringe removal was not prioritised in the early stages of the KiDS-i-800 data reduction as the plan for this dataset did not include cosmic shear studies.} As the fringe patterns are uncorrelated with the PSF, it is thought that fringing is unlikely to be the root cause of the PSF contamination, but this will be explored further in future analyses.

%As fringing hasn't been ruled out suggest to remove
%"With the most obvious differences between the datasets ruled out by our investigations and 

\am{As the primary science goals for KiDS-$i$-800 is this demonstrative comparison, we decided to defer further studies of the origin of the $i$-band PSF contamination. For the galaxy-galaxy lensing comparison, any PSF contamination is effectively removed when azimuthal averages are taken around foreground lens structures.  Additive biases are also accounted for by correcting the signal using the measured signal around random points (see Section~\ref{sec:ds}).  This level of PSF contamination renders KiDS-$i$-800 unsuitable for cosmic shear studies.}

\subsection{Matched $ri$ catalogue}
\label{sec:matchcat}
We create a matched $r$ and $i$-band catalogue, limited to galaxies that have a shape measurement in both KiDS-$i$-800 and KiDS-$r$-450, using a \SI{1}{\arcsec} matching window.  The overlapping $ri$ survey footprint has an effective area of \SI{302}{\square\degree}, taking into account the area lost to masks.  Only 39\% of the $r$-band shape catalogue in this area is matched, which is expected as the effective number density of the $r$-band shear catalogues is more than double the effective number density of the $i$-band shear catalogues (see Section~\ref{sec:ir}).  78\% of the $i$-band shape catalogue is matched, however, and this number increases to 89\% when an accurate $r$-band shape measurement is not required.  We made a visual inspection of a sample of the remaining unmatched $i$-band objects revealing different de-blending choices between the $r$-band and $i$-band images, where the {\sc SExtractor} object detection algorithm has chosen different centroids owing to the differing data quality between the two images. We also found differences in low signal-to-noise peaks, and a small fraction of objects with significant flux in the $i$-band but no significant $r$-band flux counterpart. We define a new weight for each member of this matched sample as a combination of the \emph{lens}fit weights of the galaxy, assigned in the KiDS-$i$-800 sample, $w_i$ and in KiDS-$r$-450, $w_r$, with, $w^{ir} = \sqrt{w_i w_r}$.   By combining the weights in this way we ensure that the effective weighted redshift distribution of the two matched samples is the same.

\section{Redshift data} \label{sec:zdata} 
\subsection{The spectroscopic lens samples}
\label{sec:lenses}

In our comparison study we present a galaxy-galaxy lensing analysis, where we select samples of lens galaxies from spectroscopic redshift surveys. As KiDS overlaps with a number of wide-field spectroscopic surveys, this choice reduces the error associated with the alternative approach of defining a photometric redshift selected lens sample \citep[see for example][]{Kleinheinrich2004, Nakajima2012}. The surveys employed as the lens samples are BOSS \citep{Eisenstein2011}, GAMA \citep{Driver/etal:2011} and 2dFLenS \citep{Blake/etal:2016}. The overlapping survey coverage is illustrated in Figure~\ref{fig:map}. 

BOSS is a spectroscopic follow-up of the SDSS imaging survey, which used the Sloan Telescope to obtain redshifts for over a million galaxies spanning \SI{10000}{\square\degree}. BOSS used colour and magnitude cuts to select two classes of galaxy: the `LOWZ' sample, which contains Luminous Red Galaxies (LRGs) at $z < 0.43$, and the `CMASS' sample, which is designed to be approximately stellar-mass limited for $z > 0.43$.  We used the data catalogues provided by the SDSS 12th Data Release (DR12); full details of these catalogues are given by \cite{Alam/etal:2015}.  Following standard practice, we select objects from the LOWZ and CMASS datasets with $0.15 < z < 0.43$ and $0.43 < z < 0.7$, respectively, to create homogeneous galaxy samples. In order to correct for the effects of redshift failures, fibre collisions and other known systematics affecting the angular completeness, we use the completeness weights assigned to the BOSS galaxies \citep{Ross2012}.
%%54000 CMASS, 15200 LOW-Z

2dFLenS is a spectroscopic survey conducted by the Anglo-Australian Telescope with the AAOmega spectrograph, spanning an area of \SI{731}{\square\degree}, principally located in the KiDS regions, in order to expand the overlap area between galaxy redshift samples and gravitational lensing imaging surveys.  The 2dFLenS spectroscopic dataset contains two main target classes: $\sim$\SI{40000} LRGs across a range of redshifts $z < 0.9$, selected by SDSS-inspired cuts \citep{Dawson/etal:2013}, as well as a magnitude-limited sample of $\sim$\SI{30000} objects in the range $17 < r < 19.5$, to assist with direct photometric calibration \citep{wolf/etal:2017}.  In our study we analyse the 2dFLenS LRG sample, selecting redshift ranges $0.15 < z < 0.43$ (`2dFLOZ') and $0.43 < z < 0.7$ (`2dFHIZ'), mirroring the selection of the BOSS sample.  We refer the reader to \cite{Blake/etal:2016} for a full description of the construction of the 2dFLenS selection function and random catalogues.
%10000 LZ 15000 HZ

GAMA is a spectroscopic survey carried out on the Anglo-Australian Telescope with the AAOmega spectrograph. We use the GAMA galaxies from three equatorial regions, G9, G12 and G15 from the 3rd GAMA data release \citep{Liske2015}. These equatorial regions encompass roughly \SI{180}{\square\degree}, containing $\sim$\SI{180000} galaxies with sufficient quality redshifts. The magnitude-limited sample is essentially complete down to a magnitude of $r$ = 19.8. For our weak lensing measurements, we use all GAMA galaxies in the three equatorial regions in the redshift range $0.15< z <0.51$ (as selected from TilingCatv45).
%7500G9 24700G12 16000G15

\begin{figure}
	\includegraphics[width=\columnwidth]{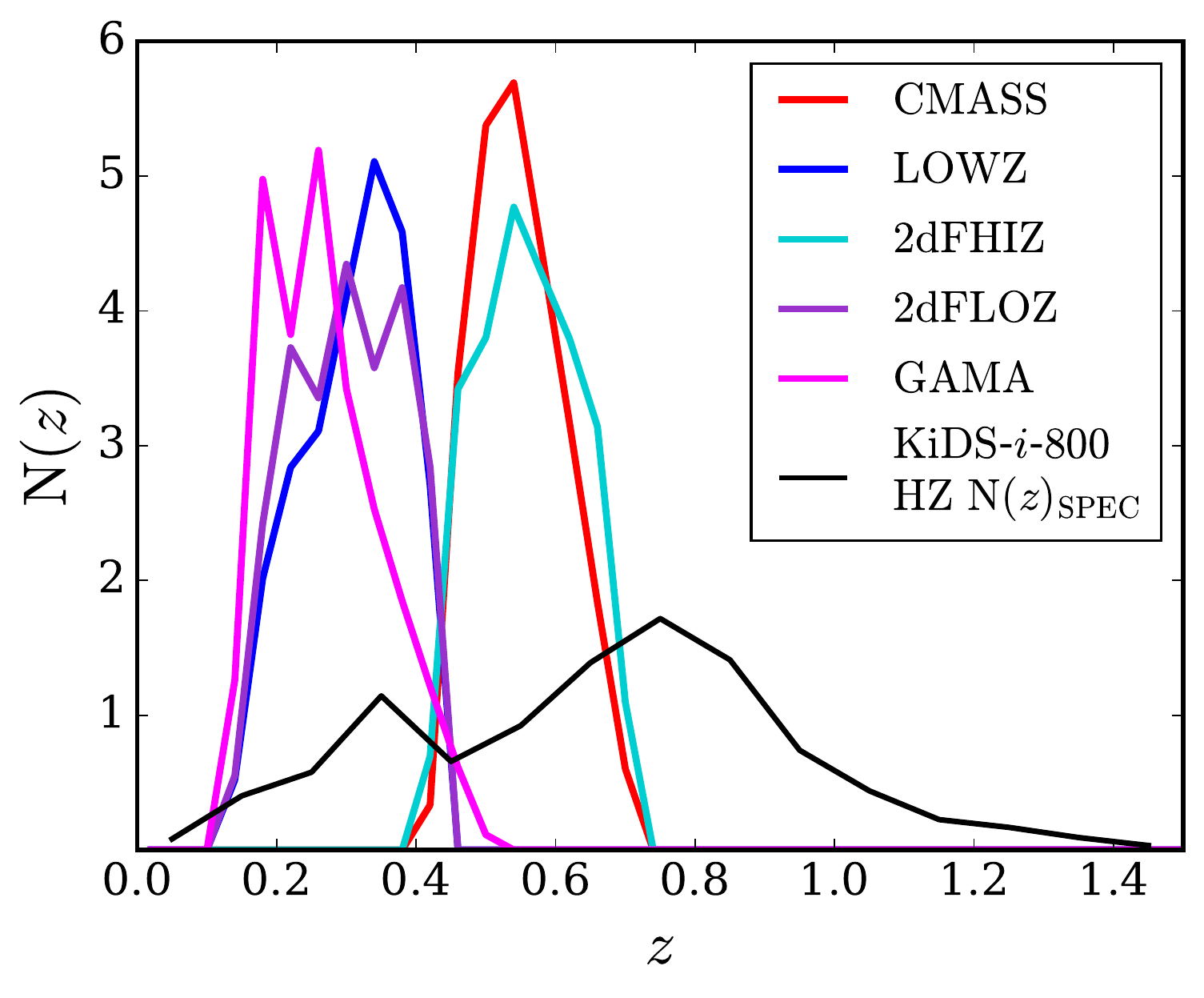}
    \caption{\label{fig:nzlens}The redshift distributions for the five spectroscopic lens samples used in the analysis, plotted alongside the estimated redshift distribution of the KiDS-$i$-800 faint (HZ) sample, obtained using the overlap of deep spectroscopic redshifts described in Section~\ref{sec:specz}.}
\end{figure}

In the galaxy-galaxy lensing analysis that follows, we group our lens samples into a `HZ' case, containing the two high-redshift lens samples, BOSS-CMASS and 2dFHIZ, and a `LZ'  case, containing the low-redshift samples, BOSS-LOWZ, 2dFLOZ and GAMA. The redshift distributions of the spec-z lens samples are presented in Figure~\ref{fig:nzlens}.

\subsection{The $r$-band redshift distribution}\label{sec:nzr}

In KiDS-$r$-450, the multi-band observations allow us to determine a Bayesian point estimate of the photometric redshift, $z_{\rm B}$, for each galaxy using the photometric redshift code {\sc BPZ} \citep{Benitez:2000}.    We use this information to select source galaxies that are most likely to be behind our `LZ' and `HZ' lens samples.   

The redshift distribution for these $z_{\rm B}$ selected KiDS-$r$-450 source samples is calibrated with the weighting technique of \cite{Lima2008}, named `DIR'.  Here we match $r$-band selected $ugri$ VST observations with deep spectroscopic redshifts from the COSMOS field \citep{Lilly/etal:2009}, the Chandra Deep Field South (CDFS) \citep{Vaccari/etal:2012} and two DEEP2 fields \citep{Newman/etal:2013}.  This matched spectroscopic redshift catalogue is then re-weighted in multi-dimensional magnitude-space such that the weighted density of spectroscopic objects is as similar as possible to the \emph{lens}fit-weighted density of the KiDS-$r$-450 lensing catalogue in each position in magnitude-space.  It was shown in \citet{Hildebrandt/etal:2017} that this `DIR' method produced reliable redshift distributions, with small bootstrap errors on the mean redshift, in the photometric redshift range $0.1 <z_{\rm B} \leqslant 0.9$.  As such, we adopt this DIR method and selection for our KiDS-$r$-450 galaxy-galaxy lensing analysis.

\subsection{Estimating the $i$-band redshift distribution}\label{sec:specz}

To estimate a redshift distribution for KiDS-$i$-800 we choose not to adopt the `DIR' method for a number of practical reasons.  As discussed in Section~\ref{sec:matchcat}, an $i$-band detected object catalogue differs from an $r$-band detected object catalogue, with $\sim 10$ percent of the $i$-band objects not present in the $r$-band catalogue.  To create a weighted $i$-band spectroscopic sample would have required a full re-analysis of the VST imaging of the spectroscopic fields using the $i$-band imaging as the detection band.  Furthermore, the DIR method was shown to be accurate in the photometric redshift range $0.1 <z_{\rm{B}} \leqslant 0.9$ and as the majority of KiDS-$i$-800 only has single-band photometric information, it is not clear whether one can define a safe sample for which this method works reliably.

\begin{figure}
	\includegraphics[width=\columnwidth]{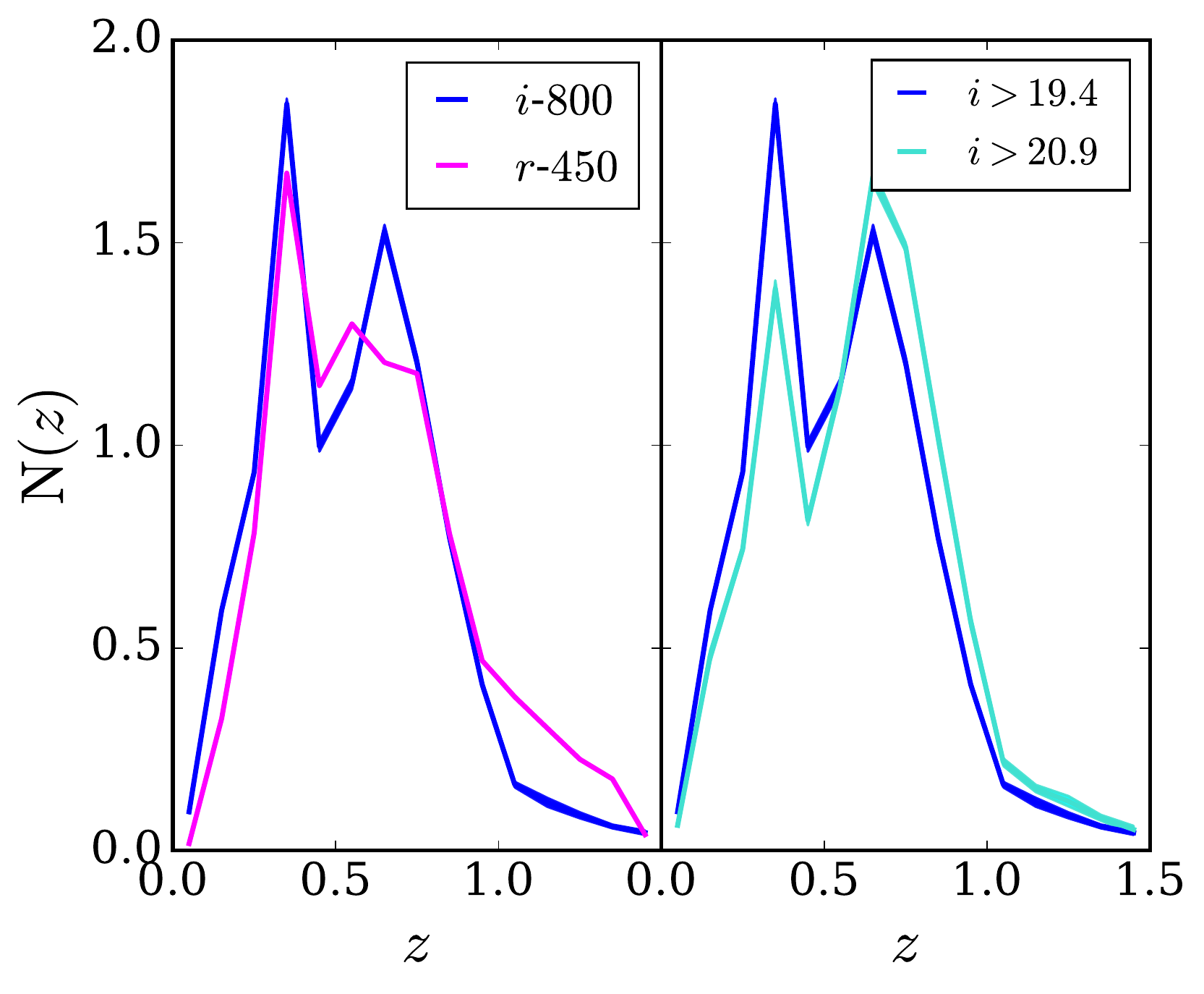}
	\caption{\label{fig:inz}The estimated redshift distributions obtained using the overlapping spectroscopic data. Left: $N(z)$ for KiDS-$i$-800 (blue) estimated using the SPEC method, described in Section~\ref{sec:specz} and the KiDS-$r$-450 (pink) estimated via the DIR method.  The median redshifts are comparable at \am{0.50 and 0.57}, for KiDS-$i$-800 and KiDS-$r$-450 respectively. The sampling of the distribution is bootstrapped for an error, indicated by the thickness of the lines. Right: The estimated $N(z)$ for KiDS-$i$-800 for a brighter (blue) and fainter (cyan) magnitude limit.}
\end{figure}

Our first estimate of the $i$-band redshift distribution, named `SPEC', instead comes from using the COSMOS and CDFS spectroscopic catalogues directly as they are fairly complete at the relatively shallow magnitude limits of the KiDS-$i$-band imaging. %As an example, Figure 31 of \cite{Newman/etal:2013} indicates an $\sim 80\%$ completeness of the DEEP2 spectroscopic catalogue at the depth of KIDS-$i$-800.  
In this case, we estimate the total redshift distribution, $N(z)$, by drawing a sample of spectroscopic galaxies such that their $i$-band magnitude distribution matches the \textit{lens}fit weighted $i$-band magnitude distribution for all KiDS-$i$-800 galaxies.  \ch{Given this methodology we do not include the DEEP2 catalogues used for the `DIR' calibration of the $r$-band redshifts, as these have been colour-selected and therefore are not representative of the $i$-band magnitude limited sample.
The resulting redshift distribution is shown} in the left-hand panel of Figure~\ref{fig:inz}, along with the average $r$-band DIR $N(z)$ with the $z_{\rm{B}}$ selection imposed.  A bootstrap analysis determined the small statistical error in these redshift distributions and is illustrated by the thickness of the line. Any systematic error, due to sample variance or incompleteness in the spectroscopic catalogue, is not represented by the bootstrap error analysis. 

As the KiDS-$i$-800 dataset lacks multi-band information and hence photometric redshift information per galaxy we choose to select galaxies based on their $i$-band magnitude to increase the average redshift of the source sample.   Using our chosen bright magnitude limit of $i>19.4$ (see Section~\ref{sec:selection}), the \emph{lens}fit weighted source sample corresponds to a median redshift above $z_{\rm{med}}=0.43$.    This magnitude selection is therefore suitable as a source sample for our `LZ' lens analysis.  Adopting a magnitude limit of $i>20.9$, we find that the faint $i$-band sample has a median redshift $z_{\rm{med}}=0.7$, thus making a suitable source sample for our `HZ' lens sample (see Figure~\ref{fig:imagzmed} in Appendix~\ref{app:cuts} for further details).
The right-hand panel of Figure~\ref{fig:inz} shows the SPEC estimated redshift distributions for the KiDS-$i$-800 bright (LZ) and faint (HZ) source galaxy samples. The median redshifts of these samples are \am{0.50 and 0.57}, respectively.  

Figure~\ref{fig:nzlens} compares the predicted redshift distribution of the $i>20.8$ KiDS-$i$-800 HZ source sample with the redshift distributions of the lens samples.  This demonstrates that even with the imposed magnitude cut on the KiDS-$i$-800 source galaxies, a significant fraction of source galaxies are still positioned in front of lenses thus diluting the signal.  In the case of galaxy-galaxy lensing, uncertainty in the redshift distributions can therefore contribute significantly to the error budget and we seek to quantify this uncertainty by investigating two additional methods to estimate the KiDS-$i$-800 redshift distribution, using 30-band photometric redshifts (Section~\ref{sec:photoz}) and a cross-correlation technique (Section~\ref{sec:cc}).

\subsection{Magnitude-weighted COSMOS-30 redshifts }
\label{sec:photoz}
One pointing in the KiDS-$r$-450 dataset overlaps with the well studied Hubble Space Telescope COSMOS field \citep{scoville/etal:2007}.   This field has been imaged using a combination of 30 broad, intermediate, and narrow photometric bands ranging from UV (GALEX) to mid-IR (Spitzer-IRAC), and this photometry has been used to determine accurate photometric redshifts \citep[COSMOS-30][]{Ilbert2009,Laigle2016}.   Comparison with the spectroscopic zCOSMOS-bright sample shows that for $i < 22.5$, the COSMOS-30 photometric redshift error $\sigma_{\Delta z/(1+z)} = 0.007$. For the full sample with $z<1.25$, the estimates on photo-z accuracy are $\sigma_{\Delta z} = 0.02, 0.04, 0.07$ for $i \sim 24.0$, $i \sim 25.0$, $i \sim 25.5$ respectively \citep{Ilbert2009}.  As the COSMOS-30 photo-z catalogue is complete at the magnitude limits of KiDS-$i$-800, it provides a complementary estimate of the $i$-band redshift distribution. 

We first match the multi-band KiDS-$r$-450 catalogue, in terms of both position and magnitude, with the COSMOS Advanced Camera for Surveys General Catalog \citep[ACS-GC][]{Griffith2012} which includes the 30-band photometric redshifts from \cite{Ilbert2009}.    These catalogues contain both stars and galaxies, which were labelled manually after the matching, by looking at the magnitude-size plot using the HST data where the separation was clean [see Hildebrandt et al. (in prep) for further details].   Once matched we sample the catalogue such that the $i$-band magnitude distribution of the selected COSMOS-30 galaxies matches the KiDS-$i$-800 \emph{lens}fit weighted magnitude distribution. Similar to the case of using a spectroscopic reference catalogue, the bootstrap analysis of the resulting $i$-band redshift distribution shows a negligible statistical error. 

\subsection{Cross-correlation (CC)} 
\label{sec:cc}

The third redshift distribution estimate is constructed by measuring the angular clustering between the KiDS-$i$-800 photometric sample and the overlapping GAMA and SDSS spectroscopic samples. Clustering redshifts are based on the fact that galaxies in photometric and spectroscopic samples of overlapping redshift distributions reside in the same structures, thereby allowing for spatial cross-correlations to be used to estimate the degree to which the redshift distributions overlap and therefore, the unknown redshift distribution. Our approach is detailed in \cite{Schmidt2013} and \cite{Menard2013} and further developed in \cite{Morrison2017}, who describe \textsc{the-wizz}\footnote{Available at: \url{http://github.com/morriscb/the-wizz/}}, the software we employ to estimate our redshifts from clustering.  A similar clustering redshift technique was employed in \cite{Choi2015}, \cite{Johnson2017} as well as \cite{Hildebrandt/etal:2017}, but in the latter case the angular clustering was measured between the KiDS-$r$-450 galaxies and COSMOS and DEEP2 spectroscopic galaxies. 

We exploit the overlapping lower-redshift SDSS and GAMA spectroscopy, the same surveys used in \cite{Morrison2017}.  The bulk of the spectroscopic sample is at a low redshift, limiting the redshift range that can be precisely constrained to $z<1.0$. This is because the high-redshift cross-correlations rely on the low density of  spectroscopic quasars from SDSS. As the $i$-band galaxies comprise a shallower dataset than KiDS-$r$-450, these spectroscopic samples were deemed appropriate. The correlation functions are estimated over a fixed range of proper separation $100-1000 \, \mathrm{kpc}$. 

\begin{figure}
	\includegraphics[width=\columnwidth]{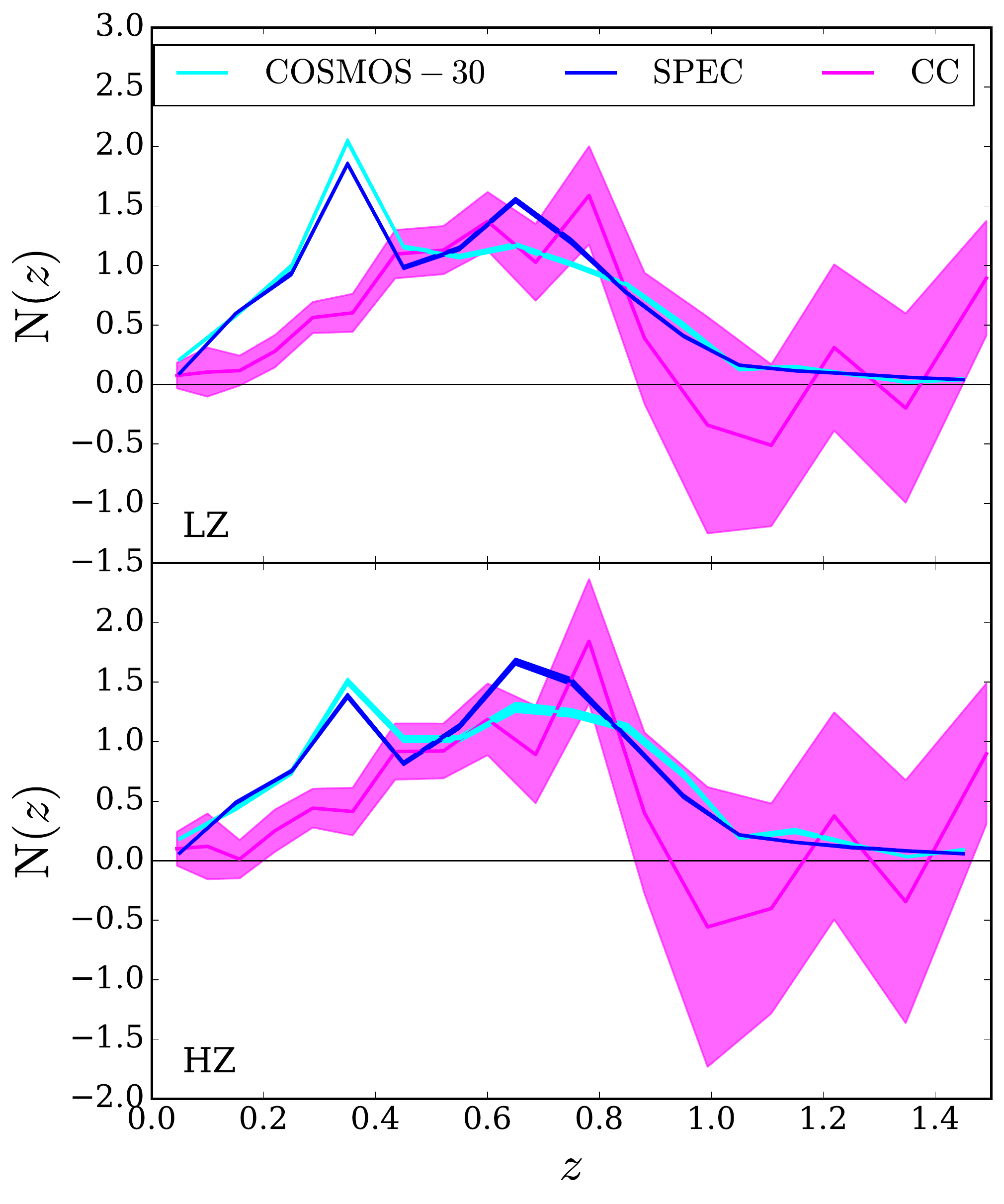}
	\caption{\label{fig:inzall}Comparison of the normalised redshift distributions for the LZ bright sample of KiDS-$i$-800 galaxies (upper panel) and the HZ faint sample (lower panel).  The distributions shown are estimated using the spectroscopic catalogue (SPEC, Section~\ref{sec:specz} ), plotted in blue, the COSMOS-30 photometric redshift catalogue (COSMOS-30, Section~\ref{sec:photoz}) in cyan and from angular cross-correlations (CC, Section~\ref{sec:cc}) in pink. }  
\end{figure}

The amplitude of the redshift estimated from spatial cross-correlations is degenerate with galaxy bias. We employ a simple strategy to mitigate for this effect by splitting the unknown-redshift sample in order to narrow the redshift distribution a priori, in the absence of a photometric redshift estimate \citep{Schmidt2013, Menard2013, Rahman2016}. This renders a more homogeneous unknown sample with a narrower redshift span, thereby minimising the effect of galaxy bias evolution as a function of redshift. As we have only the $i$-band magnitude available to us, a separation in redshift for this analysis would be imperfect. The KiDS-$i$-800 galaxies are divided by $i$-band magnitude into bins of width $\Delta i=0.5$ and the clustering redshift estimated for each subsample. The combination of these, with each subsample weighted by its number of galaxies, is shown in Figure~\ref{fig:inzall}. We conduct a bootstrap re-sampling analysis of the spectroscopic training set over the KiDS and GAMA overlapping area, where each sampled region is roughly the size of a KiDS pointing, for each magnitude subsample, in order to mitigate spatially-varying systematics in the cross-correlation. This revealed large statistical errors in the high-redshift tail of the distribution, represented by the large extent of the confidence contours in Figure~\ref{fig:inzall}. With the noisy high-redshift tail, it is possible for the cross-correlation method to produce negative, and therefore unphysical values in the full redshift distribution $N(z)$. In such cases, the final distribution is re-binned with a coarser redshift resolution in order to attain positive values in each redshift bin.

\subsection{Comparison of $i$-band redshift distributions}
\label{sec:nzcompare}

\begin{table*}
\caption{\label{tab:zstats}Values for the mean and median of the source redshift distributions, as well as the lensing efficiency, $\eta$.  The redshift distribution for the KiDS-$r$-450 subsamples is estimated using the DIR 
method. For KiDS-$i$-800 galaxies, redshifts are estimated using overlapping, deep spectroscopic surveys (SPEC), the COSMOS photometric catalogue (COSMOS-30) and the cross-correlations method (CC). The quoted errors are determined from a bootstrap resampling. } 
\begin{center}
\begin{tabular}{lccccc}
  %\hline
 %{ p{0.2cm}|p{2.5cm}|p{1.0cm}|p{2.0cm}|p{2.0cm}  }
 %\multicolumn{4}{|c|}{xyz} \\
 \hline
	Range &	Dataset &	   Method &	$z_{\rm{med}} $ &	$\bar{z} $ &	$\eta $ \\
 \hline
LZ &KiDS-$r$-450 ($0.1<z_{\rm{B}}<0.9$)	& DIR & 0.57 &  0.65 & 0.428\\
&  KiDS-$i$-800 ($i>19.4$)  &  SPEC   &	$0.501 \pm 0.002$	&	$0.555 \pm 0.001$	&	0.361\\
&&   COSMOS-30  & $0.452 \pm 0.003$   & $0.538 \pm 0.002$ & 0.344\\ 
&& CC & $0.6 \pm 0.2$ &  $0.6 \pm 0.2$  & 0.449\\ %\cline{2-5}

 \hline
HZ &KiDS-$r$-450 ($0.43<z_{\rm{B}}<0.9$) &DIR & 0.66 &  0.73 & 0.177\\
&  KiDS-$i$-800 ($i>20.8$) & SPEC &  $0.574 \pm 0.002$ &  $0.607 \pm 0.002$ & 0.126\\ 
&& COSMOS-30 & $0.545 \pm 0.005$  &  $0.594 \pm 0.003$ & 0.121\\ 
& & CC & $0.6 \pm 0.3$ &  $0.6 \pm 0.2$ & 0.117\\ %\cline{2-5}
 \hline
\end{tabular}
\end{center}
\end{table*}

We illustrate the three estimated redshift distributions for the KiDS-$i$-800 HZ and LZ samples in Figure~\ref{fig:inzall}, and compare the mean and median redshifts for each estimate with that of KiDS-$r$-450 in Table~\ref{tab:zstats}. This table also includes an estimate of the lensing efficiency $\eta(z_{\rm l})$ for each estimated source redshift distribution, with
\be
\eta(z_{\rm l}) =  \int^\infty_{z_{\rm l}} {\rm d}z_{\rm s} \, N(z_{\rm s}) \left(\frac{\chi(z_{\rm l}, z_{\rm s})}{\chi(z_{\rm s})}\right) \, ,
\label{eqn:eta}
\ee
where the source sample is characterised by a normalised redshift distribution $N(z_{\rm s})$ and $z_{\rm l}$ is set to 0.29 and 0.56 for the LZ and HZ case, respectively.  \ch{Here the lensing efficiency, for a flat geometry Universe, scales with the comoving distances to the source galaxy, $\chi(z_{\rm s})$ and the comoving distance between the lens and the source $\chi(z_{\rm l},z_{\rm s}) = \chi(z_{\rm s}) - \chi(z_{\rm l})$.}  

As already seen in Figure~\ref{fig:inzall}, the different methods used to estimate the $i$-band redshifts result in quite different source redshift distributions.  In Table~\ref{tab:zstats} we see that the resulting mean and median redshift can differ by up to 15 percent, with the COSMOS-30 method favouring a shallower redshift distribution and the \am{CC estimate generally preferring the deepest distribution.  These differences are particularly pronounced for the low-redshift galaxy sample (with mean redshifts of 0.54 and 0.56 for the COSMOS-30 and SPEC methods and 0.6 for the CC technique)}. For galaxy-galaxy lensing studies, the impact of these differences in the estimated redshift distributions can be determined from the value of the lensing efficiency term $\eta$, in the final column of Table~\ref{tab:zstats}, which differs by up to 30 percent.  This demonstrates the limitations of single-band imaging for weak lensing surveys and the importance of determining accurate source redshift distributions for weak lensing studies.

The drawback of using the SPEC method is that it is only a one-dimensional re-weighting of the magnitude-redshift relation. Section C3 of \citet{Hildebrandt/etal:2017} highlights the differences in the population in different colour spaces between the spectroscopic sample and the KiDS sample. As these differences are essentially unaccounted for in our SPEC method we expect that it could bias our estimation of the redshift distribution systematically.  In contrast the COSMOS-30 catalogue provides a complete and representative sample for the KiDS-$i$-800 data, with the drawback that redshifts are photometrically estimated. 

\am{A drawback of both the COSMOS-30 method and the SPEC method, is that the calibration samples represent small patches in the Universe. COSMOS imaging spans \SI{2}{\square\degree} while the spectroscopic data, z-COSMOS and CDFS collectively, span roughly \SI{1.2}{\square\degree} with two independent lines-of-sight.  We compute the variance between ten instances of randomly sub-sampling the $i$-band magnitude distribution from the SPEC or COSMOS-30 catalogue.  This `bootstrap' error analysis will not however include sampling variance errors.  The resulting redshift distributions} can be compared to the more representative \SI{343}{\square\degree} of homogenous spectroscopic data used in the cross-correlation technique.  The depleted number density of galaxies with redshifts $0.2<z<0.4$ determined using the cross-correlation technique, in comparison to source redshift distributions determined using the SPEC and COSMOS-30 estimates, could be an indication that the SPEC and COSMOS-30 methods are subject to sampling variance in this redshift range. 

Aside from suppressing sample variance, the cross-correlation method (CC) bypasses the need for a complete spectroscopic catalogue. On the other hand, however, the cross-correlation method (CC) is hindered by the impact of unknown galaxy bias, which tends to skew the clustering-redshifts to higher values if galaxy bias increases with redshift. One caveat of this method is that linear, deterministic galaxy bias may not apply on small scales. Our method to mitigate this effect using the $i$-band magnitude is reasonable given the level of accuracy required in this analysis, but for future studies this uncertainty will need to be addressed. In addition, the limited number of high-redshift objects in the spectroscopic catalogues that we have used makes it difficult for the clustering analysis to constrain the high-redshift tail of the distribution. 

As there are pros and cons associated with each of the methods that we employ to determine the source redshift distribution, we present the galaxy-galaxy lensing analysis that follows using all three estimations. While we can constrain the statistical uncertainty of each of the estimates using our bootstrap analyses, we rely on the spread between the resulting lensing signals to reflect our systematic uncertainty in the $i$-band redshift distribution.

\section{Comparison of i-band and r-band shape catalogues} \label{sec:ir}
We define the effective number density of galaxies following \citet{Heymans2012}, as
\begin{equation}
\centering
    n_{\mathrm{eff}}=\frac{1}{A}\frac{(\Sigma_jw_j)^2}{\Sigma_jw_j^2} \, ,
	\label{eq:neff}
\end{equation}
where $A$ is the total unmasked area and $w_j$ the \textit{lens}fit weight for galaxy $j$. This definition gives the equivalent number density of unit-weight sources with a total ellipticity dispersion, per component, $\sigma_{\epsilon}$, that would create a shear measurement of the same precision as the weighted data. We define the observed ellipticity dispersion as,
\begin{equation}
    \sigma_{\epsilon}^2=\frac{1}{2}\frac{\Sigma_jw_j^2\epsilon_j\bar{\epsilon}_j}{\Sigma_jw_j^2} \, ,
	\label{eq:sige}
\end{equation}
where $\epsilon$ is the observed complex galaxy ellipticity (see equation~\ref{eqn:ellipticity}).  For KiDS-$i$-800 we find $n_{\rm eff} = 3.80$ galaxies arcmin$^{-2}$ with an ellipticity dispersion of $\sigma_\epsilon=0.289$.  This can be compared to KiDS-$r$-450 with $n_{\rm eff} =8.5$ galaxies arcmin$^{-2}$ and $\sigma_\epsilon = 0.290$. 

\begin{figure}
	\includegraphics[width=\columnwidth]{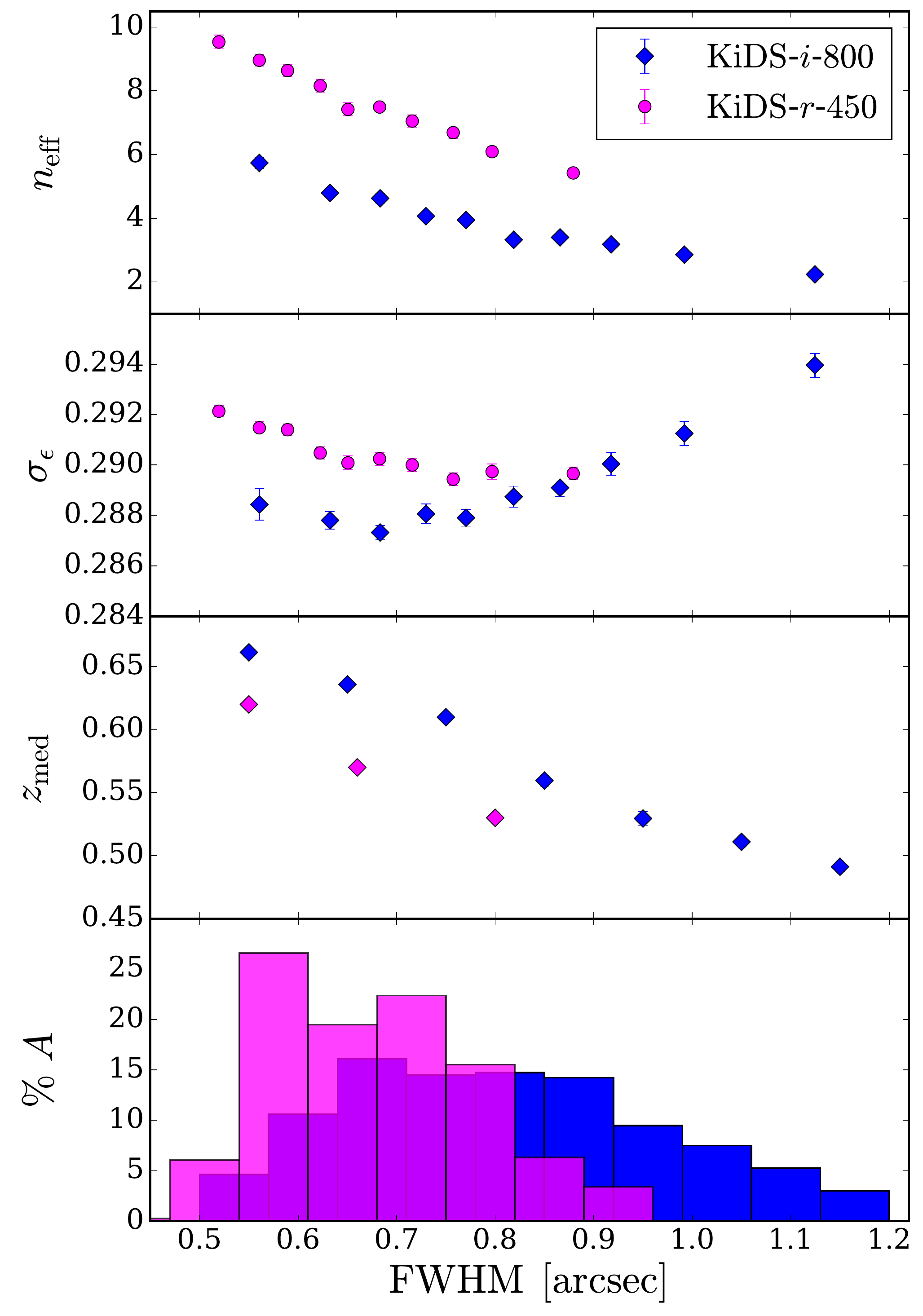}
    \caption{\label{fig:neff}The variation of the effective number density, $n_{\mathrm{eff}}$, (measured in galaxies arcmin$^{-2}$), the observed ellipticity dispersion per component, $\sigma_\epsilon$, the median redshift of the estimated redshift distribution, $z_{\mathrm{med}}$ and the percentage area of the survey, $A$, with the seeing of the data. The KiDS-$r$-450 data is plotted in pink and the KiDS-$i$-800 in blue. Note that the KiDS-$r$-450 data has the high photometric redshift limit imposed at $z_B < 0.9$. Error bars plotted for the upper three panels are the outcome of a bootstrap analysis.}
\end{figure}

In Figure~\ref{fig:neff}  we compare the effective number density, $n_{\mathrm{eff}}$, the ellipticity dispersion, $\sigma_{\epsilon}$, the median redshift and the percentage areal coverage to the observed $r$- and $i$-band seeing.  The upper panel of Figure~\ref{fig:neff} shows that the KiDS-$i$-800 data have a lower effective number density than that of the KiDS-$r$-450 sample by a factor of roughly two over the full seeing range. This reflects the different depths of the KiDS $r$- and $i$-band observations. The second panel demonstrates that as the seeing in the $i$-band degrades, the observed ellipticity dispersion remains constant to a few percent. We see a very small effect of an increase in shape measurement noise ($\epsilon^{\rm n}$ in equation~\ref{eq:eobs}) as the fraction of galaxies with a size that is comparable with the PSF grows. 
Overall, we see that the total effective number of galaxies in each of the two datasets are roughly comparable with 10.0 million in KiDS-$i$-800 and 10.8 million in KiDS-$r$-450, after applying the photometric redshift limitations of $0.1<z_{\rm B}<0.9$.  Therefore, the large-scale area of KiDS-$i$-800 still qualifies it as a competitive dataset. 

%\newpage
\begin{table*}
\caption{\label{tab:surveys}Number densities of weak lensing source galaxies drawn from KiDS \citep{Kuijken/etal:2015,Hildebrandt/etal:2017}, HSC \citep{Mandelbaum2017}, RCSLenS \citep{Hildebrandt/etal:2016}, CFHTLenS \citep{Heymans2012}, DLS \citep{Jee/etal:2013}  and DES \citep{jarvis/etal:2016}. The second column shows the effective area that the dataset spans in deg$^2$ (equation~\ref{eq:neff}), although we note that the numbers quoted from DLS and HSC may have been defined differently in comparison to the other surveys in this table, the third shows the median FWHM seeing of the data, measured in arcsec, the fourth shows the weighted effective number density of galaxies arcmin$^{-2}$, the fifth column details the observed ellipticity dispersion per component and the sixth column shows the estimated median redshift of the galaxy sample. The DES measurements correspond to their primary shape measurement algorithm, NGMIX. The bracketed numbers for RCSLenS correspond to the reduced area where $griz$-band coverage exists, as opposed to their single-band dataset.} 
\begin{center}
\begin{tabular}{lccccc}
 \hline
 Sample & A [deg$^2$] & FWHM [arcsec] & $n_{\rm{eff}}$ [galaxies arcmin$^{-2}$] & $\sigma_{\epsilon}$ &  $z_{\rm{med}}$\\%& FoM \\
 \hline
DLS & 20 & 0.88 &  $\sim$21.0 &  & $\sim1.0$\\%&\\
HSC Y1 & 137 & 0.58 & 21.8 & 0.24 & $\sim0.85$\\%&\\
DES SV & 139 & 1.08 & 6.8 & 0.265& $\sim0.65$\\% & 442/598\\ %\cline{2-5}
CFHTLenS & 126 & <0.8 & 15.1 & 0.280 & 0.7 \\%& 1400\\ %\cline{2-5}
RCSLenS & 572(384) & <1.0 & 5.5(4.9) & 0.251 & $\sim0.6$ \\%& 708\\ %\cline{2-5}
KiDS-$r$-450 & 360 & 0.66 & 8.5 & 0.290 &0.57\\%& 923\\
KiDS-$i$-800 & 733 & 0.79 &  3.8 & 0.289 & $\sim0.5$ \\%& 501\\
\hline
\end{tabular}
\end{center}
\end{table*}

Using the magnitude-weighted spectroscopic method (SPEC, Section~\ref{sec:specz}) to estimate the $i$-band redshift distribution, we show, in the third panel of Figure~\ref{fig:neff}, how the variable seeing KiDS-$i$-800 observations changes the depth of the sample of galaxies, with a higher median redshift for the better-seeing data.   The same trend can be seen for the DIR $r$-band median redshift for three seeing samples, noting that a high photometric redshift limit of $z_B < 0.9$ has been imposed for KiDS-$r$-450, lowering the overall median redshift in comparison to KiDS-$i$-800.

Finally, the lowest panel of Figure~\ref{fig:neff} presents the seeing distribution of the KiDS data, with the poorest seeing for KiDS-$r$-450 at a sub-arcsec level, while the KiDS-$i$-800 data extends to a FWHM of 1.2 arcsec. This figure illustrates that the KiDS-$i$-800 is a conglomerate of widely-varying quality data, in terms of seeing, and as a result, in terms of galaxy number density and depth.   In Table~\ref{tab:surveys} the survey parameters of KiDS-$i$-800 can be compared to other existing surveys: KiDS-$r$-450, HSC Y1, DES SV, RCSLenS, CFHTLenS and DLS. We order the surveys by their unmasked area and quote the median FWHM and median redshift of the data. We quote values for the number of galaxies arcmin$^{-2}$ using the definition given in equation~\ref{eq:neff} and the ellipticity dispersion as in equation~\ref{eq:sige}. 

To compare the shear measurement in KiDS-$i$-800 and KiDS-$r$-450, the most straightforward analysis would appear to be a direct galaxy-by-galaxy test \citep[see for example][]{heymans/etal:2005}. This would only be appropriate, however, if we had an unbiased shear measurement per galaxy. Even with perfect modelling and correction for the PSF, each shape catalogue consists of a noisy ellipticity estimate per galaxy, $\epsilon^{\rm n}$(equation~\ref{eq:eobs}).  As ellipticity is a bounded quantity $|\epsilon| < 1$, the presence of noise will always result in an overall reduction in the measured average galaxy ellipticity of a sample, an effect that has been termed `noise bias' \citep{melchior/viola:2012}. The impact of noise bias when using observed galaxy ellipticities as a shear estimate can be calibrated and accounted for \citep[see for example][]{fenech-conti/etal:2016}. This calibration correction, however, only applies when considering an ensemble of galaxies.    A secondary issue for a galaxy-by-galaxy comparison of two catalogues from different filters arises from colour gradients in galaxies \citep{voigt/etal:2012}.  With a strong colour gradient, the intrinsic ellipticity of the object, when imaged in a blue filter, could be rather different from the intrinsic ellipticity of the same object when viewed in a red filter \citep[see for example][]{Schrabback2017}. For these two reasons we do not perform any direct galaxy-by-galaxy comparisons, favouring instead tests where we should recover the same shear measurement from the ensemble of galaxies.

In this section we subject the $i$- and $r$-band shape catalogues to two different tests; a `nulled' two-point shear correlation function which tests the difference in the shear recovered for a sample of galaxies with shape measurements in both bands, and a galaxy-galaxy lensing analysis which provides a joint-test of the shape and photometric redshift measurements for the full catalogue in each band.

\subsection{The `nulled' two-point shear correlation function}

Using the matched $ri$ catalogue described in Section~\ref{sec:matchcat}, we calculate the uncalibrated (the multiplicative calibrations are applied later to the ensemble) two-point shear correlation function, $\xi_\pm$, as a function of angular separation $\theta$, for three combinations of the $i$ and $r$-band filters, $({\rm fg}) = (ii), (ir), (rr)$, with 
\be
\xi_\pm^{\rm fg}(\theta) = \frac{\Sigma w^{ir}(\bm{x_a}) w^{ir}(\bm{x_b})  [ \epsilon_{\rm t}^{\rm f}(\bm{x_a}) \epsilon_{\rm t}^{\rm g}(\bm{x_b}) \pm \epsilon_{\times}^{\rm f} (\bm{x_a})\epsilon_{\times}^{\rm g} (\bm{x_b})]}{\Sigma w^{ir}(\bm{x_a}) w^{ir}(\bm{x_b})  } \, .
\label{eqn:xipm}
\ee
Here the weighted sum is taken over galaxy pairs with $|\bm{x_a} - \bm{x_b}|$ within the interval $\Delta\theta$ around $\theta$.
The tangential and rotated ellipticity, $\epsilon_{\rm t}$ and $\epsilon_{\times}$, are determined via a tangential projection of the ellipticity components relative to the vector connecting each galaxy pair \citep{Bartelmann2001}.    For all filter combinations the weights, $w^{ir}= \sqrt{w_i w_r}$, use information from both the $i$ and $r$-band analyses such that the effective redshift distribution of the matched sample is \am{identical} for each measurement.

We calculate empirically any additive bias terms for our matched $ri$ catalogues using $c_i = \langle \epsilon_i \rangle$, where the average now takes into account the combined weight $w^{ir}$.  We apply this calibration correction to both the $i$ and $r$-band shapes, per patch on the sky, in the matched catalogue where on average, $c_1^r=0.0001 \pm 0.0001$, $c_2^r=0.0008 \pm 0.0001$, $c_1^i=0.0009 \pm 0.0001$, $c_2^i=0.0010 \pm 0.0001$.  This level of additive bias is similar to that of the full KiDS-$i$-800 and KiDS-$r$-450 samples.
 
Following \citet{miller/etal:2013}, the ensemble `noise bias' calibration correction for each filter combination is given by
\be
1 + {\rm K}^{\rm fg}(\theta) =  \frac{\Sigma w^{ir}(\bm{x_a}) w^{ir}(\bm{x_b})  [ 1+m^{\rm f}(\bm{x_a})][ 1+m^{\rm g}(\bm{x_b})]}{\Sigma w^{ir}(\bm{x_a}) w^{ir}(\bm{x_b}) } \, ,
\label{eqn:Kcal}
\ee
where $m^f(\bm{x_a})$ is the multiplicative correction for the galaxy at position $(\bm{x_a})$ imaged with filter $f$.  These multiplicative corrections are calibrated as a function of signal-to-noise and relative galaxy-to-PSF size using image simulations \citep{fenech-conti/etal:2016}.   For this matched $ri$ sample the \citet{fenech-conti/etal:2016} calibration corrections are found to be small and independent of scale, with $1 + {\rm K}^{rr} = 0.996$,  $1 + {\rm K}^{ir} = 0.987$ and $1 + {\rm K}^{ii} = 0.978$.

We define two `nulled' two-point shear correlation functions as 
\be
\xi_\pm^{\rm null}(\theta) = \frac{\xi_\pm^{ii}(\theta)}{1 + {\rm K}^{ii}(\theta)}  - \frac{\xi_\pm^{rr}(\theta)}{1 + {\rm K}^{rr}(\theta)}  \, ,
\label{eqn:null}
\ee
\be
\xi_\pm^{\rm x-null}(\theta) = \frac{\xi_\pm^{ir}(\theta)}{1 + {\rm K}^{ir}(\theta)}  - \frac{\xi_\pm^{rr}(\theta)}{1 + {\rm K}^{rr}(\theta)} \, ,
\label{eqn:xnull}
\ee
which, for a matched catalogue in the absence of unaccounted sources of systematic error, would be consistent with zero.  
The three different matched-catalogue measurements of $\xi_\pm^{\rm fg}$ will be subject to the same cosmological sampling variance error.  The covariance matrix for our `nulled' two-point statistics therefore, derives only from noise on the shape measurement in addition to noise arising from differences in the source intrinsic ellipticity when imaged in the $r$- or $i$-band (see Appendix~\ref{app:covderiv}).  As such the covariance is only non-zero on the diagonal and given by 
\be
C^{\rm null}_{\xi}(\theta_j,\theta_j) = \frac{4}{N_{\rm p}(\theta_j)}(\sigma^4_i + \sigma^4_r - 2 \sigma^4_{\rm int}) \, ,
\ee
\be
C^{\rm x-null}_{\xi}(\theta_j,\theta_j) = \frac{2}{N_{\rm p}(\theta_j)}[2\sigma^4_r + \sigma^4_{\rm int} + \sigma^2_r (\sigma^2_i-4\sigma^2_{\rm int})] \, .
\label{eqn:Cxnull}
\ee
Here $\sigma_i^2$ and $\sigma_r^2$ are the measured weighted ellipticity variance, per component (as defined in equation~\ref{eq:sige}), of the matched catalogue in the $i$- and $r$-band, respectively. For a single ellipticity component, $\sigma^2_{\rm int}$ is the variance of the part of the intrinsic ellipticity distribution that is correlated between the $i$- and the $r$-band and $N_{\rm p}(\theta)$ counts the number of pairs in each angular bin which is given by
\be
N_{\rm p}(\theta) = \pi (\theta_u^2 - \theta_l^2) A \, n_{\rm eff}^2 \,.
\label{eqn:Np}
\ee
Here $n_{\rm eff}$ is the effective number density as given in equation~\ref{eq:neff}, $\theta_u$ and $\theta_l$ are the angular scales of the upper and lower bin boundaries and $A$ is the effective survey area \citep[][see also Appendix~\ref{app:covderiv}]{Schneider/etal:2002}.  For the $ri$ matched catalogue, we measure $\sigma_i=0.296$, $\sigma_r=0.265$, $n_{\rm eff}=3.64$ arcmin$^{-2}$ and we make an educated guess for $\sigma_{\rm int}=0.255$, based on SDSS measurements of the low-redshift intrinsic ellipticity distribution \citep[see the discussion in][]{miller/etal:2013, chang/etal:2013, Kuijken/etal:2015}.  Note that we choose not to include the uncertainty in the additive or multiplicative calibration corrections from equation~\ref{eqn:Kcal} into our analytical error estimate for the nulled shear correlation functions, as this is smaller than our uncertainty on the value of the intrinsic ellipticity distribution $\sigma_{\rm int}$.

\begin{figure}
	\includegraphics[width=\columnwidth]{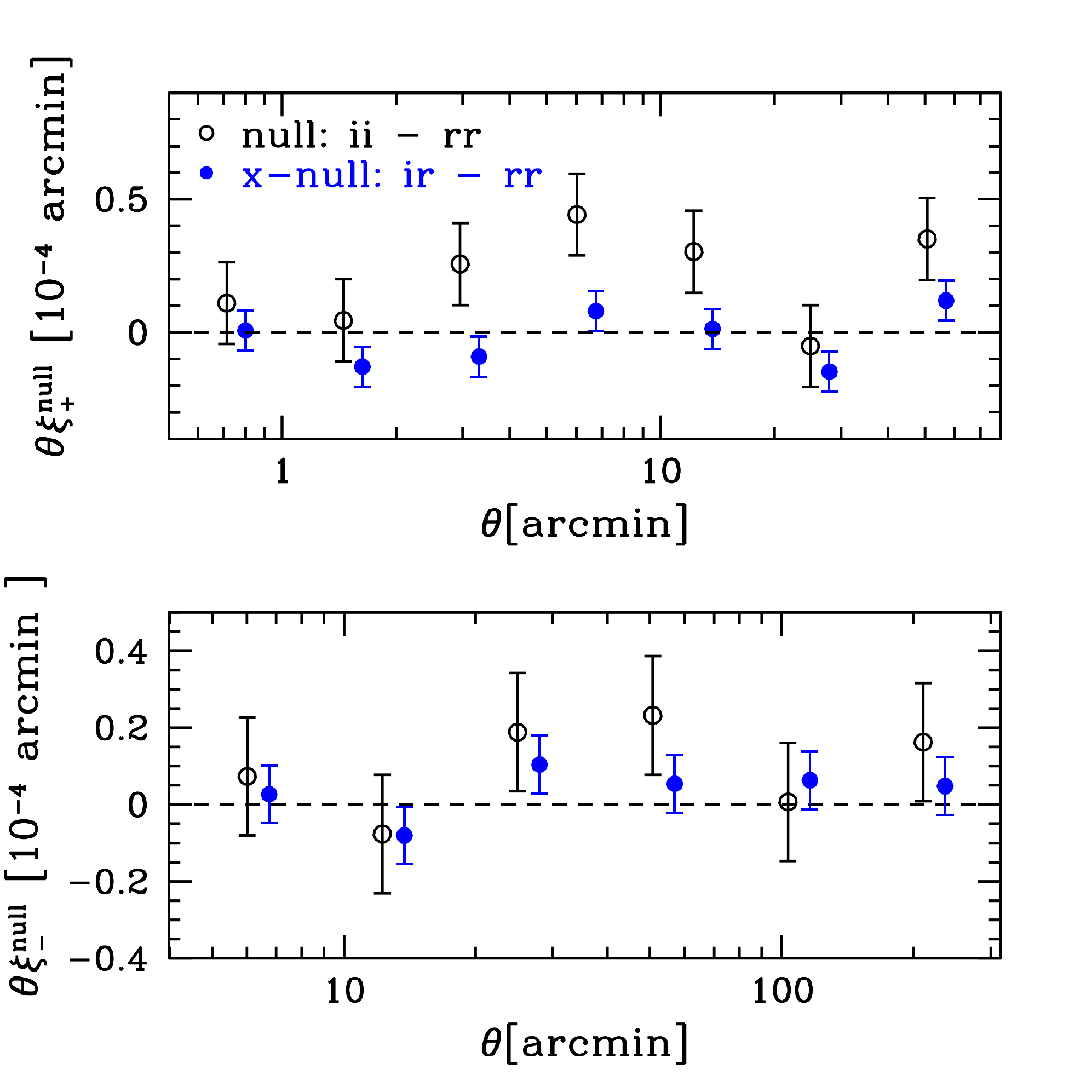}
    \caption{\label{fig:nulltwopt}The `nulled' two-point shear correlation functions $\xi_\pm^{\rm null}$ (open) and  $\xi_\pm^{\rm x-null}$ (closed).  Both the upper panel, $\xi_+$, and the lower panel, $\xi_-$, are scaled by $\theta$ to highlight any differences from zero on large scales. }
\end{figure}

Figure~\ref{fig:nulltwopt} presents measurements of $\xi_\pm^{\rm null}$ and $\xi_\pm^{\rm x-null}$.  In the upper panel of Figure~\ref{fig:nulltwopt} we find $\xi_+^{\rm null}$ to be significantly different from zero on scales $\theta >$ \SI{2}{\arcmin}.  
\ch{
Defining $\chi^2_{\rm null}$ as 
\be
\chi^2_{\rm null} = \sum_i \frac{\xi^{\rm null}_{\pm}(\theta_i)^2}{C^{\rm null}_{\xi}(\theta_i, \theta_i)} \, ,
\label{eqn:chinull}
\ee 
we find the `nulled' two point shear correlation function to be inconsistent with zero with 99 percent probability ($\chi^2_{\rm null} = 26.2$ for $13$ data points).  These results allow us to conclude that unaccounted sources of systematics exist, which have a scale dependence; this is not surprising given the non-zero PSF contamination ($\alpha$), described in Section~\ref{sec:sys}.  This null-test therefore supports our conclusion that KiDS-$i$-800 is not suitable for cosmic shear studies. Interestingly these systematics appear to contribute roughly equally to tangential and rotated correlations, such that they approximately null themselves in the $\xi_-$ statistic in the lower panel of Figure~\ref{fig:nulltwopt}.  Limiting the $\chi^2_{\rm null}$ calculation to only the $\xi^{\rm null}_{-}$ measurements we find that $\xi^{\rm null}_{-}$ is consistent with zero with $\chi^2_{\rm null} = 5.3$ for $6$ data points.} 

\ch{
We define $\chi^2_{\rm x-null}$ by replacing the `nulled' two point shear correlation function with the `cross-null' statistic $\xi_\pm^{\rm x-null}$ in equation~\ref{eqn:chinull}.   We find $\xi_\pm^{\rm x-null}$ to be consistent with zero with 16 percent probability ($\chi^2_{\rm x-null} = 16.8$ for $13$ data points).  It has an average value over angular scales, using inverse variance weights, of $\langle \xi_\pm^{\rm x-null} \rangle=(3.9 \pm 3.0) \times 10^{-8}$.  From this we can conclude that the unaccounted sources of systematics highlighted by the $\xi_\pm^{\rm null}$ statistic are uncorrelated with the $r$-band catalogue.  Importantly, finding a null result with this `cross-null' statistic demonstrates that the multiplicative shear calibration corrections for the $i$ and $r$ catalogues in equation~\ref{eqn:Kcal} produce consistent results.  The inverse variance weighted average value of $\langle \xi_\pm^{\rm x-null}/\xi_\pm^{\rm rr}\rangle = 0.010 \pm 0.035$.  In this `cross-null' analysis that isolates the impact of multiplicative shear bias, the amplitude of the shear correlation functions for the matched $r$ and $i$ catalogues therefore agree at the level of $1 \pm 4$ percent.
}

\am{In this analysis we used the KiDS-$r$-450 auto-correlation function $\xi^{rr}_\pm$ as the `truth' in the `cross-null' statistic of equation~\ref{eqn:xnull}.  Given the significant KiDS-$i$-800 PSF residuals uncovered in Section~\ref{sec:sys} this was a sensible choice to make.  In future cases, however, where both surveys have demonstrated low-levels of additive bias before the comparison analysis, the `cross-null' statistic should be defined using both auto-correlations.  This `cross-null' test can then be used as a diagnostic to isolate which survey is the most trustworthy, in the event that the initial `null' test of equation~\ref{eqn:null} has failed.  }

\am{In the example where a `cross-nulled' $ir-rr$ signal is consistent with zero, but the companion `cross-nulled' $ir-ii$ signal is significant, we can conclude that unaccounted additive sources of systematics exist in the $i$-survey, which are uncorrelated with the $r$-survey.}

\am{In the example where both `cross-nulled' signals are significant, the angular scale-dependence of these null-statistics can be analysed in order to distinguish between additive bias, which usually becomes significant on large scales,  and multiplicative bias, which impacts all scales equivalently.}

\am{The `nulled' two-point statistics defined in this section differ from the `differential shear correlation' proposed by \cite{jarvis/etal:2016}. The differential statistic derives from a galaxy-by-galaxy comparison of the ellipticities in contrast to our chosen statistic which compares the calibrated ensemble averaged shear.  We would argue that the \citet{jarvis/etal:2016} approach is only appropriate when one is able to determine an unbiased shear measurement per galaxy.  Our methodology also differs from the `split' cosmic shear analyses conducted in \citet{becker2016} and \citet{troxel2017}.  Here the source sample is divided into groups based on a number of observational properties such as seeing, PSF ellipticity, sky brightness and observed object size and signal-to-noise.  These groupings change the effective redshift distribution of each sample and introduce object selection bias \citep[see for example][]{fenech-conti/etal:2016}, adding an extra layer of complexity when interpreting the results of this split-analysis.   If these changes can be accurately modelled, however, the `split' cosmic shear analyses can provide a powerful tool to investigate the dependence of different systematic biases on a range of observational properties. Our approach of matching catalogues before conducting our `null'  cosmic shear tests removes this layer of complexity.}

Finally we note the reduced \am{uncertainties} for the `cross-null' two-point statistic in comparison to the `null' statistic, in addition to the reduction of systematic errors.  These two features of this multi-band cosmic shear analysis supports the \citet{jarvis/jain:2008} proposal to combine shear information from multiple filters to gain in effective number density, particularly if there are unknown, but uncorrelated systematic errors in each band.  This idea will be explored further in future work.

\subsection{Galaxy-galaxy Lensing Signal} \label{sec:ds}
Statistically, galaxy-galaxy lensing can be viewed as a 2-D measurement of the cross-correlation, $\xi_{\rm gm}$, of a baryonic tracer, such as a galaxy, as a relative overdensity with the fractional overdensity in the matter density field, separated by a comoving separation in 3-D space, ${\bf r}$, expressed as,
\be
\xi_{\rm gm}({\bf r})= \langle \delta_{\rm g} ({\bf x})  \delta_{\rm m} ({\bf x}+{\bf r}) \rangle_{\rm {\bf x}} \, .
\ee
For a given cosmology, the relative amplitude of the galaxy-galaxy signals from different source samples will reflect their redshift distribution and any shear calibration systematics. As such, this measurement is commonly used to test the redshift scaling of weak lensing shear measurements \citep{Hoekstra/etal:2005,Mandelbaum/etal:2005,Heymans2012, Kuijken/etal:2015, Schneider2016, Hildebrandt/etal:2017}.   In this section we compare measurements of the galaxy-galaxy lensing signal from KiDS-$i$-800 and KiDS-$r$-450 using a common set of lens samples from GAMA, BOSS and 2dFLenS, as described in Section~\ref{sec:lenses}. This comparison provides an opportunity to assess the impact of using the different estimations of the $i$-band redshift distributions from Section~\ref{sec:zdata} and the shear calibration, $m$, of the variable seeing KiDS-$i$-800 background galaxies, in comparison to the same measurement using KiDS-$r$-450 shapes.  

\subsubsection{Theory}
The galaxy-galaxy lensing cross-correlation, $\xi_{\rm gm}(\bf{r})$ can be related to the comoving projected surface mass density around galaxies, $\Sigma$ with a comoving projected separation, $R$, as
\am{
\be
\Sigma (R) = \rho_{\rm m,0} \int_{0}^{\chi(z_{\rm s})}  \ \xi_{\rm gm} \big( \sqrt{R^2+[\chi-\chi(z_{\rm l})]^2} \big) \ \rm{d}\chi\, ,
\ee
where $\chi(z_{\rm l}),\chi(z_{\rm s})$ are the comoving distances to the lens and source galaxies, respectively, $ \rho_{\rm m,0}$ is the matter density of the Universe today and $\chi$ is the comoving line-of-sight separation}. The shear is a measurement of the over-density in the matter distribution, therefore, it is a measure of the excess or differential surface density \citep{Mandelbaum/etal:2005},
\be
\Delta \Sigma (R)=\overline{\Sigma}(\leq R) - \Sigma(R) \, ,
\ee
where 
\be
\overline{\Sigma}(\leq R)=\frac{2}{R^2} \int^R_0 \Sigma(R') R' \rm{d}R' \, .
\ee
The comoving differential surface mass density\am{\footnote{We note here that $\Sigma$ and $\Sigma_{\rm c}$ refer to the comoving quantities, which differ from the respective quantities expressed in physical units by factors of $(1+z_{\rm l})^2$ \citep[see the discussion in][]{AmonEG, Dvornik2018}. In previous analyses of the KiDS survey \citep[for example, ][]{vU2016, Dvornik2017}, these symbols have denoted the physical quantities.}} can be related to the tangential shear distortion $\gamma_{\rm t}$ of the background sources as 
\be
\Delta\Sigma(R)=\gamma_{\rm t} \Sigma_{c} \, ,
\ee
in terms of a geometrical factor that accounts for the lensing efficiency, the comoving critical surface mass density, which is defined as
\be
\Sigma_{c}=\frac{c^2}{4\pi G} \frac{\chi(z_{\rm s})}{\chi(z_{\rm l}) \, \chi(z_{\rm l},z_{\rm s}) \, (1+z_{\rm l})} \, ,
\ee
where $z_{\rm l}$ is the redshift of the lens, $\chi(z_{\rm l})$ is the comoving radial co-ordinate of the lens at redshift $z_{\rm l}$, $\chi(z_{\rm s})$ is that of the source at redshift $z_{\rm s}$ and  $\chi(z_{\rm l},z_{\rm s})$ is the comoving distance between the source and the lens. Comoving separations are determined assuming a flat $\Lambda$CDM cosmology with a Hubble parameter of $H_0=100\, h \,\rm{km \, Mpc^{-1}\,  s^{-1}}$, fixing the matter density to $\Omega_{\rm m}=0.277$ \citep{Komatsu2011}.  Given our statistical power, the galaxy-galaxy lensing measurements are fairly insensitive to the choice of fiducial cosmology, and as such, using a \cite{Planck2016} cosmology would not significantly impact our analysis. 

\subsubsection{Estimators}

The azimuthal average of the tangential ellipticity of a large number of galaxies in the same area of the sky is an unbiased estimate of the shear, in the absence of systematics. Following this, the galaxy-galaxy lensing estimator is calculated as a function of angular separation, $\theta$, as the weighted sum of the tangential ellipticity of the source-lens pairs, $\epsilon_{\rm t}$  as,
\be
\gamma_{\rm t}(\theta)=\frac{\sum^{\rm N_{pairs}}_{jk}  w_{\rm s}^j w_{\rm l}^k \epsilon_{\rm t}^{jk}}{\sum^{\rm N_{pairs}}_{jk} w_{\rm s}^j w_{\rm l}^k} \, ,
\ee
where $w_{\rm s}$ are the \textit{lens}fit weights of the sources and $w_{\rm l}$ are the weights of the lenses. For this measurement we employ the \textsc{athena} software of \citet{Kilbinger2014}.

The estimator for the excess surface mass density is defined as a function of the projected radius, $R$, from the lens and the spectroscopic redshift of the lens, $z_{\rm l}$, in terms of the inverse critical surface mass density,
\be
\overline{\Delta\Sigma}(R,z_{\rm l})=\frac{\gamma_{\rm t}(R/\chi_{\rm l})}{\overline{\Sigma_c^{-1}}(z_{\rm l})} \, .
\ee
Lens galaxy samples are split by their spectroscopic redshifts into finely defined `slices' of width $\Delta z_{\rm l}=0.01$ and the inverse critical surface mass density is calculated per source-lens slice as,
\be
\overline{\Sigma_c^{-1}}(z_{\rm l}) =  \frac{4\pi G}{c^2} \, (1+z_{\rm l}) \, \chi(z_{\rm l}) \, \eta(z_{\rm l}) \, ,
%\begin{split}
%\overline{\Sigma_c^{-1}}(z_{\rm l},N(z_{\rm s})) =  \frac{4\pi G (1+z_{\rm l})\chi(z_{\rm l})}{c^2} \\
 %\int^\infty_{z_{\rm l}} dz_{\rm s} N(z_{\rm s}) & \left(1- \frac{\chi(z_{\rm l})}{\chi(z_{\rm s})}\right) \, .
%\end{split}
\ee
where $\eta(z_{\rm l})$, the lensing efficiency, is defined in equation~\ref{eqn:eta}. This geometric term accounts for the dilution in the lensing signal caused by the non-zero probability that a source is situated in front of the lens \citep{Miyatake:/etal2015, Blake2016eg}. It is computed for each lens using its spectroscopic redshift with the entire normalised source redshift probability distribution, $N(z)$. The tangential shear was measured in 7 logarithmic angular bins where the minimum and maximum $\theta$ angles were determined for each lens redshift via $R=\theta \, \chi(z_{\rm l})$, in order to satisfy a minimum and maximum comoving radii of $R=0.05$ and $R=2 \, h^{-1}$Mpc.

For the case of KiDS-$r$-450, with the availability of the $z_{\rm B}$ photometric redshift information per galaxy, the source sample could be further limited to those behind each lens slice, in order to minimise the dilution of the lensing signal due to sources correlated with the lens.  The stringency of this source redshift selection is investigated in Appendix~\ref{app:gglcorr} and a limit of $z_{\rm B}>z_{\rm l}+0.1$, is deemed optimal. We calculate the tangential shear and the differential surface mass density, $\Delta\Sigma (R)$, for each of the $N$ lens slices and stack these signals to obtain an average differential surface mass density, weighted by the number of pairs in each slice as,
\be
\centering
\overline{\Delta\Sigma}(R)=\frac{\sum^{N}_i (\gamma_{\rm t}(R/\chi_{\rm l})/\overline{\Sigma_c^{-1}})^i n_{\rm{pairs}}^i}{\sum^{N}_i n_{\rm{pairs}}^i} \frac{1}{1+K} \, ,
\ee
where 
\begin{equation}
\centering
K=\frac{\sum_{\rm s} w_{\rm s} m_{\rm s}}{\sum_{\rm s} w_{\rm s}} \, .
\end{equation}
This factor accounts for the multiplicative noise bias determined for each source galaxy, $m_{\rm s}$, weighted by its \textit{lens}fit weight $w_{\rm s}$. Note that we assume that there is no significant dependence of the multiplicative calibration on the source redshift and therefore $\overline{\Sigma_c^{-1}}$. This was deemed suitable as this calibration is at the percent level for the ensemble. 

Two corrections were made to the galaxy-galaxy lensing signal. Firstly, the excess surface mass density was computed around random points in the areal overlap. Random catalogues were generated following the angular selection function of the spectroscopic surveys, where we used a random sample 40 times bigger than the data sample. This signal has an expectation value of zero in the absence of systematics. As demonstrated by \cite{Singh2016}, it is important that a random signal, $\Delta\Sigma_{\rm{rand}}(R)$, is subtracted from the measurement in order to account for any small but non-negligible coherent additive bias of the galaxy shapes and to decrease large-scale sampling variance. The random signals determined for both KiDS-$i$-800 and KiDS-$r$-450 were found to be consistent with zero for each lens sample.  We present the random signals for each lens sample in Appendix~\ref{app:gglcorr}.

Secondly, as the estimates of the redshift distributions of the source galaxies have an associated level of uncertainty, it is necessary to account for the contamination of the clustering of source galaxies with the lens galaxies. Any sources that are physically associated with the lenses would not themselves be lensed and would therefore bias the lensing signal low at small transverse separations. To correct for this, we determine the `boost factor' for each lens-source sample and amplify the excess surface mass density measurement by it, multiplicatively. We investigate the implication of redshift cuts on this factor in Appendix~\ref{app:gglcorr}.  We assume that the boost signal originates from source-lens clustering and ignore any contribution from weak lensing magnification, which can also alter the number of sources behind the lens, as \cite{Schrabback2017} showed that this is only a small net effect. The overdensity of source galaxies around the lenses is estimated as the ratio of the weighted number of source-lens galaxy pairs for real lenses to that of the same number of randomly positioned lenses (again, where the weights and the redshift distribution of the lens sample is preserved), following \citet{Mandelbaum2006} as,
 \begin{equation}
\centering
B(R)=\frac{\sum^{\rm N_{pairs}}_{jk} w_{\rm s}^j w_{\rm l}^k}{\sum^{\rm N_{pairs}}_{jk} w_{\rm s}^j w_{\rm l}^k({\rm rand)}} \, .
\label{eqn:boost}
\end{equation}
This prescription is determined for each lens slice and the average boost, $\overline{B}(R)$ computed, weighted by the number of source-lens pairs in each slice. Hence, the corrected excess surface mass density is measured as,
\be
\centering
\overline{\Delta\Sigma}_{\rm{corr}}(R)=[\overline{\Delta\Sigma}(R)-\Delta\Sigma_{\rm{rand}}(R)] \, \overline{B}(R) \, .
\label{eqn:dscorr}
\ee
We present the boost factors that we apply to each measurement in Appendix~\ref{app:gglcorr}.

\subsubsection{Results}
\begin{figure}
	\includegraphics[width=\columnwidth]{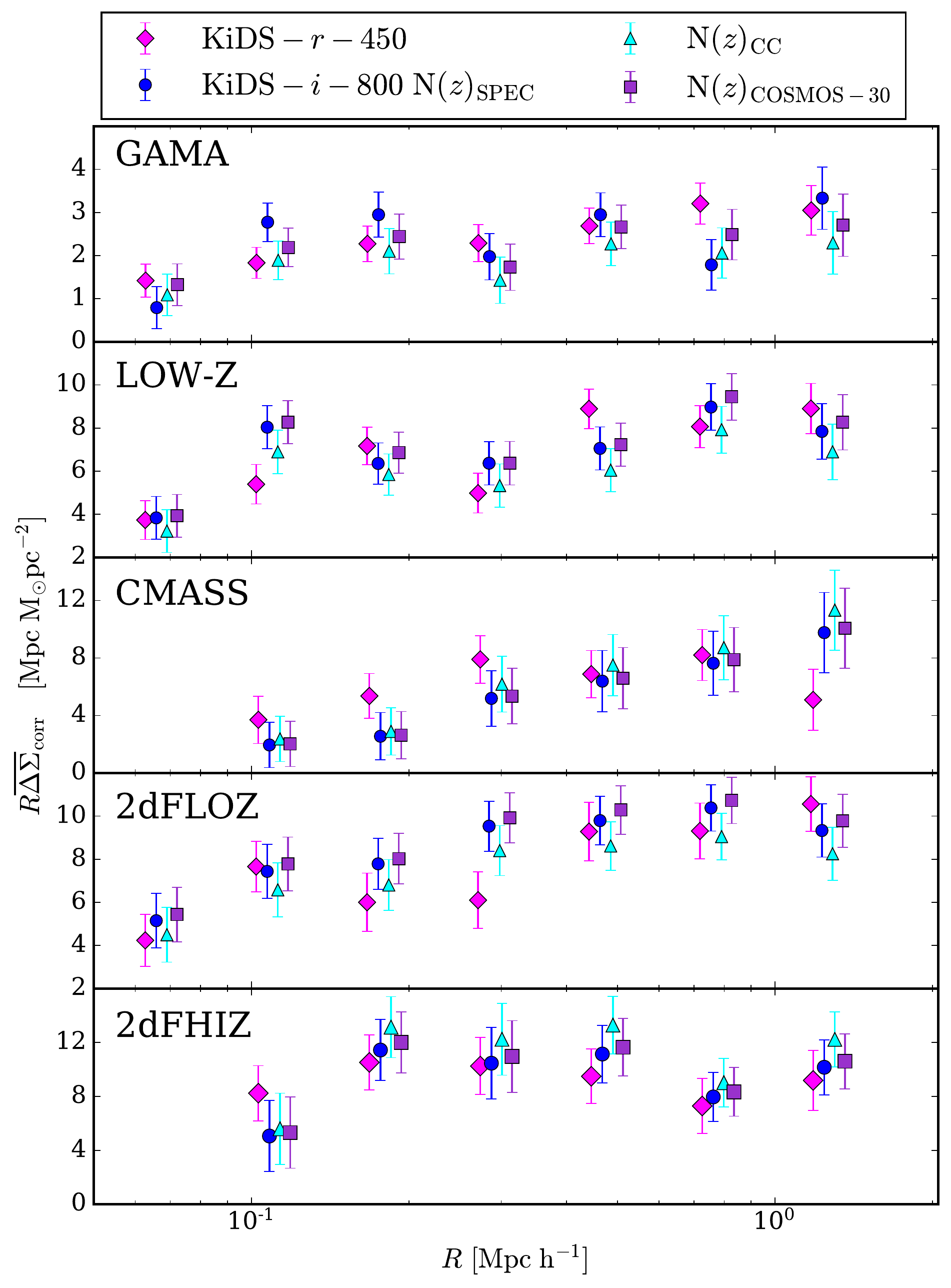}
    \caption{\label{fig:ds}The stacked differential surface mass density measurements $\overline{\Delta\Sigma}_{\rm{corr}}(R)$ for KiDS-$i$-800 (blue) and KiDS-$r$-450 (pink) galaxies with GAMA, LOWZ, CMASS, 2dFLOZ and 2dFHIZ lens galaxy samples, from top to bottom. Three KiDS-$i$-800 signals are shown- one for each of the three redshift distributions. Jack-knifed errors are determined and plotted in combination with the random signal error. Note that the errors here do not include our uncertainty on the redshift distributions. Random signals have been subtracted and measurements have had `boost' correction applied. \am{In order to distinguish between the measurements on large-scales, all signals are scaled by $R$} and data points are offset on the $R$-axis for clarity. }
\end{figure}

Figure~\ref{fig:ds} compares the KiDS-$i$-800 galaxy-galaxy lensing measurements, with the KiDS-$r$-450 measurement, for the five lens samples detailed in Section~\ref{sec:lenses}.  KiDS-$i$-800 measurements are made using each of the three estimated redshift distributions described in Sections~\ref{sec:specz}, \ref{sec:photoz} and \ref{sec:cc}.  The error bars are estimated using a Jackknife technique, where each Jackknife sample estimate is obtained by removing a single KiDS-$i$-800 pointing, such that the number of estimates corresponds to the number of pointings with a spectroscopic overlap. The signals were measured for projected separations of 0.05 $h^{-1} \rm{Mpc}$ up to 2.0 $h^{-1} \rm{Mpc}$, limited by the size of the Jackknife sample following \cite{Singh2016}. For the two high-redshift lens samples, CMASS and 2dFHIZ, we only consider scales of $R>0.08 \, h^{-1} \rm{Mpc}$ as for our high-redshift lens sample, the projected separation $0.08 \, h^{-1} \rm{Mpc}$ corresponds to an angular size smaller than the size of the \emph{lens}fit galaxy shape measurement postage stamp \citep{miller/etal:2013}.

As expected we see that the signal from the GAMA galaxies has the lowest amplitude as this lens sample is entirely magnitude limited, whereas the BOSS and 2dFLenS galaxies are samples of Luminous Red Galaxies (LRGs).  A magnitude-limited sample includes galaxies of a lower luminosity or higher number density.  These galaxies have a correspondingly lower bias factor and give rise to a lower amplitude lensing signal than the LRGs which tend to live in more massive halos. The 2dFLenS signals are higher than the BOSS counterparts as this luminosity-selected sample has a lower number density than BOSS and so preferentially selects dark matter halos of higher mass and hence a higher bias factor.

\begin{figure}
	\includegraphics[width=\columnwidth]{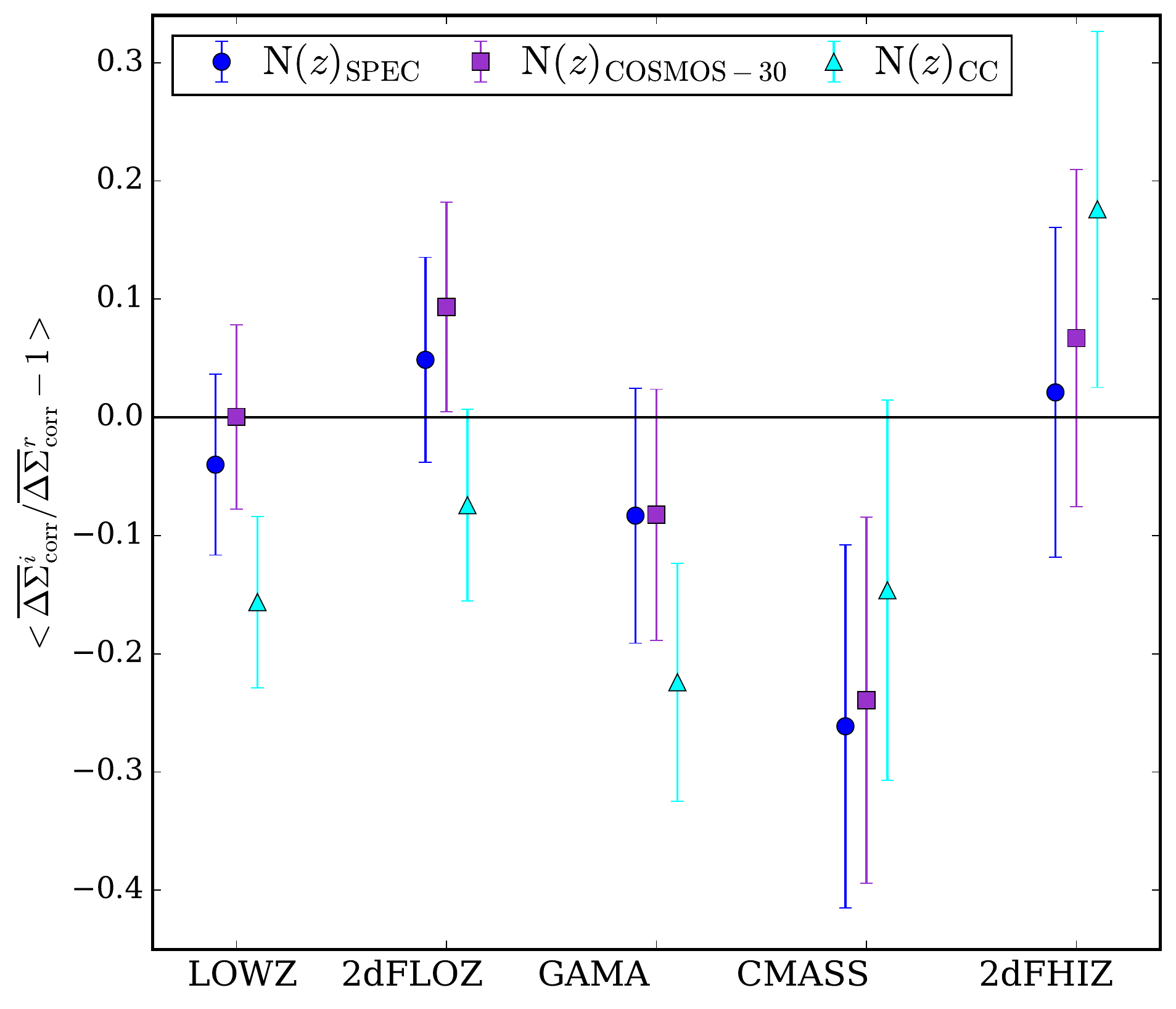}
    \caption{\label{fig:chisq}The average fractional difference between the KiDS-$i$-800 and KiDS-$r$-450 galaxy-galaxy lensing measurements, $\langle \varphi \rangle$ (an inverse variance-weighted combination of equation~\ref{eqn:varphi} over all scales), for each spectroscopic lens dataset using three different methods to estimate the redshift distribution of the $i$-band source galaxies. These measurements are inverse variance-weighted and do not include any uncertainty on the redshift distribution.}
\end{figure}

Figure~\ref{fig:chisq} shows the inverse variance-weighted average fractional difference over all scales, $ \langle \varphi \rangle$, between the KiDS-$i$-800 and KiDS-$r$-450 galaxy-galaxy lensing measurements, for each lens sample where
\be
\varphi(R)=\frac{\overline{\Delta\Sigma}^i_{\rm{corr}}(R)-\overline{\Delta\Sigma}^r_{\rm{corr}}(R)}{\overline{\Delta\Sigma}^r_{\rm{corr}}(R)} \, ,
\label{eqn:varphi}
\ee
with associated variance
\be
\sigma^2_{\varphi}(R)= \frac{\overline{\Delta\Sigma}^i_{\rm{corr}}(R)^2}{\overline{\Delta\Sigma}^r_{\rm{corr}}(R)^2}  \left(\frac{\sigma^2_{\overline{\Delta\Sigma}^i(R)}}{{\overline{\Delta\Sigma}^i_{\rm{corr}}}(R)^2}+\frac{\sigma^2_{\overline{\Delta\Sigma}^r(R)}}{{\overline{\Delta\Sigma}^r_{\rm{corr}}}(R)^2} \right) \, .
\ee
Here $\sigma^2_{\overline{\Delta\Sigma}}(R)$ is the error on the measurement of $\overline{\Delta\Sigma}_{\rm{corr}}(R)$ which is estimated using a Jackknife analysis.  For the purposes of this comparison we make the approximation that radial bins are uncorrelated, which is a reasonable approximation to make for scales $R< 1 \, h^{-1} \, {\rm Mpc} $ \citep[see Figure 5 in][]{viola/etal:2015}.   We also ignore the covariance between the KiDS-$i$-800 and KiDS-$r$-450 measurements which is appropriate given that the errors are dominated by intrinsic and measured ellipticity noise.  Furthermore, the $i$ and $r$-band catalogues contain at most 40 percent of the same source galaxies in the case of our GAMA analysis, where the entire area analysed has overlapping KiDS-$i$-800 and KiDS-$r$-450 data and  these overlapping galaxies have different weights in the different datasets (see Section~\ref{sec:matchcat}).  Typically the overlap of source galaxies is significantly less than this given the different on-sky distribution of the two surveys.   We present an inverse variance-weighted average over all angular scales as there is little angular dependence in the measured fractional difference $\varphi(R)$.

Figure~\ref{fig:chisq} shows that for each of the low-redshift lens samples, LOWZ, 2dFLOZ and GAMA, using the three different redshift distributions results in KiDS-$i$-800 measurements that are consistent with each other and with that of KiDS-$r$-450. For CMASS and 2dFHIZ, the scatter between the three KiDS-$i$-800 measurements is larger, because these high-redshift lens samples are more sensitive to the tail of the source redshift distribution, which differs the most between each redshift estimation method. \am{For these HZ lens samples, the measurement derived using both the SPEC and COSMOS-30 redshift distributions deviated the least from KiDS-$r$-450.} As discussed in Section~\ref{sec:nzcompare}, the SPEC method, as a 1-D DIR method,  is biased in comparison to the full DIR analysis in \cite{Hildebrandt/etal:2017}, as it cannot take into account the difference in population of colour space between the KiDS sample and the deep spectroscopic samples. On the other hand, the spectroscopic sample employed in the cross-correlation redshift estimation method has a limited selection of high-redshift objects and therefore little constraining power at $z>1$.

\am{Averaging over all lens samples and assuming independent measurements, both the COSMOS-30 and SPEC KiDS-$i$-800 measurements are found to be consistent with the KiDS-$r$-450 measurement at the level of $7\pm5$\%.  For the low-redshift lens samples only, the results are consistent at the level of $5\pm5$\%. The KiDS-$i$-800 measurements using  the cross-correlation redshift estimation (CC), are, on average, inconsistent with the KiDS-$r$-450 measurements.  Combining all lens samples together we find that the KiDS-$i$-800 and KiDS-$r$-450 analyses differ by $14\pm4$\% for the CC analysis. }

In this comparison we note that we have not accounted for the uncertainty in the estimate of each redshift distribution.  For the cross-correlation (CC) result, in particular, the errors are significant at high redshift (see Figure~\ref{fig:inzall}) and this hinders the method from constraining the high-redshift tail. It is therefore unsurprising that this method nominally has the worst agreement as these significant errors have not been accounted for in this analysis.  For the SPEC and COSMOS-30 redshift distributions, the bootstrap error is negligible but we have not been able to quantify likely systematic errors associated with sampling variance and incompleteness.  The spread of the galaxy-galaxy lensing measurements for each lens sample therefore provides some indication of the impact on our analysis of our systematic uncertainty in the $i$-band source redshift distribution.

We find that our measurements of the excess surface mass density profiles are not only consistent between KiDS-$i$-800 and KiDS-$r$-450, but also with previous measurements, namely \citet{Miyatake2015, Leauthaud2017,vU2016}. While the purpose of this study is to compare the two source galaxy datasets, these KiDS-$r$-450 measurements are interesting in their own right \citep{AmonEG}.

\subsection{PSF and seeing dependence}

\begin{figure}
	\includegraphics[width=\columnwidth]{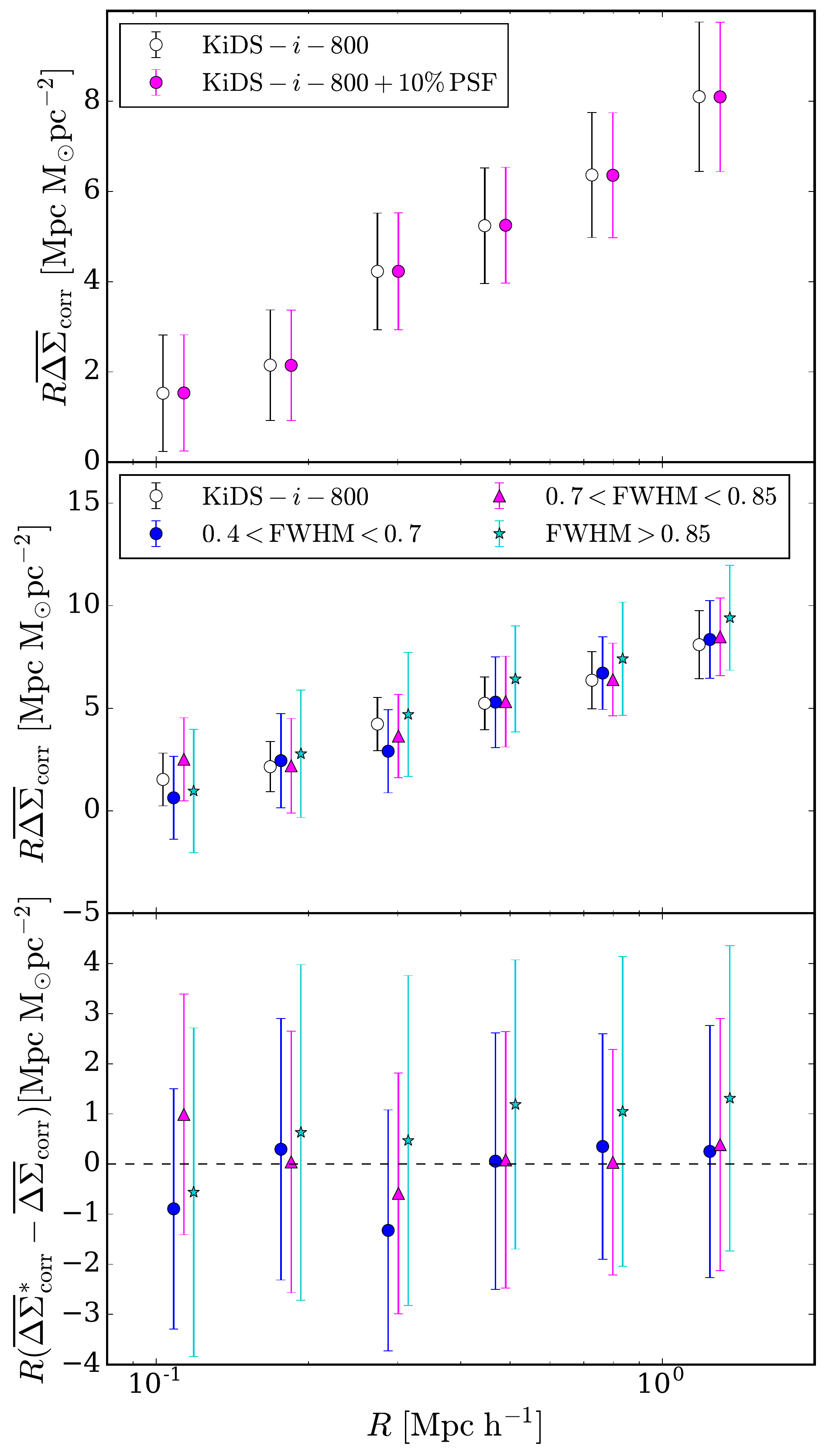}
    \caption{\label{fig:dspsf}Stress-testing the galaxy-galaxy lensing measurement: the dependence of PSF contamination and seeing on the observed stacked differential surface mass density $\Delta\Sigma$ for KiDS-$i$-800 with CMASS lenses. The upper panel compares the lensing signal obtained with a false additional 10\% PSF contamination added to the ellipticities of the galaxies (pink) with the untampered measurement (black). The middle panel shows the `seeing test': the galaxy-galaxy lensing signal obtained when the KiDS-$i$-800 sample is split into three samples by observed FWHM. The lower panel shows the difference between the galaxy-galaxy lensing measurements made with each of the seeing subsamples and the original measurement. All signals are scaled by the comoving separation, $R$ and offset for clarity. }
\end{figure}

In this section we investigate the sensitivity of the measured CMASS galaxy-galaxy lensing signal to changes in PSF contamination and the observed seeing, as these are two of the primary differences between the KiDS-$i$-800 and KiDS-$r$-450 shape catalogues.  

Motivated by the presence of the $\sim 10$\% PSF contamination detailed in Section~\ref{sec:sys}, we modify the KiDS-$i$-800 galaxy shapes to $\epsilon^{\rm cont}$, which includes an additional PSF component of $\alpha_1=\alpha_2 = 0.1$ where
\begin{equation}
\centering
\epsilon^{\rm cont}=\epsilon+\alpha \epsilon^{\mathrm{*}} \,  ,
\end{equation}
and $\alpha \epsilon^{\mathrm{*}} = \alpha_1 \epsilon_1^{\mathrm{*}} + \mathrm{i} \alpha_2 \epsilon_2^{\mathrm{*}}$.  We then re-measure the CMASS galaxy-galaxy lensing signal using the tampered HZ source sample, subtracting a `random signal' as discussed in Appendix~\ref{app:gglcorr}.  We determine the redshift distribution using the SPEC method, noting that this choice is unimportant as it scales the fiducial and PSF contaminated signal in the same way.

\begin {table*}
\caption{\label{tab:zseeing}Mean and median values of the redshift distributions for the three seeing selections of KiDS-$i$-800 and KiDS-$r$-450 galaxies. The redshift distributions for the KiDS-$i$-800 samples are estimated using the spectroscopic catalogue method (SPEC) and for the KiDS-$r$-450 samples via the DIR method. } 
\begin{center}
\begin{tabular}{lccccc}

 \hline
Dataset & FWHM (arcseconds) & $ z_{\rm{med}}$ & $ \bar{z}$ & $ \eta $\\
 \hline
 &$0.40-0.70$   &  $0.647\pm0.001$ &   $0.665\pm0.001$ & 0.171 \\
KiDS-$i$-800  ($i>20.8$)& $0.70-0.85$  & $0.593\pm0.003$ & $0.615\pm0.002$ & 0.153 \\
 &$0.85-2.00$   & $0.523\pm0.004$ & $0.578\pm0.003$ & 0.137\\
 \hline
 &$0.40-0.61$   &  0.62 & 0.68  & 0.196\\
 KiDS-$r$-450 ($0.43<z_B<0.9$) & $0.61-0.71$   & 0.57 & 0.64 & 0.179\\
 &$0.71-0.90$   & 0.53 & 0.61  & 0.162\\
 \hline
\end{tabular}
\end{center}
\end {table*}

The upper panel of Figure~\ref{fig:dspsf} compares the CMASS galaxy-galaxy lensing measurement with our fiducial KiDS-$i$-800 HZ dataset and the false PSF contamination dataset. 

\am{The additional PSF component is found to have a negligible effect on the lensing signal. This is expected given the azimuthal averaging of the signal over the smoothly varying OmegaCAM PSF.  It would not have been expected, however for surveys with rapid spatial variation in the PSF which could arise, for example, from atmospheric turbulence \citep{HeymansRowe2012}.  Survey geometry can also impact how sensitive galaxy-galaxy lensing measurements are to PSF contamination \citep{Hirata2004}. It is therefore a valid test to make.}

As demonstrated in Figure~\ref{fig:psf}, the KiDS-$i$-800 observations were taken over a wider range of seeing conditions than KIDS-$r$-450. To assess sensitivity of the galaxy-galaxy lensing measurement to variations of data quality, we split both the KiDS-$i$-800 and KiDS-$r$-450 data into three sub-samples based on the average seeing of each KiDS pointing and compare their lensing signal around the CMASS lenses. We ensure that there were roughly the same number of galaxies in each sample.  

A unique source redshift distribution is determined for each seeing sample.  For KiDS-$i$-800 we used the SPEC method, as described in Section~\ref{sec:specz}.  For the KiDS-$r$-450 subsamples, redshift distributions are computed using the DIR method with spectroscopic catalogues optimally derived for each of the seeing constraints. The mean and median redshifts for these distributions are detailed in Table~\ref{tab:zseeing}.  As one would expect, lower-seeing data results in redshift distributions with a higher mean redshift. 
\begin{figure}
	\includegraphics[width=\columnwidth]{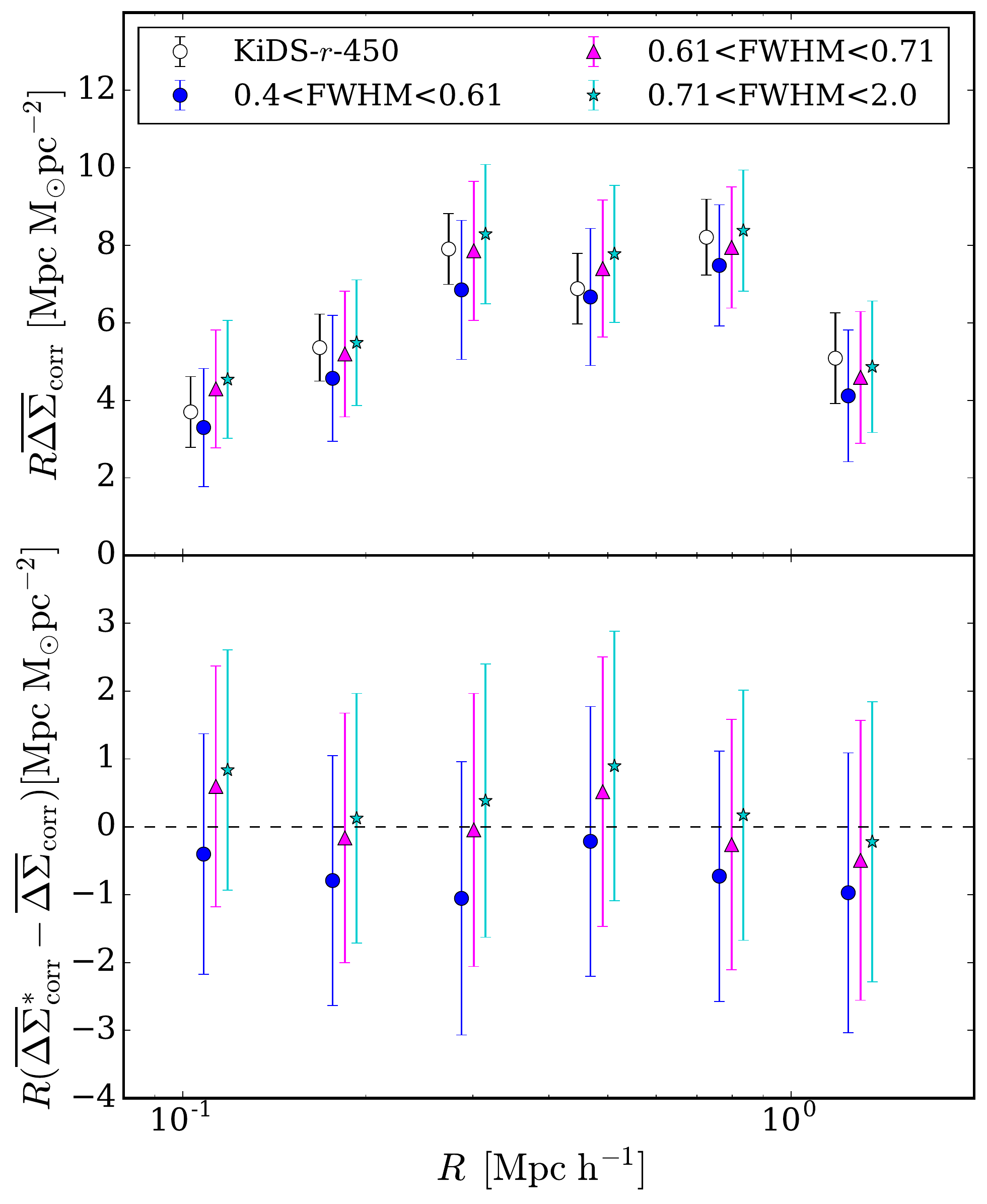}
    \caption{\label{fig:dsrband}The stacked differential surface mass density measurement $\Delta\Sigma$ for KiDS-$r$-450 with CMASS lenses. The upper panel shows the galaxy-galaxy lensing signal obtained when the the KiDS-$r$-450 sample is split into three samples by seeing. The lower panel shows the difference between these test measurements and the original measurement. All signals are scaled by the comoving separation, $R$, and offset horizontally for clarity.}
\end{figure}

The middle panel of Figure~\ref{fig:dspsf} shows the CMASS excess surface density profiles measured using the full KiDS-$i$-800 dataset, as well as the three seeing subsamples. In each case, a unique boost factor was also calculated and applied, and a corresponding random signal removed. The error is estimated from a Jackknife analysis. The lowest panel highlights the consistency between the measurements made with the different seeing subsamples and the fiducial measurement, noting that the errors will be correlated as a sub-sample is being compared to the entire KiDS-$i$-800 sample.  For consistency, Figure~\ref{fig:dsrband} shows the results for the equivalent seeing test with the KiDS-$r$-450 galaxies. Again, we find no evidence that variations in the seeing influences the galaxy-galaxy lensing measurement.  From these studies we can conclude that our galaxy-galaxy lensing measurements are insensitive to changes in the levels of PSF contamination and seeing.

\section{Summary and Outlook}  \label{sec:conc}
This paper presents $i$-band imaging data from the Kilo-Degree Survey (KiDS-$i$-800). This new lensing dataset spans \SI{815}{\square\degree}, with an average effective source density of 3.8 galaxies per square arcmin and a median redshift of $z_{\mathrm{med}} \sim 0.5$. In contrast to the deep $r$-band KiDS observations that make up the homogeneous KiDS multi-band cosmic shear dataset (KiDS-$r$-450), the $i$-band data span a wide range of observing conditions. The less-strict seeing constraints give rise to a very wide range of depth and variation in data quality, weighed against a higher data acquisition rate. We adopt the KiDS data analysis pipeline \citep{Kuijken/etal:2015}.  This includes the {\sc theli} software package for data reduction and the self-calibrating version of \emph{lens}fit for shear measurements, with an adapted methodology for star-galaxy selection in order to reliably select stars in the poorer-seeing $i$-band data.
%The unmasked effective survey area of \SI{733}{\square\degree}, with complete overlap with spectroscopic surveys, makes KiDS-$i$-800 ideal for lensing cross-correlation studies out to very large cosmological scales.

The combination of KiDS-$i$-800 and KiDS-$r$-450 allows for the first large-scale lensing analysis of two overlapping imaging surveys. We present two `null' tests to uniquely assess the consistency between the weak lensing measurements in both surveys.  

In the first test we analyse a matched $ri$ catalogue using a `nulled' two-point shear correlation function.  By limiting both surveys to the same matched source sample, we isolate differences between each survey's shear calibration.  By using a two-point correlation function we are sensitive to both additive and multiplicative shear biases.    In this study we uncovered significant additive shear bias in the KiDS-$i$-800 survey motivating an extension called the `cross-nulled' two-point shear correlation function.  This statistic also nulls additive shear bias if it is uncorrelated between the two surveys.  In this `cross-nulled' case \am{ we found the two shear calibrations for the matched samples agree at a level of $1 \pm 4$ percent}.  This demonstrates that the differing multiplicative shear calibration corrections for both surveys, produce consistent results between the very different quality KiDS-$i$-800 and KiDS-$r$-450 images.

In the second test we use overlapping spectroscopic surveys to measure a `nulled' galaxy-galaxy lensing signal, for five different lens samples, with the two KiDS lensing datasets.  By using the full depth of each lensing survey in this analysis, our results are more applicable in understanding how biases propagate through to final survey results.  The consequence, however, is that the `nulled' galaxy-galaxy lensing result can only provide a joint analysis of the shear calibration in combination with the redshift determination.  It is also only sensitive to multiplicative shear biases. We therefore recommend the combination of the two null tests that we present in this paper for future shear comparison projects, as they scrutinise the two surveys in different but complementary ways. 

%We analyse a matched $ri$ catalogue using a `nulled' two-point shear correlation function to show that the shear correlation functions from the two datasets agree at a level of $1 \pm 4$ percent.  This demonstrates that the differing shear calibration corrections, that have been applied result in the same signal between the very different quality KiDS-$i$-800 and KiDS-$r$-450 images.

%This paragraph says the same as above
%In addition, we use the overlapping spectroscopic surveys to determine five different lens samples, measuring a `nulled' galaxy-galaxy lensing signal with the two KiDS lensing datasets. The galaxy-galaxy lensing measurement extends our `nulled' two-point shear correlation function analysis as it probes the shear calibration in combination with the redshift determination to the full depth of both lensing samples. 

One obvious drawback for the single-band KiDS-$i$-800 survey is the inability to determine photometric redshifts for the individual galaxies.  Any photometric redshift distribution for the sample therefore cannot be directly inferred.
We therefore adopted three different methods to determine an average source redshift distribution for our $i$-band imaging, using overlapping spectroscopy, 30-band photometric redshifts from the COSMOS survey, (COSMOS-30) and a cross-correlation clustering technique.  \am{When using the overlapping spectroscopy or the COSMOS-30 data, the KiDS-$i$-800 and KiDS-$r$-450 galaxy-galaxy lensing signals are found to be consistent with a measured excess surface mass density for the two datasets consistent at the level of $7 \pm 5$ percent in both cases.  When adopting an $i$-band redshift distribution determined through the cross-clustering technique, however, the two datasets differ by $14 \pm 4$ percent.}  These results do not include any uncertainty on the estimates of the redshift distributions. For the cross correlation clustering redshift, in particular, we find this error to be significant at high redshifts. For the other two redshift estimations, while the statistical bootstrap error on the estimation is negligible, we have not quantified any systematic bias of the method resulting from incompleteness in the sample or sampling variance.  The spread of the galaxy-galaxy lensing measurements for each lens sample provides some indication of this uncertainty and future scientific analysis of the KiDS-$i$-800 survey will require studies to improve our knowledge of the $i$-band source redshift distribution. We also `stress test' our galaxy-galaxy lensing measurements to highlight that our two datasets are robust against variations in data quality (FWHM) as well as PSF contamination.

The susceptibility of weak lensing measurements to various sources of systematic error provides a strong motivation for the comparison of the lensing signals derived from unique datasets. 
The $\sim 5$ percent precision afforded by the `null' tests presented in this analysis are unfortunately still far from the percent level precision offered by image simulation studies of shear calibration biases \citep[e.g][]{fenech-conti/etal:2016}.  They do however already provide a meaningful and fully independent confirmation of the results from these studies.  As overlapping survey area grows these `null'  statistics will become an increasingly stringent and required test for weak lensing surveys.

While in this paper we have carried out an internal comparison with our two datasets undergoing a similar data processing procedure, the methodology we have presented can be extended to tests between any set of overlapping surveys.  With three on-going lensing surveys that partially overlap, KiDS, the Dark Energy Survey \citep[DES;][]{jarvis/etal:2016} and Hyper Supreme-Cam \citep[HSC;][]{HSC2017}, as well as the advent of next-generation surveys like LSST\footnote{\url{http://www.lsst.org/}}, Euclid\footnote{\url{http://sci.esa.int/euclid/}} and WFIRST\footnote{\url{http://wfirst.gsfc.nasa.gov/}}, our proposed comparison `null' tests will be an important new addition to ensure robust weak lensing results in the future.

\section*{Acknowledgements}
We thank members of the KiDS $i$-band eyeballing team Margot Brouwer, Marcello Cacciato, Martin Eriksen, Fabian Kohlinger and Cristobal Sif\'on for all their ground-work in road-testing the data verification checks.   We also thank Alex Hall for helpful discussions and the zCOSMOS team for making their full, non-public redshift catalogue available. Furthermore, we thank the referee for a thorough and careful report.

AA thanks the LSSTC Data Science Fellowship Program,her time as a Fellow has benefited this work.
CH acknowledges support from the European Research Council under grant numbers 647112. 
DK acknowledges support from the Deutsche Forschungsgemeinschaft in the framework of the TR33 `The Dark Universe'.
CB acknowledges the support of the Australian Research Council through the award of a Future Fellowship.
HHi acknowledges support from an Emmy Noether grant (No. Hi 1495/2-1) of the Deutsche Forschungsgemeinschaft.
HHo acknowledges support from Vici grant 639.043.512, financed by the Netherlands Organisation for Scientific Research (NWO).
CM acknowledges support from the National Science Foundation through Cooperative Agreement 1258333 managed by the Association of Universities for Research in Astronomy(AURA), and the Department of Energy under Contract No. DE-AC02-76SF00515 with the SLAC National Accelerator Laboratory. 
KK acknowledges support by the Alexander von Humboldt Foundation research award.
EvU acknowledges support from an STFC Ernest Rutherford Research Grant, grant reference ST/L00285X/1, and BJ from an STFC Ernest Rutherford Fellowship, grant reference ST/J004421/1.
MV acknowledges support from the European Research Council under FP7 grant number 279396 and the Netherlands Organisation for Scientific Research (NWO) through grants 614.001.103. 
JdJ is supported by the Netherlands Organisation for Scientific Research (NWO) through grant 621.016.402.

This work is based on data products from observations made with ESO Telescopes at the La Silla Paranal Observatory under programme IDs 177.A-3016, 177.A-3017 and 177.A-3018, and on data products produced by Target/OmegaCEN, INAF-OACN, INAF-OAPD and the KiDS production team, on behalf of the KiDS consortium.

This work has also made use of data from the European Space Agency (ESA)
mission {\it Gaia} (\url{http://www.cosmos.esa.int/gaia}), processed by the {\it Gaia} Data Processing and Analysis Consortium (DPAC, \url{http://www.cosmos.esa.int/web/gaia/dpac/consortium}). Funding for the DPAC has been provided by national institutions, in particular the institutions participating in the {\it Gaia} Multilateral Agreement. 

2dFLenS is based on data acquired through the Australian Astronomical 
Observatory, under program A/2014B/008.  It would not have been possible 
without the dedicated work of the staff of the AAO in the development 
and support of the 2dF-AAOmega system, and the running of the AAT.

We thank that GAMA consortium for providing access to their third data release. GAMA is a joint European-Australasian project based around a spectroscopic campaign using the Anglo-Australian Telescope. The GAMA input catalogue is based on data taken from the Sloan Digital Sky Survey and the UKIRT Infrared Deep Sky Survey. Complementary imaging of the GAMA regions is being obtained by a number of independent survey programs including GALEX MIS, VST KiDS, VISTA VIKING, WISE, Herschel-ATLAS, GMRT and ASKAP providing UV to radio coverage. GAMA is funded by the STFC (UK), the ARC (Australia), the AAO, and the participating institutions. The GAMA website is http://www.gama-survey.org/.

{\it Author contributions:}  All authors contributed to the development and writing of this paper.   The authorship list is given in three groups: the lead authors (AA, CH, DK, TE), followed by two alphabetical groups.  The first alphabetical group includes those who are key contributors to both the scientific analysis and the data products.   The second group covers those who have either made a significant contribution to the data products or to the scientific analysis.

\bibliographystyle{mnras}
\bibliography{iband_tech} % if your bibtex file is called example.bib

\begin{thebibliography}{}
\makeatletter
\relax
\def\mn@urlcharsother{\let\do\@makeother \do\$\do\&\do\#\do\^\do\_\do\%\do\~}
\def\mn@doi{\begingroup\mn@urlcharsother \@ifnextchar [ {\mn@doi@}
  {\mn@doi@[]}}
\def\mn@doi@[#1]#2{\def\@tempa{#1}\ifx\@tempa\@empty \href
  {http://dx.doi.org/#2} {doi:#2}\else \href {http://dx.doi.org/#2} {#1}\fi
  \endgroup}
\def\mn@eprint#1#2{\mn@eprint@#1:#2::\@nil}
\def\mn@eprint@arXiv#1{\href {http://arxiv.org/abs/#1} {{\tt arXiv:#1}}}
\def\mn@eprint@dblp#1{\href {http://dblp.uni-trier.de/rec/bibtex/#1.xml}
  {dblp:#1}}
\def\mn@eprint@#1:#2:#3:#4\@nil{\def\@tempa {#1}\def\@tempb {#2}\def\@tempc
  {#3}\ifx \@tempc \@empty \let \@tempc \@tempb \let \@tempb \@tempa \fi \ifx
  \@tempb \@empty \def\@tempb {arXiv}\fi \@ifundefined
  {mn@eprint@\@tempb}{\@tempb:\@tempc}{\expandafter \expandafter \csname
  mn@eprint@\@tempb\endcsname \expandafter{\@tempc}}}

\bibitem[\protect\citeauthoryear{{Aihara} et~al.,}{{Aihara}
  et~al.}{2017}]{HSC2017}
{Aihara} H.,  et~al., 2017, preprint, \href
  {http://adsabs.harvard.edu/abs/2017arXiv170405858A} {} (\mn@eprint {arXiv}
  {1704.05858})

\bibitem[\protect\citeauthoryear{{Alam} et~al.,}{{Alam}
  et~al.}{2015}]{Alam/etal:2015}
{Alam} S.,  et~al., 2015, \mn@doi [\apjs] {10.1088/0067-0049/219/1/12}, \href
  {http://adsabs.harvard.edu/abs/2015ApJS..219...12A} {219, 12}

\bibitem[\protect\citeauthoryear{{Alam}, {Miyatake}, {More}, {Ho}  \&
  {Mandelbaum}}{{Alam} et~al.}{2017}]{Alam2017}
{Alam} S.,  {Miyatake} H.,  {More} S.,  {Ho} S.,   {Mandelbaum} R.,  2017,
  \mn@doi [\mnras] {10.1093/mnras/stw3056}, \href
  {http://adsabs.harvard.edu/abs/2017MNRAS.465.4853A} {465, 4853}

\bibitem[\protect\citeauthoryear{{Amon} et~al.,}{{Amon} et~al.}{2017}]{AmonEG}
{Amon} A.,  et~al., 2017, preprint, \href
  {http://adsabs.harvard.edu/abs/2017arXiv171110999A} {} (\mn@eprint {arXiv}
  {1711.10999})

\bibitem[\protect\citeauthoryear{{Bartelmann} \& {Schneider}}{{Bartelmann} \&
  {Schneider}}{2001}]{Bartelmann2001}
{Bartelmann} M.,  {Schneider} P.,  2001, \mn@doi [\physrep]
  {10.1016/S0370-1573(00)00082-X}, \href
  {http://adsabs.harvard.edu/abs/2001PhR...340..291B} {340, 291}

\bibitem[\protect\citeauthoryear{{Becker} et~al.,}{{Becker}
  et~al.}{2016}]{becker2016}
{Becker} M.~R.,  et~al., 2016, \mn@doi [\prd] {10.1103/PhysRevD.94.022002},
  \href {http://adsabs.harvard.edu/abs/2016PhRvD..94b2002B} {94, 022002}

\bibitem[\protect\citeauthoryear{{Ben{\'{\i}}tez}}{{Ben{\'{\i}}tez}}{2000}]{Benitez:2000}
{Ben{\'{\i}}tez} N.,  2000, \mn@doi [\apj] {10.1086/308947}, \href
  {http://adsabs.harvard.edu/abs/2000ApJ...536..571B} {536, 571}

\bibitem[\protect\citeauthoryear{{Bertin} \& {Arnouts}}{{Bertin} \&
  {Arnouts}}{1996}]{Bertin/etal:1996}
{Bertin} E.,  {Arnouts} S.,  1996, \mn@doi [\aaps] {10.1051/aas:1996164}, \href
  {http://adsabs.harvard.edu/abs/1996A%26AS..117..393B} {117, 393}

\bibitem[\protect\citeauthoryear{{Blake} et~al.,}{{Blake}
  et~al.}{2016a}]{Blake2016eg}
{Blake} C.,  et~al., 2016a, \mn@doi [\mnras] {10.1093/mnras/stv2875}, \href
  {http://adsabs.harvard.edu/abs/2016MNRAS.456.2806B} {456, 2806}

\bibitem[\protect\citeauthoryear{{Blake} et~al.,}{{Blake}
  et~al.}{2016b}]{Blake/etal:2016}
{Blake} C.,  et~al., 2016b, \mn@doi [\mnras] {10.1093/mnras/stw1990}, \href
  {http://adsabs.harvard.edu/abs/2016MNRAS.462.4240B} {462, 4240}

\bibitem[\protect\citeauthoryear{{Chang} et~al.,}{{Chang}
  et~al.}{2013}]{chang/etal:2013}
{Chang} C.,  et~al., 2013, \mn@doi [\mnras] {10.1093/mnras/stt1156}, \href
  {http://adsabs.harvard.edu/abs/2013MNRAS.434.2121C} {434, 2121}

\bibitem[\protect\citeauthoryear{{Choi} et~al.,}{{Choi}
  et~al.}{2016}]{Choi2015}
{Choi} A.,  et~al., 2016, \mn@doi [\mnras] {10.1093/mnras/stw2241}, \href
  {http://adsabs.harvard.edu/abs/2016MNRAS.463.3737C} {463, 3737}

\bibitem[\protect\citeauthoryear{{Dawson} et~al.,}{{Dawson}
  et~al.}{2013}]{Dawson/etal:2013}
{Dawson} K.~S.,  et~al., 2013, \mn@doi [\aj] {10.1088/0004-6256/145/1/10},
  \href {http://adsabs.harvard.edu/abs/2013AJ....145...10D} {145, 10}

\bibitem[\protect\citeauthoryear{{Driver} et~al.,}{{Driver}
  et~al.}{2011}]{Driver/etal:2011}
{Driver} S.~P.,  et~al., 2011, \mn@doi [\mnras]
  {10.1111/j.1365-2966.2010.18188.x}, \href
  {http://adsabs.harvard.edu/abs/2011MNRAS.413..971D} {413, 971}

\bibitem[\protect\citeauthoryear{{Dvornik} et~al.,}{{Dvornik}
  et~al.}{2017}]{Dvornik2017}
{Dvornik} A.,  et~al., 2017, \mn@doi [\mnras] {10.1093/mnras/stx705}, \href
  {http://adsabs.harvard.edu/abs/2017MNRAS.468.3251D} {468, 3251}

\bibitem[\protect\citeauthoryear{{Dvornik} et~al.,}{{Dvornik}
  et~al.}{2018}]{Dvornik2018}
{Dvornik} A.,  et~al., 2018, preprint, \href
  {http://adsabs.harvard.edu/abs/2018arXiv180200734D} {} (\mn@eprint {arXiv}
  {1802.00734})

\bibitem[\protect\citeauthoryear{{Eisenstein} et~al.,}{{Eisenstein}
  et~al.}{2011}]{Eisenstein2011}
{Eisenstein} D.~J.,  et~al., 2011, \mn@doi [\aj] {10.1088/0004-6256/142/3/72},
  \href {http://adsabs.harvard.edu/abs/2011AJ....142...72E} {142, 72}

\bibitem[\protect\citeauthoryear{{Erben} et~al.,}{{Erben}
  et~al.}{2005}]{Erben2005}
{Erben} T.,  et~al., 2005, \mn@doi [Astronomische Nachrichten]
  {10.1002/asna.200510396}, \href
  {http://adsabs.harvard.edu/abs/2005AN....326..432E} {326, 432}

\bibitem[\protect\citeauthoryear{{Erben} et~al.,}{{Erben}
  et~al.}{2009}]{Erben/etal:2009}
{Erben} T.,  et~al., 2009, \mn@doi [\aap] {10.1051/0004-6361:200810426}, \href
  {http://adsabs.harvard.edu/abs/2009A%26A...493.1197E} {493, 1197}

\bibitem[\protect\citeauthoryear{{Erben} et~al.,}{{Erben}
  et~al.}{2013}]{Erben/etal:2013}
{Erben} T.,  et~al., 2013, \mn@doi [\mnras] {10.1093/mnras/stt928}, \href
  {http://adsabs.harvard.edu/abs/2013MNRAS.433.2545E} {433, 2545}

\bibitem[\protect\citeauthoryear{{Fenech Conti}, {Herbonnet}, {Hoekstra},
  {Merten}, {Miller}  \& {Viola}}{{Fenech Conti}
  et~al.}{2017}]{fenech-conti/etal:2016}
{Fenech Conti} I.,  {Herbonnet} R.,  {Hoekstra} H.,  {Merten} J.,  {Miller} L.,
    {Viola} M.,  2017, \mn@doi [\mnras] {10.1093/mnras/stx200}, \href
  {http://adsabs.harvard.edu/abs/2017MNRAS.467.1627F} {467, 1627}

\bibitem[\protect\citeauthoryear{{Gaia Collaboration}, {Brown}, {Vallenari},
  {Prusti}, {de Bruijne}, {Mignard}, {Drimmel}  \& {co-authors}}{{Gaia
  Collaboration} et~al.}{2016}]{Gaia/etal:2016}
{Gaia Collaboration} {Brown} A.~G.~A.,  {Vallenari} A.,  {Prusti} T.,  {de
  Bruijne} J.,  {Mignard} F.,  {Drimmel} R.,   {co-authors} .,  2016, preprint,
  \href {http://adsabs.harvard.edu/abs/2016arXiv160904172G} {} (\mn@eprint
  {arXiv} {1609.04172})

\bibitem[\protect\citeauthoryear{{Griffith} et~al.,}{{Griffith}
  et~al.}{2012}]{Griffith2012}
{Griffith} R.~L.,  et~al., 2012, \mn@doi [\apjs] {10.1088/0067-0049/200/1/9},
  \href {http://adsabs.harvard.edu/abs/2012ApJS..200....9G} {200, 9}

\bibitem[\protect\citeauthoryear{{Hand} et~al.,}{{Hand}
  et~al.}{2015}]{Hand/etal:2013}
{Hand} N.,  et~al., 2015, \mn@doi [\prd] {10.1103/PhysRevD.91.062001}, \href
  {http://adsabs.harvard.edu/abs/2015PhRvD..91f2001H} {91, 062001}

\bibitem[\protect\citeauthoryear{{Heymans} et~al.,}{{Heymans}
  et~al.}{2005}]{heymans/etal:2005}
{Heymans} C.,  et~al., 2005, \mn@doi [\mnras]
  {10.1111/j.1365-2966.2005.09152.x}, \href
  {http://adsabs.harvard.edu/abs/2005MNRAS.361..160H} {361, 160}

\bibitem[\protect\citeauthoryear{{Heymans} et~al.,}{{Heymans}
  et~al.}{2006}]{Heymans2006}
{Heymans} C.,  et~al., 2006, \mn@doi [\mnras]
  {10.1111/j.1365-2966.2006.10198.x}, \href
  {http://adsabs.harvard.edu/abs/2006MNRAS.368.1323H} {368, 1323}

\bibitem[\protect\citeauthoryear{{Heymans}, {Rowe}, {Hoekstra}, {Miller},
  {Erben}, {Kitching}  \& {van Waerbeke}}{{Heymans}
  et~al.}{2012a}]{HeymansRowe2012}
{Heymans} C.,  {Rowe} B.,  {Hoekstra} H.,  {Miller} L.,  {Erben} T.,
  {Kitching} T.,   {van Waerbeke} L.,  2012a, \mn@doi [\mnras]
  {10.1111/j.1365-2966.2011.20312.x}, \href
  {http://adsabs.harvard.edu/abs/2012MNRAS.421..381H} {421, 381}

\bibitem[\protect\citeauthoryear{{Heymans} et~al.,}{{Heymans}
  et~al.}{2012b}]{Heymans2012}
{Heymans} C.,  et~al., 2012b, \mn@doi [\mnras]
  {10.1111/j.1365-2966.2012.21952.x}, \href
  {http://adsabs.harvard.edu/abs/2012MNRAS.427..146H} {427, 146}

\bibitem[\protect\citeauthoryear{{Hildebrandt} et~al.,}{{Hildebrandt}
  et~al.}{2016}]{Hildebrandt/etal:2016}
{Hildebrandt} H.,  et~al., 2016, \mn@doi [\mnras] {10.1093/mnras/stw2013},
  \href {http://adsabs.harvard.edu/abs/2016MNRAS.tmp.1132H} {}

\bibitem[\protect\citeauthoryear{{Hildebrandt} et~al.,}{{Hildebrandt}
  et~al.}{2017}]{Hildebrandt/etal:2017}
{Hildebrandt} H.,  et~al., 2017, \mn@doi [\mnras] {10.1093/mnras/stw2805},
  \href {http://adsabs.harvard.edu/abs/2017MNRAS.465.1454H} {465, 1454}

\bibitem[\protect\citeauthoryear{{Hirata} et~al.,}{{Hirata}
  et~al.}{2004}]{Hirata2004}
{Hirata} C.~M.,  et~al., 2004, \mn@doi [\mnras]
  {10.1111/j.1365-2966.2004.08090.x}, \href
  {http://adsabs.harvard.edu/abs/2004MNRAS.353..529H} {353, 529}

\bibitem[\protect\citeauthoryear{{Hoekstra}}{{Hoekstra}}{2004}]{Hoekstra2004}
{Hoekstra} H.,  2004, \mn@doi [\mnras] {10.1111/j.1365-2966.2004.07327.x},
  \href {http://adsabs.harvard.edu/abs/2004MNRAS.347.1337H} {347, 1337}

\bibitem[\protect\citeauthoryear{{Hoekstra} \& {Jain}}{{Hoekstra} \&
  {Jain}}{2008}]{HoekstraJain2008}
{Hoekstra} H.,  {Jain} B.,  2008, \mn@doi [Annual Review of Nuclear and
  Particle Science] {10.1146/annurev.nucl.58.110707.171151}, \href
  {http://adsabs.harvard.edu/abs/2008ARNPS..58...99H} {58, 99}

\bibitem[\protect\citeauthoryear{{Hoekstra}, {Yee}  \& {Gladders}}{{Hoekstra}
  et~al.}{2004}]{HoekstraYeeGladder2004}
{Hoekstra} H.,  {Yee} H.~K.~C.,   {Gladders} M.~D.,  2004, \mn@doi [\apj]
  {10.1086/382726}, \href {http://adsabs.harvard.edu/abs/2004ApJ...606...67H}
  {606, 67}

\bibitem[\protect\citeauthoryear{{Hoekstra}, {Hsieh}, {Yee}, {Lin}  \&
  {Gladders}}{{Hoekstra} et~al.}{2005}]{Hoekstra/etal:2005}
{Hoekstra} H.,  {Hsieh} B.~C.,  {Yee} H.~K.~C.,  {Lin} H.,   {Gladders} M.~D.,
  2005, \mn@doi [\apj] {10.1086/496913}, \href
  {http://adsabs.harvard.edu/abs/2005ApJ...635...73H} {635, 73}

\bibitem[\protect\citeauthoryear{{Ilbert} et~al.,}{{Ilbert}
  et~al.}{2009}]{Ilbert2009}
{Ilbert} O.,  et~al., 2009, \mn@doi [\apj] {10.1088/0004-637X/690/2/1236},
  \href {http://adsabs.harvard.edu/abs/2009ApJ...690.1236I} {690, 1236}

\bibitem[\protect\citeauthoryear{{Jarvis} \& {Jain}}{{Jarvis} \&
  {Jain}}{2008}]{jarvis/jain:2008}
{Jarvis} M.,  {Jain} B.,  2008, \mn@doi [\jcap]
  {10.1088/1475-7516/2008/01/003}, \href
  {http://adsabs.harvard.edu/abs/2008JCAP...01..003J} {1, 003}

\bibitem[\protect\citeauthoryear{{Jarvis} et~al.,}{{Jarvis}
  et~al.}{2016}]{jarvis/etal:2016}
{Jarvis} M.,  et~al., 2016, \mn@doi [\mnras] {10.1093/mnras/stw990}, \href
  {http://adsabs.harvard.edu/abs/2016MNRAS.460.2245J} {460, 2245}

\bibitem[\protect\citeauthoryear{{Jee}, {Tyson}, {Schneider}, {Wittman},
  {Schmidt}  \& {Hilbert}}{{Jee} et~al.}{2013}]{Jee/etal:2013}
{Jee} M.~J.,  {Tyson} J.~A.,  {Schneider} M.~D.,  {Wittman} D.,  {Schmidt} S.,
   {Hilbert} S.,  2013, \mn@doi [\apj] {10.1088/0004-637X/765/1/74}, \href
  {http://adsabs.harvard.edu/abs/2013ApJ...765...74J} {765, 74}

\bibitem[\protect\citeauthoryear{{Johnson} et~al.,}{{Johnson}
  et~al.}{2017}]{Johnson2017}
{Johnson} A.,  et~al., 2017, \mn@doi [\mnras] {10.1093/mnras/stw3033}, \href
  {http://adsabs.harvard.edu/abs/2017MNRAS.465.4118J} {465, 4118}

\bibitem[\protect\citeauthoryear{{Kilbinger}}{{Kilbinger}}{2015}]{Kilbinger2015}
{Kilbinger} M.,  2015, \mn@doi [Reports on Progress in Physics]
  {10.1088/0034-4885/78/8/086901}, \href
  {http://adsabs.harvard.edu/abs/2015RPPh...78h6901K} {78, 086901}

\bibitem[\protect\citeauthoryear{{Kilbinger}, {Bonnett}  \&
  {Coupon}}{{Kilbinger} et~al.}{2014}]{Kilbinger2014}
{Kilbinger} M.,  {Bonnett} C.,   {Coupon} J.,  2014, {athena: Tree code for
  second-order correlation functions}, Astrophysics Source Code Library
  (\mn@eprint {ascl} {1402.026})

\bibitem[\protect\citeauthoryear{{Kleinheinrich} et~al.,}{{Kleinheinrich}
  et~al.}{2004}]{Kleinheinrich2004}
{Kleinheinrich} M.,  et~al., 2004, ArXiv Astrophysics e-prints, \href
  {http://adsabs.harvard.edu/abs/2004astro.ph.12615K} {}

\bibitem[\protect\citeauthoryear{{Komatsu} et~al.,}{{Komatsu}
  et~al.}{2011}]{Komatsu2011}
{Komatsu} E.,  et~al., 2011, \mn@doi [\apjs] {10.1088/0067-0049/192/2/18},
  \href {http://adsabs.harvard.edu/abs/2011ApJS..192...18K} {192, 18}

\bibitem[\protect\citeauthoryear{{Kuijken} et~al.,}{{Kuijken}
  et~al.}{2015}]{Kuijken/etal:2015}
{Kuijken} K.,  et~al., 2015, \mn@doi [\mnras] {10.1093/mnras/stv2140}, \href
  {http://adsabs.harvard.edu/abs/2015MNRAS.454.3500K} {454, 3500}

\bibitem[\protect\citeauthoryear{{Kwan} et~al.,}{{Kwan}
  et~al.}{2017}]{Kwan2017}
{Kwan} J.,  et~al., 2017, \mn@doi [\mnras] {10.1093/mnras/stw2464}, \href
  {http://adsabs.harvard.edu/abs/2017MNRAS.464.4045K} {464, 4045}

\bibitem[\protect\citeauthoryear{{Laigle} et~al.,}{{Laigle}
  et~al.}{2016}]{Laigle2016}
{Laigle} C.,  et~al., 2016, \mn@doi [\apjs] {10.3847/0067-0049/224/2/24}, \href
  {http://adsabs.harvard.edu/abs/2016ApJS..224...24L} {224, 24}

\bibitem[\protect\citeauthoryear{{Leauthaud} et~al.,}{{Leauthaud}
  et~al.}{2017}]{Leauthaud2017}
{Leauthaud} A.,  et~al., 2017, \mn@doi [\mnras] {10.1093/mnras/stx258}, \href
  {http://adsabs.harvard.edu/abs/2017MNRAS.467.3024L} {467, 3024}

\bibitem[\protect\citeauthoryear{{Lilly} et~al.,}{{Lilly}
  et~al.}{2009}]{Lilly/etal:2009}
{Lilly} S.~J.,  et~al., 2009, \mn@doi [\apjs] {10.1088/0067-0049/184/2/218},
  \href {http://adsabs.harvard.edu/abs/2009ApJS..184..218L} {184, 218}

\bibitem[\protect\citeauthoryear{{Lima}, {Cunha}, {Oyaizu}, {Frieman}, {Lin}
  \& {Sheldon}}{{Lima} et~al.}{2008}]{Lima2008}
{Lima} M.,  {Cunha} C.~E.,  {Oyaizu} H.,  {Frieman} J.,  {Lin} H.,   {Sheldon}
  E.~S.,  2008, \mn@doi [\mnras] {10.1111/j.1365-2966.2008.13510.x}, \href
  {http://adsabs.harvard.edu/abs/2008MNRAS.390..118L} {390, 118}

\bibitem[\protect\citeauthoryear{{Liske} et~al.,}{{Liske}
  et~al.}{2015}]{Liske2015}
{Liske} J.,  et~al., 2015, \mn@doi [\mnras] {10.1093/mnras/stv1436}, \href
  {http://adsabs.harvard.edu/abs/2015MNRAS.452.2087L} {452, 2087}

\bibitem[\protect\citeauthoryear{{Mandelbaum} et~al.,}{{Mandelbaum}
  et~al.}{2005}]{Mandelbaum/etal:2005}
{Mandelbaum} R.,  et~al., 2005, \mn@doi [\mnras]
  {10.1111/j.1365-2966.2005.09282.x}, \href
  {http://adsabs.harvard.edu/abs/2005MNRAS.361.1287M} {361, 1287}

\bibitem[\protect\citeauthoryear{{Mandelbaum}, {Seljak}, {Kauffmann}, {Hirata}
  \& {Brinkmann}}{{Mandelbaum} et~al.}{2006}]{Mandelbaum2006}
{Mandelbaum} R.,  {Seljak} U.,  {Kauffmann} G.,  {Hirata} C.~M.,   {Brinkmann}
  J.,  2006, \mn@doi [\mnras] {10.1111/j.1365-2966.2006.10156.x}, \href
  {http://adsabs.harvard.edu/abs/2006MNRAS.368..715M} {368, 715}

\bibitem[\protect\citeauthoryear{{Mandelbaum} et~al.,}{{Mandelbaum}
  et~al.}{2017}]{Mandelbaum2017}
{Mandelbaum} R.,  et~al., 2017, preprint, \href
  {http://adsabs.harvard.edu/abs/2017arXiv170506745M} {} (\mn@eprint {arXiv}
  {1705.06745})

\bibitem[\protect\citeauthoryear{{Melchior} \& {Viola}}{{Melchior} \&
  {Viola}}{2012}]{melchior/viola:2012}
{Melchior} P.,  {Viola} M.,  2012, \mn@doi [\mnras]
  {10.1111/j.1365-2966.2012.21381.x}, \href
  {http://adsabs.harvard.edu/abs/2012MNRAS.424.2757M} {424, 2757}

\bibitem[\protect\citeauthoryear{{M{\'e}nard}, {Scranton}, {Schmidt},
  {Morrison}, {Jeong}, {Budavari}  \& {Rahman}}{{M{\'e}nard}
  et~al.}{2013}]{Menard2013}
{M{\'e}nard} B.,  {Scranton} R.,  {Schmidt} S.,  {Morrison} C.,  {Jeong} D.,
  {Budavari} T.,   {Rahman} M.,  2013, preprint, \href
  {http://adsabs.harvard.edu/abs/2013arXiv1303.4722M} {} (\mn@eprint {arXiv}
  {1303.4722})

\bibitem[\protect\citeauthoryear{{Miller} et~al.,}{{Miller}
  et~al.}{2013}]{miller/etal:2013}
{Miller} L.,  et~al., 2013, \mn@doi [\mnras] {10.1093/mnras/sts454}, \href
  {http://adsabs.harvard.edu/abs/2013MNRAS.429.2858M} {429, 2858}

\bibitem[\protect\citeauthoryear{{Miyatake} et~al.,}{{Miyatake}
  et~al.}{2015a}]{Miyatake:/etal2015}
{Miyatake} H.,  et~al., 2015a, \mn@doi [\apj] {10.1088/0004-637X/806/1/1},
  \href {http://adsabs.harvard.edu/abs/2015ApJ...806....1M} {806, 1}

\bibitem[\protect\citeauthoryear{{Miyatake} et~al.,}{{Miyatake}
  et~al.}{2015b}]{Miyatake2015}
{Miyatake} H.,  et~al., 2015b, \mn@doi [\apj] {10.1088/0004-637X/806/1/1},
  \href {http://adsabs.harvard.edu/abs/2015ApJ...806....1M} {806, 1}

\bibitem[\protect\citeauthoryear{{Morrison}, {Hildebrandt}, {Schmidt},
  {Baldry}, {Bilicki}, {Choi}, {Erben}  \& {Schneider}}{{Morrison}
  et~al.}{2017}]{Morrison2017}
{Morrison} C.~B.,  {Hildebrandt} H.,  {Schmidt} S.~J.,  {Baldry} I.~K.,
  {Bilicki} M.,  {Choi} A.,  {Erben} T.,   {Schneider} P.,  2017, \mn@doi
  [\mnras] {10.1093/mnras/stx342}, \href
  {http://adsabs.harvard.edu/abs/2017MNRAS.467.3576M} {467, 3576}

\bibitem[\protect\citeauthoryear{{Nakajima}, {Mandelbaum}, {Seljak}, {Cohn},
  {Reyes}  \& {Cool}}{{Nakajima} et~al.}{2012}]{Nakajima2012}
{Nakajima} R.,  {Mandelbaum} R.,  {Seljak} U.,  {Cohn} J.~D.,  {Reyes} R.,
  {Cool} R.,  2012, \mn@doi [\mnras] {10.1111/j.1365-2966.2011.20249.x}, \href
  {http://adsabs.harvard.edu/abs/2012MNRAS.420.3240N} {420, 3240}

\bibitem[\protect\citeauthoryear{{Newman} et~al.,}{{Newman}
  et~al.}{2013}]{Newman/etal:2013}
{Newman} J.~A.,  et~al., 2013, \mn@doi [\apjs] {10.1088/0067-0049/208/1/5},
  \href {http://adsabs.harvard.edu/abs/2013ApJS..208....5N} {208, 5}

\bibitem[\protect\citeauthoryear{{Planck Collaboration} et~al.,}{{Planck
  Collaboration} et~al.}{2016}]{Planck2016}
{Planck Collaboration} et~al., 2016, \mn@doi [\aap]
  {10.1051/0004-6361/201525830}, \href
  {http://adsabs.harvard.edu/abs/2016A%26A...594A..13P} {594, A13}

\bibitem[\protect\citeauthoryear{{Rahman}, {Mendez}, {M{\'e}nard}, {Scranton},
  {Schmidt}, {Morrison}  \& {Budav{\'a}ri}}{{Rahman} et~al.}{2016}]{Rahman2016}
{Rahman} M.,  {Mendez} A.~J.,  {M{\'e}nard} B.,  {Scranton} R.,  {Schmidt}
  S.~J.,  {Morrison} C.~B.,   {Budav{\'a}ri} T.,  2016, \mn@doi [\mnras]
  {10.1093/mnras/stw981}, \href
  {http://adsabs.harvard.edu/abs/2016MNRAS.460..163R} {460, 163}

\bibitem[\protect\citeauthoryear{{Ross} et~al.,}{{Ross}
  et~al.}{2012}]{Ross2012}
{Ross} A.~J.,  et~al., 2012, \mn@doi [\mnras]
  {10.1111/j.1365-2966.2012.21235.x}, \href
  {http://adsabs.harvard.edu/abs/2012MNRAS.424..564R} {424, 564}

\bibitem[\protect\citeauthoryear{{Schirmer}}{{Schirmer}}{2013}]{Schirmer2013}
{Schirmer} M.,  2013, \mn@doi [\apjs] {10.1088/0067-0049/209/2/21}, \href
  {http://adsabs.harvard.edu/abs/2013ApJS..209...21S} {209, 21}

\bibitem[\protect\citeauthoryear{{Schmidt}, {M{\'e}nard}, {Scranton},
  {Morrison}  \& {McBride}}{{Schmidt} et~al.}{2013}]{Schmidt2013}
{Schmidt} S.~J.,  {M{\'e}nard} B.,  {Scranton} R.,  {Morrison} C.,   {McBride}
  C.~K.,  2013, \mn@doi [\mnras] {10.1093/mnras/stt410}, \href
  {http://adsabs.harvard.edu/abs/2013MNRAS.431.3307S} {431, 3307}

\bibitem[\protect\citeauthoryear{{Schneider}}{{Schneider}}{2016}]{Schneider2016}
{Schneider} P.,  2016, \mn@doi [\aap] {10.1051/0004-6361/201628506}, \href
  {http://adsabs.harvard.edu/abs/2016A%26A...592L...6S} {592, L6}

\bibitem[\protect\citeauthoryear{{Schneider}, {van Waerbeke}, {Kilbinger}  \&
  {Mellier}}{{Schneider} et~al.}{2002}]{Schneider/etal:2002}
{Schneider} P.,  {van Waerbeke} L.,  {Kilbinger} M.,   {Mellier} Y.,  2002,
  \mn@doi [\aap] {10.1051/0004-6361:20021341}, \href
  {http://ukads.nottingham.ac.uk/abs/2002A%26A...396....1S} {396, 1}

\bibitem[\protect\citeauthoryear{{Schrabback} et~al.,}{{Schrabback}
  et~al.}{2016}]{Schrabback2017}
{Schrabback} T.,  et~al., 2016, preprint, \href
  {http://adsabs.harvard.edu/abs/2016arXiv161103866S} {} (\mn@eprint {arXiv}
  {1611.03866})

\bibitem[\protect\citeauthoryear{{Scoville} et~al.,}{{Scoville}
  et~al.}{2007}]{scoville/etal:2007}
{Scoville} N.,  et~al., 2007, \mn@doi [\apjs] {10.1086/516585}, \href
  {http://adsabs.harvard.edu/abs/2007ApJS..172....1S} {172, 1}

\bibitem[\protect\citeauthoryear{{Seitz} \& {Schneider}}{{Seitz} \&
  {Schneider}}{1997}]{seitz/schneider:1997}
{Seitz} C.,  {Schneider} P.,  1997, \aap, \href
  {http://adsabs.harvard.edu/abs/1997A%26A...318..687S} {318, 687}

\bibitem[\protect\citeauthoryear{{Shanks} et~al.,}{{Shanks}
  et~al.}{2015}]{Shanks/etal:2015}
{Shanks} T.,  et~al., 2015, \mn@doi [\mnras] {10.1093/mnras/stv1130}, \href
  {http://adsabs.harvard.edu/abs/2015MNRAS.451.4238S} {451, 4238}

\bibitem[\protect\citeauthoryear{{Singh}, {Mandelbaum}, {Seljak}, {Slosar}  \&
  {Vazquez Gonzalez}}{{Singh} et~al.}{2016}]{Singh2016}
{Singh} S.,  {Mandelbaum} R.,  {Seljak} U.,  {Slosar} A.,   {Vazquez Gonzalez}
  J.,  2016, preprint, \href
  {http://adsabs.harvard.edu/abs/2016arXiv161100752S} {} (\mn@eprint {arXiv}
  {1611.00752})

\bibitem[\protect\citeauthoryear{{Skrutskie} et~al.,}{{Skrutskie}
  et~al.}{2006}]{Skrutskie/etal:2006}
{Skrutskie} M.~F.,  et~al., 2006, \mn@doi [\aj] {10.1086/498708}, \href
  {http://adsabs.harvard.edu/abs/2006AJ....131.1163S} {131, 1163}

\bibitem[\protect\citeauthoryear{{Troxel} et~al.,}{{Troxel}
  et~al.}{2017}]{troxel2017}
{Troxel} M.~A.,  et~al., 2017, preprint, \href
  {http://adsabs.harvard.edu/abs/2017arXiv170801538T} {} (\mn@eprint {arXiv}
  {1708.01538})

\bibitem[\protect\citeauthoryear{{Vaccari} et~al.,}{{Vaccari}
  et~al.}{2010}]{Vaccari/etal:2012}
{Vaccari} M.,  et~al., 2010, \mn@doi [\aap] {10.1051/0004-6361/201014694},
  \href {http://adsabs.harvard.edu/abs/2010A%26A...518L..20V} {518, L20}

\bibitem[\protect\citeauthoryear{{Vaccari} et~al.,}{{Vaccari}
  et~al.}{2012}]{Vaccari2012}
{Vaccari} M.,  et~al., 2012, in Science from the Next Generation Imaging and
  Spectroscopic Surveys. p.~49

\bibitem[\protect\citeauthoryear{{Viola}, {Kitching}  \& {Joachimi}}{{Viola}
  et~al.}{2014}]{Viola2014}
{Viola} M.,  {Kitching} T.~D.,   {Joachimi} B.,  2014, \mn@doi [\mnras]
  {10.1093/mnras/stu071}, \href
  {http://adsabs.harvard.edu/abs/2014MNRAS.439.1909V} {439, 1909}

\bibitem[\protect\citeauthoryear{{Viola} et~al.,}{{Viola}
  et~al.}{2015}]{viola/etal:2015}
{Viola} M.,  et~al., 2015, \mn@doi [\mnras] {10.1093/mnras/stv1447}, \href
  {http://adsabs.harvard.edu/abs/2015MNRAS.452.3529V} {452, 3529}

\bibitem[\protect\citeauthoryear{{Voigt} \& {Bridle}}{{Voigt} \&
  {Bridle}}{2010}]{Voigt2010}
{Voigt} L.~M.,  {Bridle} S.~L.,  2010, \mn@doi [\mnras]
  {10.1111/j.1365-2966.2010.16300.x}, \href
  {http://adsabs.harvard.edu/abs/2010MNRAS.404..458V} {404, 458}

\bibitem[\protect\citeauthoryear{{Voigt}, {Bridle}, {Amara}, {Cropper},
  {Kitching}, {Massey}, {Rhodes}  \& {Schrabback}}{{Voigt}
  et~al.}{2012}]{voigt/etal:2012}
{Voigt} L.~M.,  {Bridle} S.~L.,  {Amara} A.,  {Cropper} M.,  {Kitching} T.~D.,
  {Massey} R.,  {Rhodes} J.,   {Schrabback} T.,  2012, \mn@doi [\mnras]
  {10.1111/j.1365-2966.2011.20395.x}, \href
  {http://adsabs.harvard.edu/abs/2012MNRAS.421.1385V} {421, 1385}

\bibitem[\protect\citeauthoryear{{Wittman} et~al.,}{{Wittman}
  et~al.}{2002}]{Wittman2002}
{Wittman} D.~M.,  et~al., 2002, in {Tyson} J.~A.,  {Wolff} S.,  eds,  \procspie
  Vol. 4836, Survey and Other Telescope Technologies and Discoveries. pp 73--82
  (\mn@eprint {} {astro-ph/0210118}), \mn@doi{10.1117/12.457348}

\bibitem[\protect\citeauthoryear{{Wolf} et~al.,}{{Wolf}
  et~al.}{2017}]{wolf/etal:2017}
{Wolf} C.,  et~al., 2017, \mn@doi [\mnras] {10.1093/mnras/stw3151}, \href
  {http://adsabs.harvard.edu/abs/2017MNRAS.466.1582W} {466, 1582}

\bibitem[\protect\citeauthoryear{{de Jong}, {Verdoes Kleijn}, {Kuijken}  \&
  {Valentijn}}{{de Jong} et~al.}{2013}]{deJong/etal:2013}
{de Jong} J.~T.~A.,  {Verdoes Kleijn} G.~A.,  {Kuijken} K.~H.,   {Valentijn}
  E.~A.,  2013, \mn@doi [Experimental Astronomy] {10.1007/s10686-012-9306-1},
  \href {http://adsabs.harvard.edu/abs/2013ExA....35...25D} {35, 25}

\bibitem[\protect\citeauthoryear{{de Jong} et~al.,}{{de Jong}
  et~al.}{2015}]{deJong/etal:2015}
{de Jong} J.~T.~A.,  et~al., 2015, \mn@doi [\aap]
  {10.1051/0004-6361/201526601}, \href
  {http://adsabs.harvard.edu/abs/2015A%26A...582A..62D} {582, A62}

\bibitem[\protect\citeauthoryear{{de Jong} et~al.,}{{de Jong}
  et~al.}{2017}]{deJong2017}
{de Jong} J.~T.~A.,  et~al., 2017, preprint, \href
  {http://adsabs.harvard.edu/abs/2017arXiv170302991D} {} (\mn@eprint {arXiv}
  {1703.02991})

\bibitem[\protect\citeauthoryear{{van Uitert}, {Hoekstra}, {Velander},
  {Gilbank}, {Gladders}  \& {Yee}}{{van Uitert}
  et~al.}{2011}]{vanuitert/etal:2011}
{van Uitert} E.,  {Hoekstra} H.,  {Velander} M.,  {Gilbank} D.~G.,  {Gladders}
  M.~D.,   {Yee} H.~K.~C.,  2011, \mn@doi [\aap] {10.1051/0004-6361/201117308},
  \href {http://adsabs.harvard.edu/abs/2011A%26A...534A..14V} {534, A14}

\bibitem[\protect\citeauthoryear{{van Uitert} et~al.,}{{van Uitert}
  et~al.}{2016}]{vU2016}
{van Uitert} E.,  et~al., 2016, \mn@doi [\mnras] {10.1093/mnras/stw747}, \href
  {http://adsabs.harvard.edu/abs/2016MNRAS.459.3251V} {459, 3251}

\makeatother
\end{thebibliography}

\appendix
\section{Data Reduction and Quality Control}
\label{app:QC}
In this paper we use the {\sc theli} data reduction software to
produce lensing quality images in both the $i$- and $r$-band. The
{\sc theli} analysis of the KiDS-$r$-450 data is well
documented in \cite{Hildebrandt/etal:2016}, \cite{Kuijken/etal:2015}
and \cite{deJong/etal:2013}. In this appendix we therefore only
highlight the differences in the data reduction between the KiDS-$r$-450
data release and the KiDS-$i$-800 survey. We also discuss the web-based
quality control system that we employ to ensure that the final data product is
robust.

\begin{figure}
  \centering
  \begin{subfigure}{0.31\columnwidth}
    \includegraphics[scale=0.62]{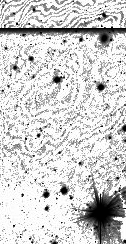}
  \end{subfigure}
  ~
  \begin{subfigure}{0.31\columnwidth}
    \includegraphics[scale=0.6]{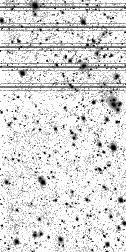}
  \end{subfigure}
  ~
  \begin{subfigure}{0.31\columnwidth}
    \includegraphics[scale=0.6]{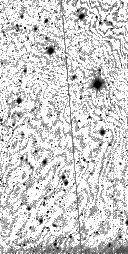}
  \end{subfigure}
  \caption{\label{fig:rejections}Three examples of chip exposures that
    would be rejected by our quality control process. The left panel
    shows sharp changes that are sometimes seen in the gain of the
    affected Chip 16 (ESO \#73). The middle panel reveals electronic
    defects which can been seen as dark lines across a chip. The right
    panel shows an example of a low-signal-to-noise satellite track
    that has not been detected in the co-added image by our automated
    routine. Once visually identified, the track is manually masked.
    Note that the grey-scale in this image is set to enhance any
    low-signal-to-noise features. This setting also highlights the residual
    low-level fringing that can occasionally  be found in the $i$-band imaging,
   after the removal of a fringe model in the data reduction process.}
\end{figure}

The two main differences between the KiDS-$r$-450 data reduction and the
KiDS-$i$-800 data processing are the need for fringe correction in
the $i$-band and the methodology behind the astrometric solution.
\am{Fringe-removal was carried out following Section 4.8 of \citet{Erben2005}.  A fringe model is first built directly from the OmegaCAM imaging, where the fringe geometry pattern has been shown to be very stable over time \citep{deJong/etal:2015}.  The amplitude of the fringes to be subtracted from the individual exposures is then assumed to be directly proportional to the sky-background level. As discussed in \citet{Erben2005}, this assumption can break down in very bright observing conditions and, as such, the fringe removal of our data is of varying quality. While our procedure works well for the majority of the i-band survey, residual fringing is apparent in some images, as illustrated in Figure~\ref{fig:rejections}.}

\am{The differences in astrometric calibration between the $r$ and
$i$-band data are as follows:} For the $r$-band data we were
motivated to adopt a global
astrometric solution, as there is almost contiguous coverage in five
patches in the $r$-band. This global solution takes into account
information from the overlaps between KiDS tiles and data from the
overlapping VST ATLAS Survey \citep{Shanks/etal:2015}. \am{Given the
level of residual fringing in parts of the $i$-band imaging}, the
KiDS-$i$-800 survey was never intended for the cosmic shear studies
that are particularly sensitive to optical camera distortions. The
$i$-band astrometry in the KiDS-South patch was therefore only
calibrated on a pointing basis including information from the five
exposures at slightly different dither positions. This non-global
astrometric solution is expected to be less accurate. In the
KiDS-North patch, the astrometry for both KiDS-$i$-800 and
KiDS-$r$-450 was tied to SDSS-DR12 \citep{Alam/etal:2015} as the
reference frame. Both reductions used 2MASS in the South for the
absolute astrometric reference frame \citep{Skrutskie/etal:2006}.

For KiDS-$i$-800 we developed a new web-based quality control system that
allows for distributed work, with two independent checks of each KiDS tile
to detect problematic exposures without the requirement to access
large FITS files. While the {\sc theli} data processing is running,
the web tool regularly checks for new reduced pointings, adds them
automatically to its database, creates a suite of `check plots' and
distributes each pointing randomly to two `eyeballers' and informs via
e-mail so that an almost instant investigation is possible.

\begin{figure}
  \centering
  \includegraphics[scale=0.22]{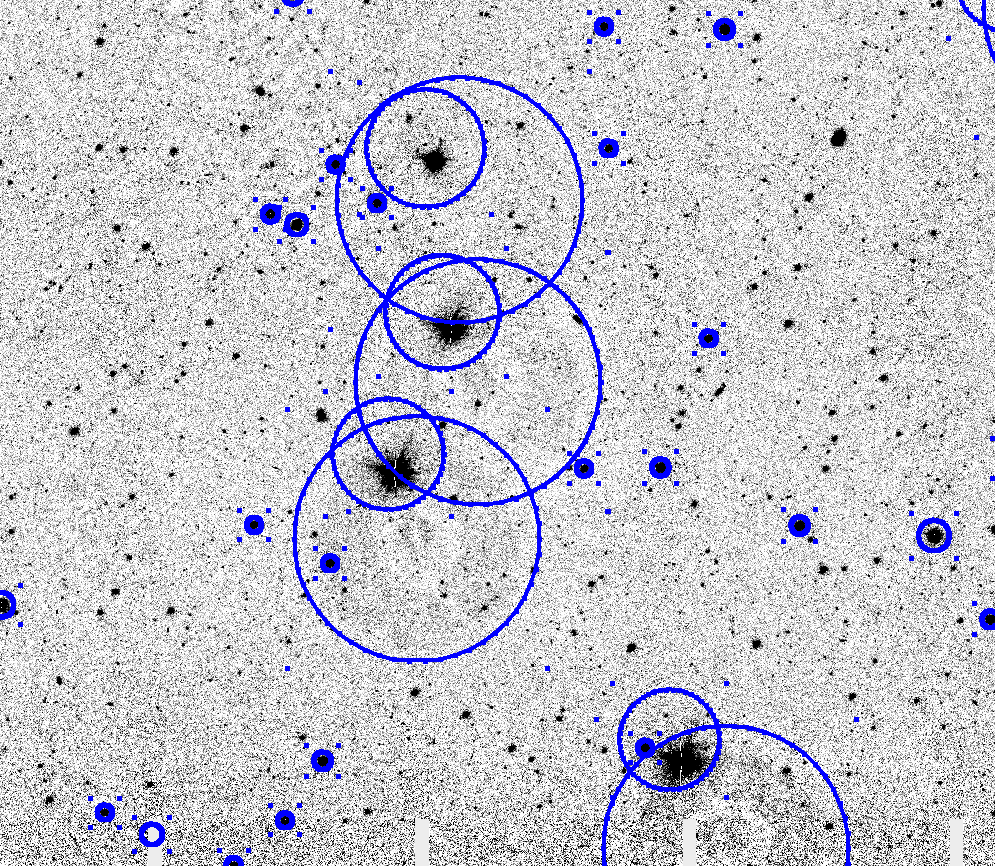}
    \caption{\label{fig:QC_masks} An image section of
    roughly 20\si{\arcminute} x 20\si{\arcminute} of a KiDS-$i$-800 pointing with our automatically generated stellar masks overlayed.
    All stars brighter than $i<14.5$ are masked. Stars with $11.5 < i
    < 14.5$ are masked with a single, central and circular mask with a radius
    that scales with object magnitude. Brighter sources are
    treated with up to three masks covering the stellar core and
    associated reflection halos. Note that the positions of the reflection
    halos with respect to the stellar core are strongly variable on
    the OmegaCAM field-of-view. We found that a two-dimensional second
    order polynomial describes these variations very accurately. }
\end{figure}

Different categories are available which all have to pass our quality
requirements before the pointing is released to the collaboration.
These categories are:
\begin{itemize}
\item{`Calibration', where the bias, dark and flat frames are inspected
    to identify any corrupted images or unusual features.}
\item{`Co-add', where the stacked image and weight image are verified.
    Here any catastrophic error in the astrometric solution can be
    immediately identified with the same objects appearing multiple
    times.}
\item{`Single exposures', where each individual exposure is checked
    and flagged if large satellite tracks are identified, or problems
    are observed on the chip-level.}
\item{`Photometry/Astrometry', where the selection of the stars that
    are used for the astrometric and photometric solution are checked.
	The automated star-galaxy separation
    method that automatically locates the stellar locus in the size-magnitude plane is verified by eye.
    The astrometric solution is then checked by inspecting the offset
    between the position of each star in the co-added image and its
    position in each exposure. Finally the photometry is checked by
    comparing the measured galaxy number counts, per magnitude bin, to
    the same galaxy number counts from deep OmegaCAM data of
    the Chandra Deep Field South \citep{Vaccari2012}}.
\item{`Masks', where the automated stellar halo masks are checked to
    ensure that all bright stars are masked \citep[see Section 3.4
    of][for further details]{Kuijken/etal:2015}. An example mask is
    shown in Figure~\ref{fig:QC_masks}.}
\item{`PSF', where exposures with a strongly elliptical or rapidly
    varying point spread function are removed from the stack.}
\end{itemize}

For each category, any possible issues are listed and can be selected. Depending on the issues found,
the pointing is released to the collaboration or is investigated and
reprocessed. Examples of cases flagged by this quality control
include exposures with electronic defects which appear as lines (see the
left panel of Figure~\ref{fig:rejections}) and problematic chips with
step-changes in the gain (see the middle panel of Figure
\ref{fig:rejections}). In addition there are satellite tracks that are
not seen in the co-added image and hence not automatically recognised
by the {\sc theli} pipeline, but are clearly visible in the single
exposure (right panel of Figure \ref{fig:rejections}). Once identified
the problematic chips can be manually excluded and the satellite
tracks can be manually masked within the web-interface using JS9\footnote{\url{http://js9.si.edu/}}. All re-processed data are re-verified
until the requirements are met in all quality-control categories, at which point the pointing
moves forward to the shape measurement stage of the data analysis
pipeline using \emph{lens}fit (Section~\ref{sec:shape}). Note that
due to different seeing conditions and PSFs in the five exposures of
one pointing, {\sc theli} does not always use all exposures for
co-addition.  \emph{lens}fit has its own
quality control criteria to judge if a single exposure is accepted or
discarded.

The seeing distribution for the resulting KiDS-$i$-800 data can be
compared with the KiDS-$r$-450 data in Figure~\ref{fig:neff}. It is mostly
sub-arcsecond and meets the survey requirements ($<$\SI{1.2}{\arcsec})
with a median of \SI{0.79}{\arcsec}. For comparison: KiDS-$r$-450
has a median seeing of \SI{0.66}{\arcsec}. However, the $i$-band distribution is very broad
and this is reflected in the broad range of limiting magnitudes with
$i_{\rm lim} = 22.7 \pm 0.3$, where the $5\sigma$ limiting magnitude
is defined within a \SI{2}{\arcsec} radius.

\section{Selection criteria}
\label{app:cuts}
 \begin{figure}
	\includegraphics[width=\columnwidth]{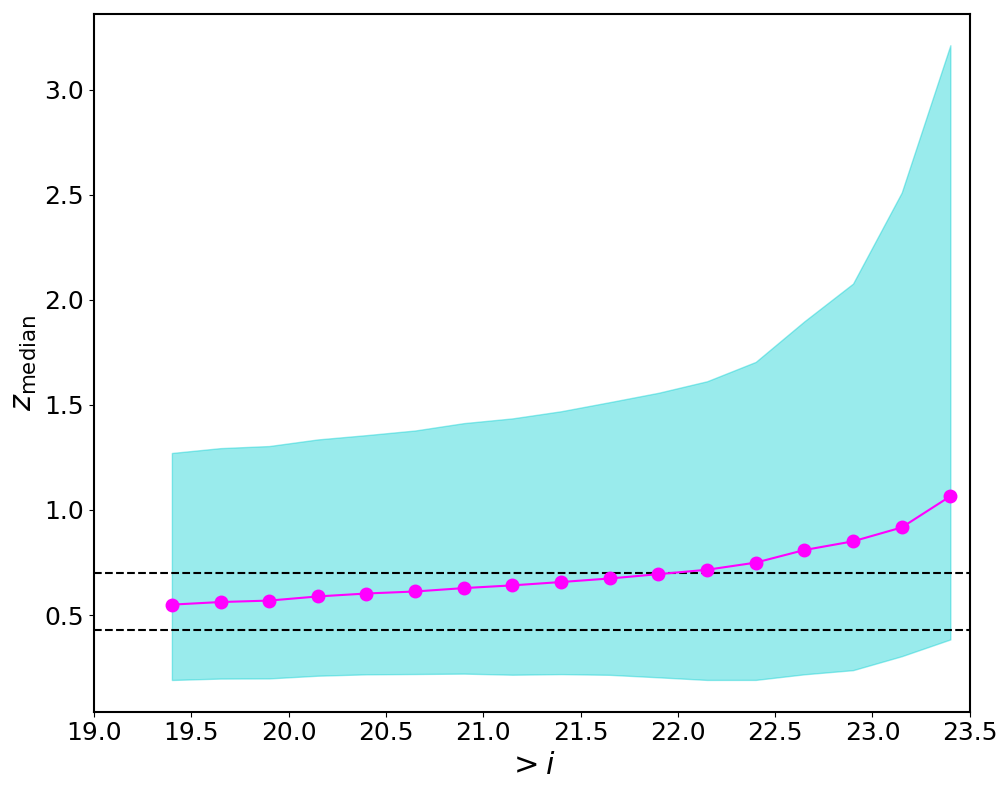}
	\caption{\label{fig:imagzmed}The median redshift for the KiDS-$i$-800 redshift distribution, estimated using the SPEC method as a function of the lower $i-$band magnitude limit. The shaded range represents the lower and upper quartile spread of the distribution. }   
\end{figure}
In section~\ref{sec:zdata} we introduce a selection on $i$-band magnitude in order to increase the average redshift of the sample.  Figure~\ref{fig:imagzmed} shows the median redshift of the distribution, estimated using the SPEC method (section~\ref{sec:specz}), as a function of the lower limit $i-$band magnitude. The turquoise shaded region corresponds to the the upper and lower quartile of the distribution. Dashed lines indicate the maximum redshift for the LZ and HZ lens samples, $z_l < 0.43$ and $z_l < 0.7$. For KiDS-$r$-450, sources are selected to be behind each lens slice according to their photometric redshift with $z_{\rm B}>z_l+0.1$. As this was not possible for KiDS-$i$-800, Figure~\ref{fig:imagzmed} was used to optimise the $i$-band magnitude selection.  In order to achieve a sample with a median redshift that was deeper than the maximum lens redshift, galaxies were selected to have $i>20.8$ for the analysis of the higher redshift HZ lenses. The \emph{lens}fit selection criteria of $i>19.4$, see Section~\ref{sec:selection}, is shown to be suitable for the analysis of the low redshift LZ lenses.

\section{Comparison of selected KiDS stars with Gaia DR1}
\label{app:Gaia}
The {\it Gaia} mission is on course to create a fully three-dimensional map of the objects in our Milky-Way galaxy, measuring accurate $G$-band photometry, astrometry and proper motion for almost a billion point-like sources.  In this appendix we use data from the first year of observations \citep[DR1;][]{Gaia/etal:2016} to explore the levels of galaxy contamination in the star catalogue that is used to create PSF models in Section~\ref{sec:psf}.    We match the {\it Gaia}-DR1 source list with KiDS object catalogues using a \SI{0.2}{arcsec} search radius.  We do not apply any filtering to the {\it Gaia} catalogue as problematic cases and spurious sources have already been removed from the sample \citep{Gaia/etal:2016}.   This first-look at the {\it Gaia} data shows that, once fully complete, {\it Gaia} will provide a very promising resource for future accurate PSF modelling.

\begin{figure}
\begin{tabular}{ll}
  	\includegraphics[width=0.48\columnwidth]{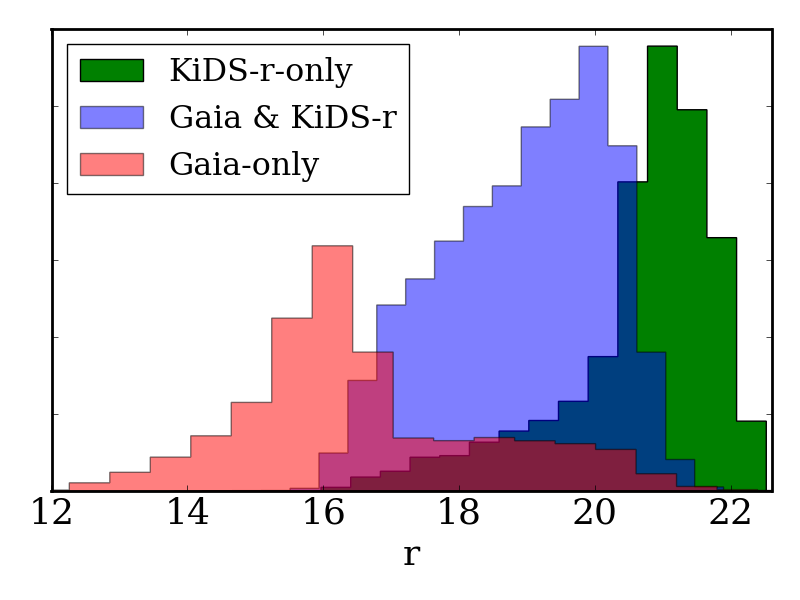} &
          \includegraphics[width=0.48\columnwidth]{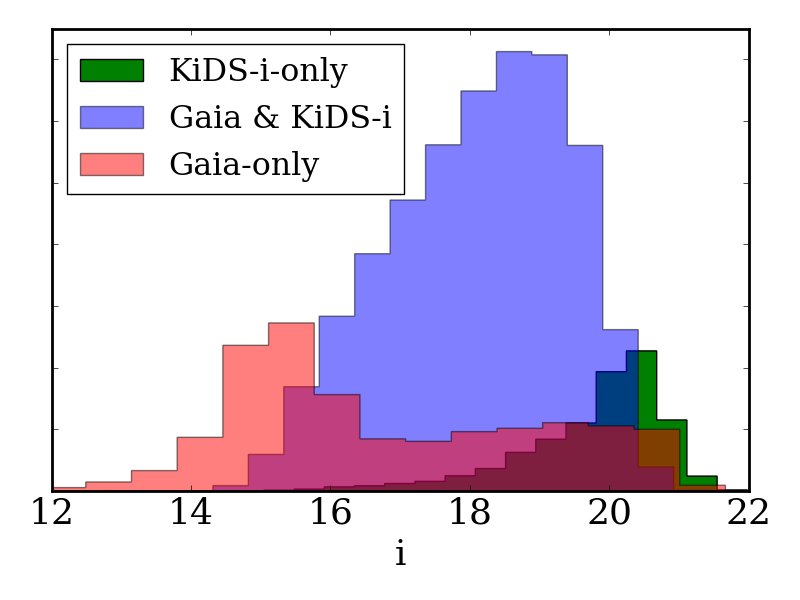}
\end{tabular}
 \caption{\label{fig:starhist}The magnitude distribution of three different stellar samples as a function of their KiDS-measured magnitude, $r$-band (left) and $i$-band (right).  In red, {\it Gaia} objects that are not used to determine the KiDS PSF model.   In blue, {\it Gaia} objects that are used in the KiDS PSF model.  In green, objects that are used to determine the KiDS PSF model, but are not present in the {\it Gaia} catalogues.}
\end{figure}

Figure~\ref{fig:starhist} compares the magnitude distribution of three different samples as a function of their KiDS-measured magnitude, $r$-band (left) and $i$-band (right).  The brightest sample contains {\it Gaia} objects that are not used to determine the KiDS PSF model (shown in red).   The majority of these objects are saturated in the KiDS-imaging.   The faintest sample contains objects that are used to determine the KiDS PSF model, but are not present in the {\it Gaia} catalogues (shown in green).  The majority of these objects are too faint to be detected given the {\it Gaia} flux limit $G<20$.  The centre sample (shown in blue) are {\it Gaia} objects that are used in the KiDS PSF model.   Where the objects have been used in the KiDS PSF model,  we present a weighted distribution determined from the weight given to each object in the PSF model.  Comparing the results for KiDS-$r$-450 (left) and KiDS-$i$-800 (right) we find that the star catalogue is significantly deeper in the $r$-band, as expected given the exposure time.  Both star catalogues, however, show a tail of brighter objects that are not detected by {\it Gaia} but are within the nominal {\it Gaia} magnitude limits.   Both also show a tail of fainter {\it Gaia} objects that are not used in the PSF modelling but are within the KiDS saturation limits.

\begin{figure}
	\includegraphics[width=\columnwidth]{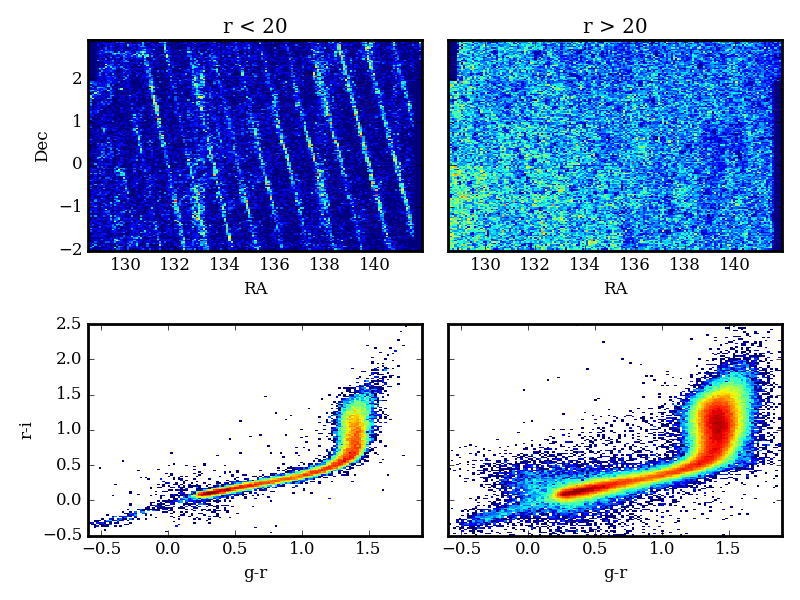}
    \caption{\label{fig:LFnotGaia}Upper panels: The on-sky distribution of the KiDS-$r$-only stars for the `G9' field, split by magnitude, with $r<20.5$ (left) and $r>20.5$ (right), revealing the {\it Gaia}-DR1 scanning pattern.   Lower panels: $(r-i) - (g-r)$ colour-colour diagrams that follow the expected stellar locus.  In all panels the colour scale increases from dark blue (few objects) through to red.  The upper panels use a linear scale, the lower panels use a log-scale.}
\end{figure}

Figure~\ref{fig:LFnotGaia} shows the on-sky distribution of the KiDS-$r$-only stars for the `G9' field, split by magnitude with a bright sample ($r<20.5$, shown left) and a faint sample ($r>20.5$, shown right). 
\am{In this first {\it Gaia} data release the scanning pattern is such that some stripes of sky have been visited more often than others.  This is seen in the left panel of Figure~\ref{fig:LFnotGaia} where the variable on-sky incompleteness is reflected as stripes in the faint star number density.  On the completion of Gaia, the survey depth will be homogenous.}
For the faint sample, in contrast, the on-sky distribution is relatively smooth.  The same patterns are found when inspecting the on-sky distribution of the KiDS-$i$-only stars.   The lower panels of Figure~\ref{fig:LFnotGaia} show that the faint and bright samples from the selected KiDS-$r$-only stars closely follow a stellar-locus in this colour-colour diagram.  We therefore conclude that our star selection is robust and not subject to significant galaxy contamination.   

We investigated the {\it Gaia} sources that were not used as input to constrain our PSF model.    We carried out a visual inspection of {\it Gaia} objects with a KiDS $r$-band magnitude that was significantly brighter than the {\it Gaia} $G$-band magnitude, with $r-G > 0.5$.   This revealed that $\sim 7\%$ of the {\it Gaia} objects not used in the PSF modelling were actually bright galaxies where {\it Gaia} had only resolved the core.   This galaxy contamination contributes at the level of  $\sim 3\%$ to the full {\it Gaia}-DR1 catalogue.   Other {\it Gaia} sources were flagged as unusable in KiDS through image defects, but this effect does not account for the full missing {\it Gaia} sample in our KiDS star catalogue.  This suggests that selection bias could have been introduced during our star selection.  We note that the fraction of unsaturated {\it Gaia} sources not used in the PSF modelling is higher for our $i$-band (14\%) compared to our $r$-band (10\%).

In principle, {\it Gaia} could also be used to determine the level of stellar contamination in our galaxy sample.  For our KiDS sample we impose a bright magnitude cut at $r>20.0$ and $i>19.4$ owing to limitations in the accuracy of shear calibration correction in this regime \citep[see][]{fenech-conti/etal:2016}.   As Figure~\ref{fig:starhist} therefore shows, the overlap between this galaxy sample and the {\it Gaia} depth is minimal and so what we can learn about stellar contamination of our galaxy catalogue, from {\it Gaia} is limited.

\section{Corrections to the galaxy-galaxy lensing signal}
\label{app:gglcorr}
\begin{figure}
	\includegraphics[width=\columnwidth]{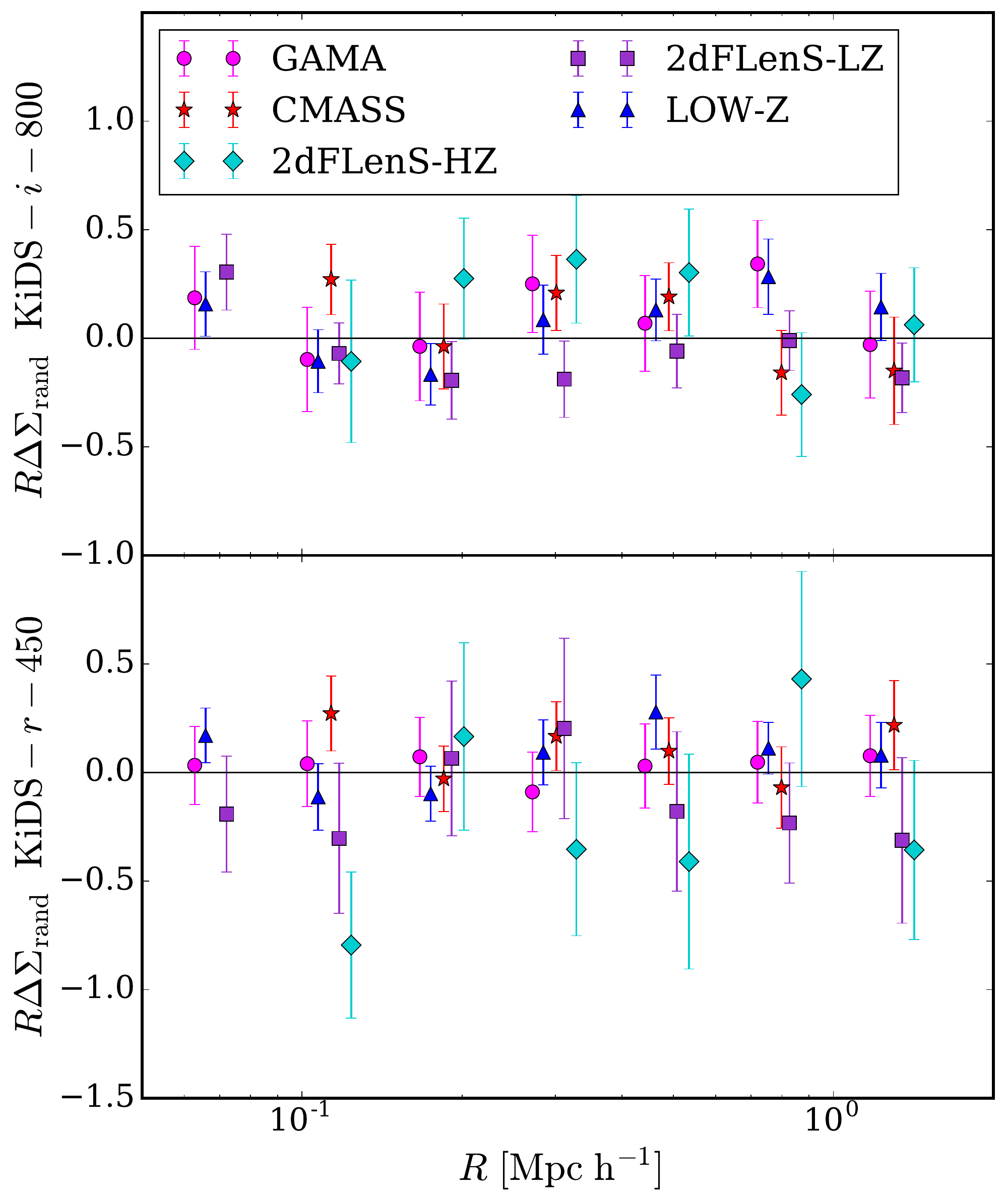}
    \caption{\label{fig:rands}The random galaxy-galaxy lensing signals $\Delta \Sigma(R)_{\rm rand}$ for KiDS-$i$-800 (upper) and KiDS-$r$-450 (lower) for each of the lens samples. The errors show the standard error on the mean of the forty random signals computed.}
\end{figure}
In this appendix we present details of the `random signal' and `boost correction' that are used to correct for errors arising from additive bias, sampling variance and lens-source clustering in our galaxy-galaxy lensing measurements (see equation~\ref{eqn:dscorr}).

Random lens catalogues are created for each spectroscopic sample by preserving the redshift distribution of the lenses but replacing their positions with random points generated with the angular mask of the survey area. Galaxy-galaxy lensing measurements are then determined for each source sample using 40 independent random catalogues for each lens sample. These `random signals', $\Delta \Sigma_{\rm rand}$, are presented for each spectroscopic sample in Figure~\ref{fig:rands} using the KiDS-$i$-800 source galaxies in the upper panel and KiDS-$r$-450 in the lower. The error bars represent the error on the mean of the signal from 40 realisations of the signals and show the `random' signal to be consistent with zero.   We still correct our galaxy-galaxy lensing measurements with this signal, however, as \citet{Singh2016} has shown, this correction reduces sampling variance errors.  The error on the mean random signal is propagated through to the final error on the corrected $\Delta \Sigma (R)^{\rm corr}$ measurement. 
\begin{figure}
	\includegraphics[width=\columnwidth]{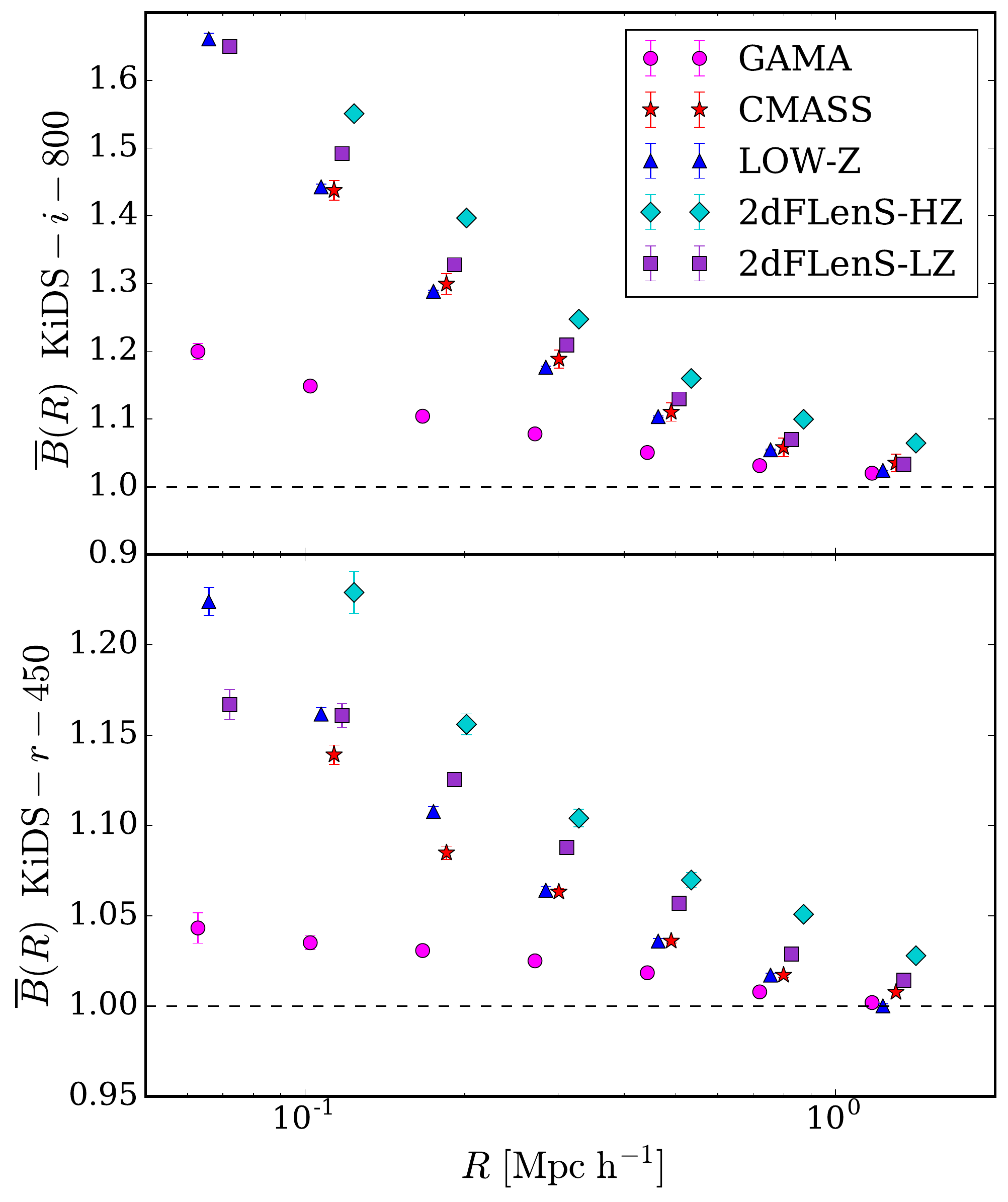}
    \caption{\label{fig:boosts}The boost factors for KiDS-$i$-800 (upper) and KiDS-$r$-450 (lower) that account for a dilution of the galaxy-galaxy lensing signal due to sources being associated with the lenses. The errors represent the standard error on the mean of the ten realisations computed and are consistent with the size of the data-points.}
\end{figure}

The boost factors $\overline{B}({R})$, are computed as a function of projected separation as an average over each of the lens slices, for each of the measurements according to equation~\ref{eqn:boost} and shown in Figure~\ref{fig:boosts}.  Contamination by sources that are associated with the lens biases the lensing signal low by a factor that is equal to the overdensity of sources around the lenses; hence we can correct the lensing signal by multiplying it by the boost factor.  We make 10 independent measurements of the boost factor for each lens sample with both KiDS-$i$-800 and the KiDS-$r$-450 using unique realisations of the random lens catalogue. The mean of these are plotted in Figure~\ref{fig:boosts} with the error on the mean of the ten realisations represented by the errorbars.  On scales $R< 2 h^{-1} {\rm Mpc}$ we find boost signals $\overline{B}(R) > 1$ showing that the source sample on these scales is contaminated by galaxies that are associated with the lenses. Note that as the errors on the boost factors are small, we do not propagate these errors through to the final error on the corrected $\Delta \Sigma (R)^{\rm corr}$ measurement. 

As expected, the corrections for the KiDS-$i$-800 galaxies, shown in the upper panel, are larger than those of KiDS-$r$-450, in the lower panel.  By limiting KiDS-$r$-450 lensing galaxies to only those with a photometric redshift behind each lens slice by $z_B>z_l+0.1$ we decrease the overall number of sources associated with the lens and therefore lower the boost correction, compared to that of KiDS-$i$-800, where this photometric redshift selection is not possible. However, as redshift distributions for the source redshifts are still broad, this `boost factor' is still non-negligible for KiDS-$r$-450. For both lensing samples, the boost factors are highest for the high redshift and then low redshift 2dFLenS galaxies. As was the case for the galaxy-galaxy lensing measurements shown in Figure \ref{fig:ds}, these lens samples contain the largest fraction of LRG galaxies, which have a higher galaxy bias and therefore the  greatest possible association compared to the BOSS samples. For both 2dFLenS and BOSS, the high redshift lens samples have a greater overlap with the redshift distributions of the source galaxies and so a higher fraction of the source sample will be associated with the lenses. For the KiDS-$i$-800 sample, the boost factor is as high as 50\% at $0.1 h^{-1}$ Mpc and for the KiDS-$r$-450 galaxies the boost correction is at the level of 20\%. In both cases, the corrections taper to one beyond $2.0 h^{-1}$ Mpc.

\begin{figure}
	\includegraphics[width=\columnwidth]{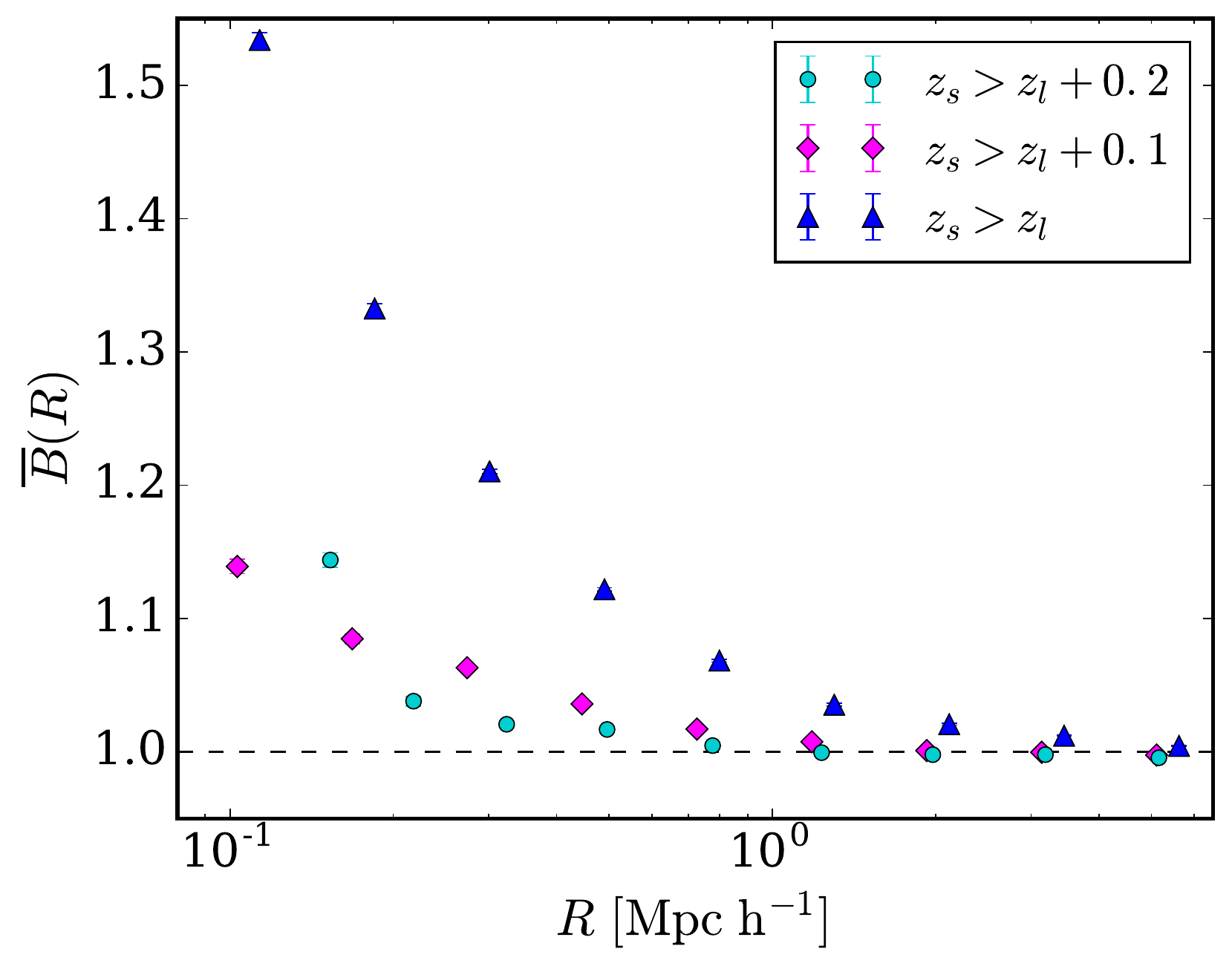}
    \caption{\label{fig:boosttest}The measured boost factors for KiDS-$r$-450 with different source galaxy selections. More stringent limitations of sources to those behind the lenses reduce the number of sources associated with the lenses, thereby reducing the boost factor. The errors represent the standard error on the mean of the ten realisations computed and are consistent with the size of the data-points. }
\end{figure}

For the KiDS-$r$-450 measurements, we investigated the optimal source redshift cut to minimise the contamination of galaxies that are physically associated with the lenses. While this contamination is entirely accounted for by an accurate boost factor, the errors are also inflated by this factor. On the other hand, the contamination of source galaxies associated with the lens can be further suppressed by applying more aggressive cuts to the source sample, but this also removes real source galaxies and undesirably decreases the lensing signal-to-noise ratio. Figure~\ref{fig:boosttest} compares the boost factor measured around CMASS galaxies for our fiducial galaxy selection, with a less and more stringent $z_{\rm B}$ selection. Limiting source galaxies to $z_B>z_l$ more than doubles the level of contamination, resembling the boost factors computed for the KiDS-$i$-800 galaxies.

%\appendix{Angular dependence of the fractional difference between the lensing signals}\label{app:GGLfrac}

%\begin{figure}
%	\includegraphics[width=\columnwidth]{Figs/deltasigma_fracdiff_to2.png}
 %   \caption{\label{fig:dsdiff}The fractional difference between the stacked differential surface mass density signals of the KiDS-$i$-800 and KiDS-$r$-450 with the various lens samples. The errors are the propagated combination of the two jackknifed analyses.  }
%\end{figure}

%Figure~\ref{fig:dsdiff} highlights the fractional difference between the galaxy-galaxy lensing measurements using the KiDS-$r$-450 galaxies and the KiDS-$i$-800 galaxies using the redshift estimation derived from the spectroscopic catalogue for all of the spectroscopic galaxy samples. \am{CH: maybe we should say we show this particular one because it's the most discrepant with r band? But to do that I need to average the frac diff over all lens samples...}. The errorbars represent a combination of the Jackknifed errors from the two measurements. This quantity, over all angular scales reveals no significant angular dependence of the difference between the two datasets.

\section{Analytical Covariance for the `nulled' two-point shear correlation function}
\label{app:covderiv}
We model the observed ellipticity in terms of a number of components (equation~\ref{eq:eobs}), which for $\gamma << 1$, in the absence of systematic errors, can be written as
\be
\epsilon^{\mathrm{obs}} = \gamma+\epsilon^{\mathrm{int}} + \epsilon^{\mathrm{n}} \, ,
\ee
where $\epsilon^{\mathrm{int}}$ is the galaxy's intrinsic ellipticity, $\epsilon^{\mathrm{n}}$ is the random noise on the measured galaxy ellipticity which will increase as the signal-to-noise of the galaxy decreases,  and $\gamma$ is the true lensing-induced shear that we wish to detect.  
With this model, the shear correlation function $\xi_\pm$ (equation~\ref{eqn:xipm}) can be expanded into a series of terms
\be
\xi_{\rm tot} = \xi_{\gamma \gamma} + \xi_{\rm int \, int} + \xi_{\rm nn} + 2 \xi_{\rm n \, int} + 2 \xi_{\rm n\gamma} + 2 \xi_{\rm \gamma \, int} \, ,
\ee
where the subscripts indicate the correlation between the different components that contribute to the observed ellipticity measurements.
The two `nulled' shear correlation functions in equations~\ref{eqn:null} and~\ref{eqn:xnull} can be written in this form with
\ba
\xi_{\rm tot}^{\rm null} &=& \xi_{\rm tot}^{ii}  -\xi_{\rm tot}^{rr}  \\
                                       &=& \xi_{\rm nn}^{ii}  -\xi_{\rm nn}^{rr} + 2(\xi^{i}_{\rm n \, int}  - \xi^{r}_{\rm n \, int}) \, , \\
\xi_{\rm tot}^{\rm x-null} &=& \xi_{\rm tot}^{ir}  -\xi_{\rm tot}^{rr}  \\
                                          &=& \xi_{\rm nn}^{ir}  -\xi_{\rm nn}^{rr} + \xi^{i}_{\rm n \, int}  - \xi^{r}_{\rm n \, int} \, ,
\ea
where the superscripts $r$ and $i$ indicate the combination of filters that are being analysed.  In this derivation we have assumed that terms which include the correlation between measurement noise and shear are sufficiently small that they can be ignored.  We also assume that the intrinsic ellipticity  $\epsilon^{\mathrm{int}}$ does not change significantly between the two filters \citep[as shown by][]{jarvis/jain:2008}. With these assumptions we see that all terms that include the shear cancel, leaving only intrinsic and shape measurement noise terms.  

The auto-noise covariance matrix for the shear correlation function is diagonal with 
\be
C_{\rm nn}(\theta_j, \theta_j) = \frac{4\sigma_{\rm n}^4}{N_p(\theta_j)} \, ,
\ee
where $\sigma_{\rm n}$ is the per component ellipticity dispersion of $\epsilon^{\mathrm{n}}$, and $N_p(\theta_j)$ is the number of pairs in angular bin, $j$, (see equation~\ref{eqn:Np}), derived analogously to \citet{Schneider/etal:2002}. The cross-covariance for the noise between the two filters is also diagonal with
\be
C_{\rm nn}^{ir}(\theta_j, \theta_j) = \frac{2\sigma_{\rm{ n},i}^2\sigma_{\rm{ n},r}^2}{N_p(\theta_j)} \, ,
\ee
where we assume that the shape measurement noise is independent of the intrinsic shape of the galaxy and hence is uncorrelated between bands.  Under this assumption, the cross covariance between the shape noise and intrinsic ellipticity noise is also diagonal with
\be
C_{\rm n\, int}(\theta_j, \theta_j) = \frac{2\sigma_{\rm n}^2 \sigma_{\rm int}^2}{N_p(\theta_j)} \, .
\ee
With an educated guess for the value of the intrinsic ellipticity dispersion $\sigma_{\rm int}$, we can calculate $\sigma_{\rm n}$ from the observed ellipticity dispersion $\sigma_{\epsilon}$ (equation~\ref{eq:sige}) as $\sigma_{\epsilon}^2 = \sigma_{\rm n}^2 + \sigma_{\rm int}^2$. We use the notation of $\sigma_i$ for the measured ellipticity dispersions in the $i$-band images and $\sigma_r$ for that of the $r$-band.  We can then derive an analytical covariance for the two `nulled' shear correlation functions with
\ba
C^{\rm null}_{\xi}(\theta_j,\theta_j) &=& C_{\rm nn}^{rr} +  C_{\rm nn}^{ii} + 4(C_{\rm n\, int}^r + C_{\rm n\, int}^i)\\
                                                             &=& \frac{4}{N_p(\theta_j)}(\sigma^4_i + \sigma^4_r - 2 \sigma^4_{\rm int}) \, ,\\
C^{\rm x-null}_{\xi}(\theta_j,\theta_j) &=& C_{\rm nn}^{rr} +  C_{\rm nn}^{ir} + C_{\rm n\, int}^r + C_{\rm n\, int}^i\\
                                                                 &=& \frac{2}{N_p(\theta_j)}[2\sigma^4_r + \sigma^4_{\rm int} + \sigma^2_r (\sigma^2_i-4\sigma^2_{\rm int})] \, ,
\ea
where again the superscripts $r$ and $i$ indicate the combination of filters that are being analysed.

% Don't change these lines
\bsp	% typesetting comment
\label{lastpage}
\end{document}